\documentclass[12pt,aip,jcp,showpacs,floatfix,nofootinbib,preprintnumbers,superscriptaddress,amsmath,amssymb]{revtex4-2}
\usepackage{graphicx}
\usepackage{dcolumn}
\usepackage{bm}
\usepackage{xcolor}
\usepackage{url}
\usepackage{verbatim} 
\usepackage{caption}
\usepackage[utf8]{inputenc}
\usepackage[T1]{fontenc}
\usepackage{mathptmx}
\usepackage{etoolbox}
\usepackage{appendix}
\usepackage{enumerate}
\usepackage{orcidlink}
\usepackage{subcaption}
\usepackage{dsfont}
\usepackage{bm}

\definecolor{myred}{rgb}{0.5, 0, 0.13}
\definecolor{mygreen}{rgb}{0.13, 0.55, 0.13}
\definecolor{myblue}{rgb}{0, 0, 1.0}
\definecolor{mycor}{rgb}{1, 0.6, 0}
\definecolor{mycom}{rgb}{1, 0.4, 0}
\definecolor{mycol}{rgb}{1, 0, 0.5}

\definecolor{myow}{rgb}{0.55, 0.2, 0.65}
\definecolor{mymh}{rgb}{0.0, 0.55, 0.65}

\usepackage{amsthm} 
\newcommand{\NZIAS}{Centre for Theoretical Chemistry and Physics,
The New Zealand Institute for Advanced Study, Massey University Auckland, Private Bag 102904, 0745 Auckland, New Zealand}

\newcommand{\GAC}{German Aerospace Center (DLR), Institute of Software Technology, High-Performance Computing Department, Linder H\"ohe, Cologne, 51147, Germany}
\newcommand{\DM}{Department of Mathematics, Saarland University, D-66041 Saarbr\"ucken, Germany}

\newcommand{\Porelab}{Porelab, Department of Chemistry, Norwegian University of Science and Technology, NO-7491 Trondheim, Norway}
\newcommand{\thermophys}{Thermophys, NO-7034 Trondheim, Norway}

\newcommand{\ETH}{Department of Mathematics, ETH Z\"urich, R\"amistrasse 101, CH-8092 Z\"urich, Switzerland}
\newcommand{\LPL}{Laboratoire de Physique des Lasers, Université Paris 13, Sorbonne Paris Cité, CNRS, 99 Avenue Jean-Baptiste Clément, Villetaneuse, 93430, France}

\begin{document}


\title{Pressure-Dependent hcp-to-fcc Phase Transition Path Derived from Exact Lattice Summations over Inverse Power Law Potentials}

\author{Tage W. Maltby\,\orcidlink{0000-0002-0290-1713}}
\affiliation{\Porelab}

\author{Andres Robles-Navarro\,\orcidlink{0000-0001-9586-1003}}
\affiliation{\NZIAS}

\author{Jonathan K. Hartlaub\,\orcidlink{0009-0001-3323-3455}}
\affiliation{\DM}
\affiliation{\GAC}

\author{Andreas A. Buchheit$^*$\,\orcidlink{0000-0003-4004-713X}}
\affiliation{\DM}
\affiliation{\ETH}

\author{{\O}ivind Wilhelmsen\,\orcidlink{0000-0003-4631-0349}}
\affiliation{\Porelab}
\affiliation{\thermophys}

\author{Morten Hammer,\orcidlink{0000-0002-0921-7527}}
\affiliation{\Porelab}
\affiliation{\thermophys}

\author{Peter Schwerdtfeger\,\orcidlink{0000-0003-4845-686X}}\email[Corresponding authors: ]{peter.schwerdtfeger@gmail.com, tagemaltby@gmail.com, andres.robles.n@gmail.com, andreas.buchheit@uni-saarland.de}
\affiliation{\NZIAS}
\affiliation{\LPL}


\date{\today}


\begin{abstract}
We investigate the phase transition between the two close-packed crystal structures, hexagonal close-packed (hcp) and face-centered cubic (fcc), along a Burgers–Bain transformation path at pressures ranging up to several hundred gigapascals. To this end, we employ a hierarchy of interatomic interaction models, from simple $(n,m)$-Lennard–Jones potentials to more sophisticated descriptions incorporating expansions in terms of inverse power law potentials, namely the extended Lennard–Jones two-body interaction, and the Axilrod–Teller–Muto three-body term, together with density functional theory (DFT) calculations for solid argon. For inverse power-law potentials, the free energy is computed using exact lattice summations combined with efficiently computable integral representations based on generalizations of the Epstein zeta function to anisotropic interaction kernels, followed by optimization of the lattice parameters. This methodology enables an accurate determination of the pressure-dependent Burgers–Bain minimum-energy path and the associated activation barrier for the phase transition. The resulting transition pathway exhibits a pronounced symmetry-breaking feature connecting the Burgers with the Bain path that is confirmed by subsequent DFT calculations. At high pressures, the activation barrier for the hcp-to-fcc transformation increases approximately linearly with pressure due to the dominant pressure-volume term. Taking entropy effects into account, we get excellent agreement with experimental results for the equation of state of the temperature-dependent volume $V(T)$ up to the melting point if anharmonicity effects are accounted for. The DFT results further highlight the growing importance of many-body interactions under compression: unlike two-body potentials, which predict that the transformation path passes through the body-centered cubic (bcc) structure only at very high pressures, DFT shows that the bcc configuration is generally bypassed. Inclusion of the Axilrod–Teller–Muto three-body term slightly corrects this discrepancy. On the other hand, DFT performs far better at higher pressures where dispersion forces are less dominant.
\end{abstract}


\maketitle 


\section{Introduction}
\label{sec:intro}

Owing to its simplicity and computational efficiency, the (12,6)-Lennard-Jones (LJ) potential is the most widely employed interaction potential in molecular dynamics (MD) simulations of extended or bulk systems.\cite{Wales2024} In its most general form, the interaction between two atoms $i$ and $j$ in a monatomic cluster or infinite system is described by the $(n,m)$-LJ potential\cite{Gruneisen1912,Jones-1924b,Jones-1925} (also known as the Gr\"uneisen-Mie potential,\cite{Mie1903,Gruneisen1912})
\begin{equation} \label{eq:VLJ}
\phi_{\textrm{LJ}}(r_{ij})=\frac{\epsilon nm}{n-m} \;  \left[ \frac{1}{n}\left( \frac{r_e}{r_{ij}} \right)^{n} - \frac{1}{m} \left( \frac{r_e}{r_{ij}} \right)^{m} \right] 
\end{equation}
with $n,m\in\mathbb{R}, n>m$ and $m>3$ for physically relevant systems,\cite{Schwerdtfeger-2006} $\epsilon$ and $r_e$ are usually taken as the binding energy and the equilibrium distance for the diatomic respectively, and $\sigma=r_e(m/n)^{1/(n-m)}$ is the zero-crossing distance. The interaction potential can easily be extended to mixed atomic systems. Both the Sutherland potential\cite{Sutherland1893} $(n\rightarrow\infty)$ and the (kissing) hard-sphere approximation\cite{Baxter1968,Stell1991} $(n,m\rightarrow\infty, n>m)$ can be obtained as limiting cases from the LJ potential.

The two-body LJ potential belongs to the broader class of many-body inverse power law potentials. It can be further generalized to the extended Lennard-Jones (ELJ) potential\cite{Schwerdtfeger-2006}
\begin{equation} \label{eq:ExtVLJ}
\phi_{\textrm{ELJ}}(r_{ij})= \sum_{n=1}^{n_\text{max}}C_n\left( \frac{r_e}{r_{ij}} \right)^{s_n}
\end{equation}
with the appropriate boundary conditions for the location of the potential energy minimum $r_e$ and for the binding energy $\epsilon$, and we have $s_n>3$ to avoid divergence (or $s_n\ge 6$ to correctly describe the long-range dispersion behavior). This partially corrects the inaccurate behavior of the LJ potential in the repulsive region and also provides an improved description of the attractive region.

Beyond the two-body approximation, the most widely used long-range three-body correction is the Axilrod–Teller–Muto (ATM) inverse power expression,\cite{AxilrodTeller1943,Muto1943}
\begin{align}\label{eq:ATM}
\phi_\text{ATM}^{(ijk)}&=  c_\text{ATM}f_{\theta}r_g^{-9} \\
\nonumber
\textrm{   with   } f_{\theta}&=\left(1+3\cos\theta_i\cos\theta_j\cos\theta_k\right) \textrm{   and   } r_g=\left(r_{ij}r_{ik}r_{jk}\right)^{1/3} .
\end{align}
where $r_{ij}$ is the distance between atoms $i$ and $j$,
$r_{jk}=|\vec{r}_{ij}-\vec{r}_{ik}|$, and $\theta_i$ the corresponding angle between the vectors $\vec{r}_{ij}$ and $\vec{r}_{ik}$. 
To more clearly illustrate its relation to inverse power law potentials, we rewrite the ATM potential in the following form,\cite{Zucker_1968}
\begin{align}\label{eq:ATM2}
f_{\theta}r_g^{-9} = r_{ij}^{-3}r_{ik}^{-3}r_{jk}^{-3}
+ \frac{3}{8}C_{ij}C_{ik}C_{jk}r_{ij}^{-5}r_{ik}^{-5}r_{jk}^{-5}
\end{align}
with
\begin{align}\label{eq:ATM2a}
\nonumber
C_{ij} = r_{ik}^{2}+r_{jk}^{2} - r_{ij}^{2}
\end{align}
In MD simulations of infinite systems with periodic boundary conditions, such as phase-transition simulations, the LJ potential suffers from slow convergence of the infinite inverse power sums appearing in the total energy term. This is evident, for example, from melting simulations using LJ potentials.\cite{Mastny2007} The slow-convergence issue is commonly addressed by introducing a long-range cut-off radius $r_c$ for the summation over inverse distances, or by using the cut-and-shift Lennard-Jones (LJTS) potential,\cite{AllenTildesley1987}
\begin{equation} \label{eq:LJTS}
\phi_{\textrm{LJTS}}(r)= \phi_{\textrm{LJ}}(r) - \phi_{\textrm{LJ}}(r_c) \quad \textrm{for} \quad r<r_c \quad \textrm{and} \quad \phi_{\textrm{LJTS}}(r)= 0 \quad \textrm{otherwise}.
\end{equation}
This approach has the undesirable feature of introducing an additional, unphysical parameter that influences the thermodynamic properties.\cite{Wiebe2020} In the case of the LJTS potential, it also leads to discontinuities\cite{Toxvaerd2011} in the energy derivatives at $r=r_c$.

John Lennard-Jones and Albert E. Ingham \cite{Jones-1925} already demonstrated in 1925 that, for a cubic lattice, the slowly converging three-dimensional lattice sums, obtained from summing over all interactions in the lattice by applying periodic boundary conditions, associated with inverse power law potentials belong to the broader class of Epstein zeta functions,\cite{Epstein-1903}
\begin{equation}\label{eq:Epstein1}
Z(s;A,\vec{v},\vec{k}) = \,\sideset{}{'}\sum_{\vec n\in \mathbb{Z}^d}\frac{e^{-2\pi i  A\vec{n}\cdot \vec k}}{\vert A\vec{n}-\vec{v}\vert^s},\
\end{equation}
with $A\in \mathbb R^{d\times d}$ positive definite, $s\in\mathbb{C}$, a shift vector $\vec{v}\in \mathbb R^d$, and a wave vector $\vec{k}\in \mathbb R^d$. The primed sum excludes the case $A\vec n=\vec v$.

The generator matrix $B$ is related to $A$ through $A=B^\top$, while atoms within the elementary lattice cell are described by fractional coordinates $\vec{\eta}$ through $\vec{v}=B^\top\vec{\eta}$. Lennard-Jones and Ingham were the first to employ Bessel-function expansions to transform the originally slowly convergent three-dimensional lattice sums for certain cubic lattices into rapidly convergent series.\cite{Jones-1925} For general lattices, such as those describing atoms in a simulation box, the total internal energy (neglecting zero-point vibrational contributions) can be expressed as a linear combination of Epstein zeta functions for inverse power law potentials. An efficient evaluation of the Epstein zeta function is therefore essential, yet this has remained a longstanding mathematical challenge, particularly for general lattices and in higher dimensions.\cite{borwein-2013,Burrows2022x}

Audrey Terras \cite{terras-1973} introduced a general method for treating Epstein zeta functions in $d$ dimensions when the shift vector $\vec{v}$ in Eq.\eqref{eq:Epstein1} vanishes, resulting in lengthy expressions involving rapidly convergent Bessel-function series. More recently, the Massey group in  Auckland developed expansions into rapidly convergent Bessel-function series for a number of lattices, including multi-lattices for Barlow structures,\cite{burrows-2020,burrows2021b,Schwerdtfeger2022jpc,Cooper2023,Cooper2025,Schwerdtfeger-2026} as well as lattices in higher dimensions.\cite{Burrows2022x} Crandall reformulated the Epstein zeta function as a Mellin transform and, by applying Riemann splitting to the resulting integral together with Poisson summation, expressed the result as two super-exponentially convergent series.\cite{crandall1998fast,crandall2012} More recently, Buchheit and Busse reformulated Crandall’s representation of the Epstein zeta function and developed an efficient evaluation algorithm to be used in computer simulations.\cite{Buchheit2026,Buchheit2026a} The method has very recently been extended to power-law interactions with general anisotropy.\cite{buchheitComputationAnisotropicSingular2026}

In our previous work, we demonstrated that the Burgers hcp-to-fcc phase transition is connected to the (cuboidal) fcc-to-bcc Bain path through a distinct bifurcation point,\cite{Schwerdtfeger-2026} where the original hcp bi-lattice becomes a cuboidal Bravais lattice, the latter connecting the bcc with the fcc phase. In this process, the bcc phase is either being bypassed or traversed depending on the hardness of the Lennard-Jones (LJ) potential. In the present work, we investigate the hcp-to-fcc transformation along the combined Burgers-Bain path\cite{Bain1924,Burgers1934,Schwerdtfeger-2026} at constant pressure ($P\ge0$) within the framework of inverse power-law potentials. To achieve near machine-precision accuracy, we combine the method for computing Epstein zeta functions \cite{Buchheit2026} with the integral representation for many-body interactions \cite{Buchheit2026a} and extend it to multi-atomic lattices based on tensors involving anisotropic Epstein zeta functions,\cite{buchheitComputationAnisotropicSingular2026} yielding high-precision results for both two- and three-body interactions. We subsequently apply the resulting formalism to solid argon. The results are then compared with dispersion-corrected density functional theory calculations to assess how many-body interactions influence the shape of the Burgers-Bain path.

\section{Theory}

\subsection{The Burgers-Bain Phase Transition Path between hcp, bcc and fcc}
\label{sec:BurgersBain}

\begin{figure}[tb!]
\centering
\includegraphics[width=.33\columnwidth]{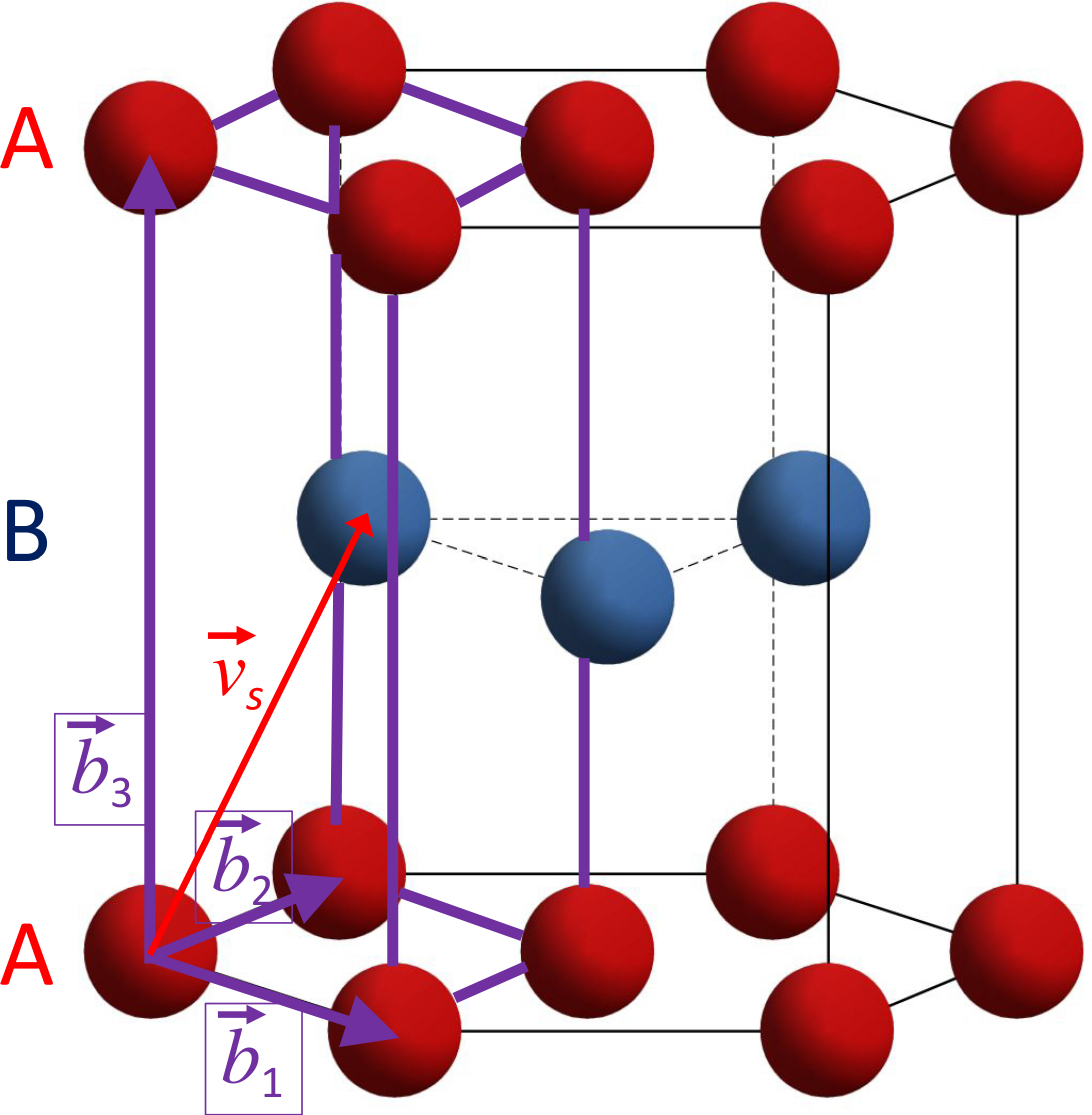}
\caption{The underlying primitive monoclinic cell in the Burgers transformation showing two distinct layers A and B (A in red and B in blue). The ideal hcp structure with a (AB)$_\infty$ sequence of hexagonal layers has the corresponding basis vectors and cell parameters $|\vec{b}_1|=|\vec{b}_2|=a$, $\angle(\vec{b}_1,\vec{b}_2)=60^\circ$ and $|\vec{b}_3|=c_3$ and $c_3/a=\sqrt{8/3}$ (cf. Ref.\onlinecite{Schwerdtfeger-2026}). For $c_3/a=\sqrt{2}$ and $\angle(\vec{b}_1,\vec{b}_2)=90^\circ$ we obtain the fcc lattice with sequence (ABC)$_\infty$, and for $c_3/a=1$ and $\angle(\vec{b}_1,\vec{b}_2)=90^\circ$ the bcc lattice.}
\label{fig:hcp}
\end{figure}

The Burgers–Bain path describes the transformation from a hcp structure to a cubic structure (fcc or bcc) within a common bi-lattice representation.\cite{Bain1924,Burgers1934,Schwerdtfeger-2026} For this purpose, we introduce a $(3 \times 3)$ generator matrix $B$, whose rows are formed by the row-stacked basis lattice vectors $\vec{b}_j$, along with an additional shift vector $\vec{v}$, so that any lattice point measured from the origin can be expressed as,
\begin{equation}
\vec{r}_{i_1i_2i_3}\left[B(\alpha),\vec{v}(\alpha)\right] = B^\top(\alpha)\vec{i}+\vec{v}(\alpha)
\end{equation}
where $\alpha\in[0,1]$ is our single phase-transition parameter and $\vec{i}\in\mathbb{Z}^3$. For convenience, we rotate the underlying unit cell such that the generator matrix $B$ assumes a triangular form, 
\begin{equation}\label{eq:Bmatrix}
B(\alpha) = 
\begin{pmatrix}
\vec{b}_1^\top \\
\vec{b}_2^\top \\
\vec{b}_3^\top
\end{pmatrix}
=
\begin{pmatrix}
a  & 0  &  0 \\
b\cos\theta_{12} &b\sin\theta_{12} &  0 \\
c_{1} & c_{2}  &  c_{3}
\end{pmatrix}
=
\begin{pmatrix}
a  & 0  &  0 \\
\frac{1}{2}b(1-\alpha)  & \frac{1}{2}b\sqrt{(1+\alpha)(3-\alpha)}  &  0 \\
c_{1} & c_{2}  &  c_{3}
\end{pmatrix}
\end{equation}
The two shift vectors required for the bilattice are given by
\begin{equation}
\vec{v}_1=\vec{0} \quad \text{and} \quad \vec{v}_2=\tfrac{a}{2}\vec{\beta},
\end{equation}
see Table \ref{tab:latticeparameters},
where the first vector positions the atoms at the corners of the unit cell, whereas the second positions an atom within the unit cell, corresponding to the middle layer as shown in Figure \ref{fig:hcp}. 
\begin{center}
\begin{table}[htbp] 
\setlength{\tabcolsep}{8pt}
\begin{tabular}{ |l|r|r|r|r| } 
 \hline
 Parameter                          & hcp                   & fcc           & bcc(1)        & bcc(2)\\ 
 \hline 
$\alpha$                            & $0$                   & $1$           & $1$           & $\frac{1}{3}$ \\
$\gamma_1=\frac{b}{a}$              & $1$                   & $1$           & $1$           & 1 \\
$c_1=c_2$                           & 0                     & 0             & 0             & 0 \\
$\gamma_2=\frac{c_3}{a}$            & $\sqrt{\frac{8}{3}}$  & $\sqrt{2}$    & $1$           & $\sqrt{\frac{8}{3}}$ \\
$(\vec \beta)_1$                           & $1$                   & $1$           & $1$           & $\frac{4}{3}$\\
$(\vec\beta)_2$                           & $\frac{1}{\sqrt{3}}$  & $1$           & $1$           & $\frac{2\sqrt{2}}{3}$\\
$(\vec\beta)_3$                           & $\sqrt{\frac{8}{3}}$  & $\sqrt{2}$    & $1$           & $\sqrt{\frac{8}{3}}$ \\
$\angle(\vec{b}_{1},\vec{b}_{2})$   & 60$^\circ$            & 90$^\circ$    & 90$^\circ$    & arccos$\left( \frac{1}{3}\right)$ \\
$V$                                 & $\sqrt{2}$            & $\sqrt{2}$    & $\frac{8}{3\sqrt{3}}$& $\frac{8}{3\sqrt{3}}$ \\
$\rho$                              & $\frac{\pi}{3\sqrt{2}}$& $\frac{\pi}{3\sqrt{2}}$& $\frac{\pi\sqrt{3}}{8}$& $\frac{\pi\sqrt{3}}{8}$ \\
\hline
\end{tabular}
\caption{Ideal lattice parameters for the three distinct lattices hcp, fcc and bcc with lattice constant $|\vec{b}_1|=a$, $|\vec{b}_2|=b$ and $|\vec{b}_3|=c_3$ as shown in Figure \ref{fig:hcp}. The values for the hard-sphere volumes $V$ for hcp, fcc and bcc (setting nearest neighbor distance to 1, i.e. $a=b=1$ for hcp, fcc and bcc(2) and $a=b=2/\sqrt{3}$ for bcc(1)) and corresponding packing densities $\rho$ are also shown. The values for the bcc structure occurring in the vicinity of the Burgers path at $\alpha=\tfrac{1}{3}$ are also shown as structure bcc(2). For more details see Ref.~\onlinecite{Schwerdtfeger-2026}.}
\label{tab:latticeparameters}
\end{table}
\end{center}

This definition of the $B$-matrix has the advantage that $|\vec{b}_2|=b$ and cos$\theta=\frac{1}{2}(1-\alpha)$, where $\theta$ denotes the angle between the vectors $\vec{b}_1$ and $\vec{b}_2$. For example, the hcp structure is described by an underlying hexagonal unit cell in which the atom in the middle layer is located at the fractional coordinates $\vec{\eta}=(\frac{1}{3},\frac{1}{3},\frac{1}{2})$, where $\vec{\eta}=A^{-1}\vec{v}_i$ with $A=B^\top$. The hcp phase belongs to the class of densest-packed Barlow multi-lattices (or $n$-lattices, for which there exist $n$ shift vectors $\vec v_1,...,\vec{v}_n$ including the zero vector).\cite{Barlow1883,Stillinger2019} This class includes the fcc lattice, which is the only Barlow lattice that can be represented by a single Bravais lattice. The common bi-lattice adopted here for hcp and fcc is the smallest unit cell which enables a smooth transformation between these two structures along the so-called Burgers phase transition path, which corresponds to a martensitic diffusionless transformation. In addition, we find that $c_1=c_2=0$ along the Burgers-Bain path (see discussion below), and therefore set $c=c_3$ in the following. The $B$-matrix used in this work then becomes
\begin{equation}\label{eq:Bmatrix2}
B(\alpha) = a
\begin{pmatrix}
1  & 0  &  0 \\
\frac{1}{2}\gamma_1(1-\alpha)  & \frac{1}{2}\gamma_1\sqrt{(1+\alpha)(3-\alpha)}  &  0 \\
0 & 0  &  \gamma_2
\end{pmatrix}.
\end{equation}
We recently showed that the $B(\alpha)$-matrix arises from a smooth linear transformation between the close packed fcc and hcp structures.\cite{Schwerdtfeger-2026} The lattice parameters corresponding to the critical points along the Burgers–Bain path are collected in Table \ref{tab:latticeparameters}.

The Gram matrix defined by $G=BB^\top$ must be positive definite, which implies $a>0$, $\alpha\in(-1,3)$ for $b>0$. The unit-cell volume is defined by 
\begin{equation}\label{eq:volume}
V_\text{UC} = |\text{det}(B)|= \sqrt{\text{det}(G)}= \frac{a^3\gamma_1\gamma_2}{2}\sqrt{(1+\alpha)(3-\alpha)},
\end{equation}
with the third condition of $V_\text{UC}>0$. Note that, for the $PV$ contribution to the Gibbs free energy, the volume must be taken per atom, such that $V=V_\text{UC}/2$. Within the kissing hard-sphere model, one readily verifies that the ideal hcp and fcc structures satisfy $V_\text{hcp}=V_\text{fcc}$ using the lattice parameters listed in Table \ref{tab:latticeparameters}, while $V_\text{bcc}/V_\text{fcc}=\frac{8}{3\sqrt{6}}=1.08866$.

Having established all the conditions required for the Burgers–Bain transformation,\cite{Schwerdtfeger-2026} we now introduce the various models and approximations employed to calculate the cohesive and free energies along the Burgers-Bain path. For a more general treatment of (Barlow) multi-lattices, we refer the reader to Refs.~\onlinecite{Stillinger2019,Cooper2025}. The formalism presented here can also be readily extended to higher dimensions in conjunction with analytical continuations of the Epstein zeta function.\cite{borwein-2013,Burrows2022x}

\subsection{The (\textit{n,m})-Lennard-Jones Potential and Corresponding Lattice Sums}

As a first approximation, we employ the $(n,m)$-LJ potential \eqref{eq:VLJ} for a mono-atomic system to obtain qualitative insight into the minimum-energy path associated with phase transitions under constant pressure.\cite{Gruneisen1912,Jones-1924b,Jones-1925} Within this model, we consider the zero-temperature limit, neglect vibrational contributions by adopting the infinite-mass approximation, and omit higher-order terms in the interaction potential of the solid. Furthermore, dimensionless units are employed by setting $r_e=1$ and $\epsilon=1$. Physical quantities are then recovered through the transformations $ar_e, E\epsilon, G\epsilon, Vr_e^3$, $P\epsilon/r_e^3$ and $T\epsilon/k_B$ (note that we used the experimentally accessible equilibrium distance $r_e$ instead of the zero-crossing distance $\sigma$ here for the definition of the reduced units). For solid argon, for example, highly accurate relativistic coupled-cluster calculations by Jaeger \textit{et al.}\cite{Jager_MolPhys107,Jaeger2011} yield $r_e=$3.7618 \AA~ and $\epsilon=$1.1900 kJ/mol (143 K) such that a reduced pressure of $P=1$ corresponds to 37.119$\times 10^6$ Pa.

In dimensionless units, the enthalpy $H$ (per atom) at constant pressure $P$ and the volume \eqref{eq:volume} for the lattice defined in the previous section is given by the following expression,\cite{Schwerdtfeger-2006,Smits2021}
\begin{align} \label{eq:GVLJ}
H(n,m;B,\vec{v};P)&=E_\text{coh}(n,m;B,\vec{v})+ E_\text{PV}(B) \\
\nonumber
&= \frac{1}{2(n-m)} \; \left[ {m}{L(n;B,\vec{v})}  - {n}{L(m;B,\vec{v})}  \right] + P \frac{abc}{4}\sqrt{(1+\alpha)(3-\alpha)},
\end{align}
where we give the explicit expression for the volume term per atom. We use the (experimentally observed low-temperature) fcc phase as our reference system,
\begin{equation} \label{eq:DeltaGibbs}
\Delta H(n,m;B,\vec{v};P)=H(n,m;B,\vec{v};P) - H_\text{fcc}(n,m;B,\vec{v};P).
\end{equation}

The lattice sums\cite{borwein-2013} $L(n,B,\vec{v})$ required for the calculation of the cohesive energy are members of the broader class of Epstein zeta functions \eqref{eq:Epstein}. For the present bi-lattice, this yields 
\begin{equation}\label{eq:LS}
L(n;B,\vec v_2) = Z(n;B^\top,\vec{0},\vec{0}) + Z(n;B^\top,\vec v_2,\vec{0})
\end{equation}
with the special Epstein zeta function (setting $\vec{k}=\vec{0}$ in Eq.\eqref{eq:Epstein1} and choosing $A=B^\top$ as the matrix argument)
\begin{align} \label{eq:Epstein}
Z(n;B^\top,\vec{v},\vec{0}) = {\sum_{\vec{i}\in\mathbb{Z}^3}}^{'} \big| B^\top\vec{i} + \vec{v} \big|^{-n} = {\sum_{\vec{i}\in\mathbb{Z}^3}}^{'} \big| B^\top\big(\vec{i} + \vec{\eta} \big)\big|^{-n},
\end{align}
choosing (without loss of generality) a positive sign in front of the shift vector $\vec{v}$ and excluding the self-interaction $\vec{i}+\vec{\eta}=0$ as indicated by the primed sum. The last expression comes from the fact that fractional coordinates $\eta\in [0,1)^3$ are used instead, and represents the general form for lattice sums adopted by Glasser and Zucker\cite{glasser-zucker-1980} if $\vec{k}\ne\vec{0}$ in Eq.\eqref{eq:Epstein1}. The Epstein zeta function can be evaluated to computer precision either through certain Bessel-function expansions \cite{Cooper2023,burrows-2020} or by means of the method introduced by Crandall\cite{crandall2012} and Buchheit\cite{Buchheit2026}. This makes the present model particularly suitable for a general high-pressure study of the Burgers–Bain phase transition. For more details see Ref.~\onlinecite{Schwerdtfeger-2026}. In this work we adopt a generalization of the formalism\cite{Buchheit2026,Andres2025a,Buchheit2026a,Schwerdtfeger-2026} to multi-atomic lattice sums, as detailed in Appendix \ref{sec:EpsteinATM}.

We consider a range of constant pressures in the interval $P\in[0,10^{4}]$ and minimize the enthalpy at fixed $P$ with respect to the intrinsic lattice parameters along the Burgers path for fixed values of $\alpha\in[0,1]$. A Newton-Raphson procedure is used to minimize the lattice parameters along the Burgers-Bain path for fixed value of $\alpha\in[0,1]$ with details given in Appendix \ref{NewtonRaphson}. We notice that the $E_\text{PV}(B)$ term exhibits a maximum at $\alpha=1$ corresponding to the fcc structure, under constant-pressure conditions. Consequently, at very high pressures this term may favor a distortion of the angle between $\vec{b}_1$ and $\vec{b}_2$ away from the ideal cuboidal value of 90$^\text{o}$.

\subsection{The Extended Lennard-Jones Potential and the Three-Body Axilrod-Teller-Muto Term for Solid Argon}

In our next, more refined approximation, we consider both two- and three-body inverse power law potentials for solid argon. The two-body interaction is described by the ELJ potential \eqref{eq:ExtVLJ}, with exponents $s_n\in\mathbb{R}_+$ satisfying $s_n>3$, together with the constraints $\phi_{\rm ELJ}(r_e)=-\epsilon$, $d\phi_{\rm ELJ}(r)/dr|_{r_e}=0$, and $d^2\phi_{\rm ELJ}(r)/dr^2|_{r_e}>0$. The cohesive energy of the bi-lattice then takes a form analogous to Eq.\eqref{eq:GVLJ},
\begin{align} \label{eq:GVLJext} 
E_\text{coh}(B,\vec{v})&=
\frac{1}{2}\sum_{n=1}^{N} C_n L(s_n,B,\vec v_2) r_e^{s_n} \quad.
\end{align}

\begin{figure}[tb!]
\centering
\includegraphics[width=.66\columnwidth]{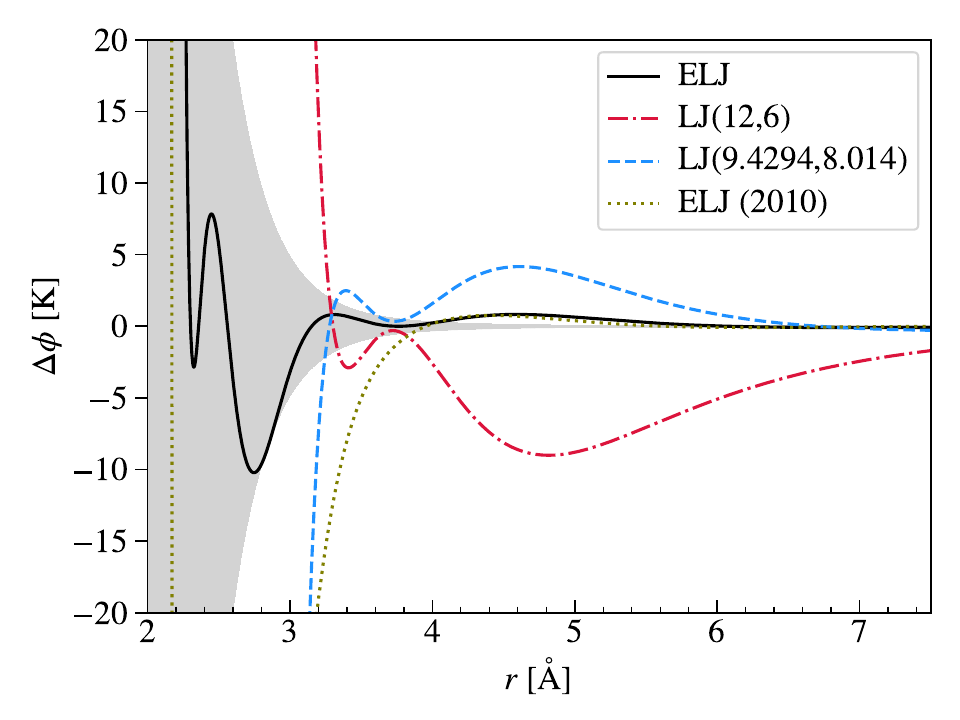}
\caption{Residual energy $\Delta \phi(r) = \phi_{\text{JHBV}}(r)-\phi(r)$ between the JHBV potential and other interaction potentials. ELJ is the newly optimized potential, for parameters see Table \ref{tab:ELJparameters}. The (9.4294,8.014)-LJ potential is the one by Walker \textit{et al.}\cite{Walker2022} ELJ(2010) refers to a previously published ELJ fit for the long-range region.\cite{Assadollahzadeh2010}}
\label{fig:potentials}
\end{figure}

Because the long-range dispersion interaction between two argon atoms is dominated by the $r^{-6}$ term, we set $s_1=6$ and impose $s_n>6, n=2,...,N$. The coefficient $C_1=-64.289$ a.u. is fixed to the argon van der Waals coefficient.\cite{PatkowskiSzalewicz2010,Jiang2015} The remaining adjustable parameters are obtained from a least-squares fit by minimizing the objective function
\begin{equation} \label{eq:LeastSquare}
f_\textrm{OBJ} =\sum_{k=1}^{k_{\max}}w_k\sum_{n=1}^{N_k}\left( \left[ \frac{\phi_{\text{ELJ}}(r_i) - \phi^{(2)}_\mathrm{JHBV}(r_i)}{e(r_i)}\right]^2 + \left[ \frac{\partial\phi_{\text{ELJ}}(r_i)/\partial r_i - \partial \phi^{(2)}_\mathrm{JHBV}(r_i)/\partial  r_i}{\partial e(r_i)/\partial r_i}\right]^2\right).
\end{equation}
Here, $\phi^{(2)}_\mathrm{JHBV}$ denotes the analytical Ar$_2$ potential of J{\"a}ger \textit{et al.},\cite{Jager_MolPhys107} and the function $e$ is the uncertainty function for BO+SR (Born-Oppenheimer plus scalar relativistic) effects, given in Eq.(38) of Lang \textit{et al.}\cite{Lang2024} The interaction potential was optimized to $k_{\max} = 4$ regions of equidistant points ($N_k=40$) representing the short range $I_1=[2.300,3.355]$ \AA, the well ($I_2=[3.355,3.764]$ \AA~ and $I_3=[3.764,4.500]$ \AA), and the long range $I_4=[4.500,7.500]$ \AA~ of the potential. The weighting factors $w_k$  assigned to these regions were 12, 1, 1, and 7, respectively. The resulting coefficients $C_n$ (multiplied by $r_e^{s_n}$) are listed in Table \ref{tab:ELJparameters}, together with the corresponding lattice sums $L(s_n;B,\vec v_2)$. A comparison of the different potentials is shown in Figure \ref{fig:potentials}.

\begin{center}
\begin{table}[tb!] 
\setlength{\tabcolsep}{8pt}
\begin{tabular}{ |c|c|c|c|c|} 
 \hline
$n$ & coefficients $C_nr_e^{s_n}$ & exponents $s_n$ & $L(s_n;B,\vec v_2)$ fcc& $L(s_n;B,\vec v_2)$ hcp\\ 
 \hline 
1 & $-6.4289\times 10^{1}$ & 6.00000000000000 & 14.45392104374446 & 14.45489727784162 \\
2 & $-7.75982394468287\times 10^{4}$ & 9.06191675970054 & 12.47839333766431 & 12.47916089973100 \\
3 & $2.40433965244871\times 10^{7}$ & 10.07004565982138 & 12.30165834529823 & 12.30230007338478 \\
4 & $-1.40411263922951\times 10^{7}$ & 10.09196916046628 & 12.29872501525676 & 12.29936397824045 \\
5 & $-1.38710083499818\times 10^{7}$ & 10.20943544201191 & 12.28353477924867 & 12.28415894834142 \\
6 & $-1.88986024037505\times 10^{6}$ & 10.77162647280785 & 12.22168382464454 & 12.22223818904406 \\
7 & $-7.16207936310204\times 10^{6}$ & 10.92212612699355 & 12.20775172964519 & 12.20828783482735 \\
8 & $1.00113954751259\times 10^{8}$ & 12.19910394274630 & 12.12148003789281 & 12.12187287265657 \\
9 & $1.25667406076729\times 10^{8}$ & 12.56904204694045 & 12.10442300512802 & 12.10477907476412 \\
10 & $3.48675055201955\times 10^{8}$ & 13.04176704787827 & 12.08626689419873 & 12.08657943532319 \\
11 & $-7.20243086527715\times 10^{8}$ & 14.53425474437267 & 12.04790177655774 & 12.04810234128407 \\
12 & $-6.1595613664798\times 10^{10}$ & 15.39319599782686 & 12.03443374720823 & 12.03458613941418 \\
13 & $3.93492815348309\times 10^{11}$ & 16.74327437220118 & 12.02069715012362 & 12.02079372565571 \\
14 & $-1.05110318323671\times 10^{13}$ & 19.83474699865386 & 12.00666583139768 & 12.00669686193151 \\
15 & $7.64207618470187\times 10^{13}$ & 21.09239436374122 & 12.00424218302175 & 12.00426117676638 \\
16 & $-7.10143345901384\times 10^{14}$ & 21.94308553268752 & 12.00313137460380 & 12.00314489620523 \\
17 & $7.5237924988463\times 10^{14}$ & 22.18242082231914 & 12.00287574909222 & 12.00288802503733 \\
\hline
\end{tabular}
\caption{Coefficients $C_nr_e^{s_n}$ ($r_e= 7.1088$ a.u.) and exponents $s_n$ for the ELJ potential of Ar$_2$, and corresponding basic lattice sums $L(s_n,B,\vec v_2)$ for the unit cells of fcc ($\alpha=1$) and hcp ($\alpha=0$). Atomic units are used. The lattice sums for the lattice parameter $a=1$ are listed to demonstrate the small differences between the two close-packed structures. The zero-crossing distance is at $\sigma=6.3431$ a.u. and the inflection point at $r_I=7.8576$ a.u.}
\label{tab:ELJparameters}
\end{table}
\end{center}

The present ELJ potential represents a significant improvement over previously published inverse power law potentials, particularly in the strongly repulsive region. For example, the original ELJ potential for argon,\cite{Schwerdtfeger-2006} which was fitted to the data of Cybulski and Toczylowski,\cite{CybulskiToczylowski1999} predicts a pressure of $P=243.9$ GPa at a molar volume of 6 cm$^3$/mol. This value differs substantially from that obtained with the more accurate ELJ-CCSD(T)/CBS potential,\cite{Assadollahzadeh2010} which predicts $P$=300.6 GPa under the same conditions. In contrast, the newly fitted potential presented here yields $P$=293.7 GPa, in excellent agreement with the highly accurate JHBV potential,\cite{Jager_MolPhys107} which gives $P$=293.6 GPa and is taken as the reference in the present work.

The dominant long-range (LR) three-body contribution arises from the triple-dipole interaction derived from the classical Drude model, namely the Axilrod-Teller-Muto (ATM) term defined in Eq.~\eqref{eq:ATM}. In the present work, we adopt the widely used value $c_\text{ATM}$=518.3 a.u. reported by Kumar and Meath.\cite{Kumar1985,Kumar2010} In reduced units, this corresponds to $c_\text{ATM}^{'}=c_\text{ATM}/\epsilon r_e^9 = 0.02466$, where the dissociation energy $\epsilon$ and equilibrium bond distance $r_e$ are taken from the JHBV potential parameters listed in Table~\ref{tab:PropPot}. These reference values include scalar-relativistic (SR) and spin-orbit (SO) effects.\cite{Schwerdtfeger2017}

The advantage is that the ATM contribution can still be treated using exact and rapidly convergent lattice-summation techniques based on anisotropic Epstein zeta functions,
the generalization of Epstein zeta functions to anisotropic interaction kernels,\cite{buchheitComputationAnisotropicSingular2026}
although the formalism is somewhat more involved than for the corresponding inverse power two-body interaction.
Details of the method are given in Appendix~\ref{sec:EpsteinATM}. Nevertheless, many-body potentials are computationally more demanding in bulk simulations. Consequently, there have been several alternative approaches to approximate these through effective potentials such as the glue or Finnis-Sinclair potential,\cite{Finnis1984,Ercoless1988} or more recently through machine-learned pair potentials to study krypton under extreme conditions.\cite{Iwasaki2025} Notably, Marcelli, Wang, and Sadus\cite{marcelli2000effective,wang2006relationship} examined the relationship between the two- and three-body energies of noble gases along the vapor-liquid equilibrium line and found an approximate linear relationship between them. We therefore introduce an effective three-body potential (E3B) based on a density- (or volume-) dependent relation, with parameters fitted to reproduce the cohesive-energy contribution of a chosen three-body potential,
\begin{equation} \label{eq:ELJ-eff3}
E_{\text{coh}}^{(\text{E3B})}(V)=\frac{\lambda_3}{V} \lim_{N\rightarrow\infty}\frac{1}{N}\sum_{i<j}^N \sum_{n=1}^{n_\text{max}} d_nr_{ij}^{-t_n} 
\end{equation}
where $\lambda_3$ is an optimized prefactor and $d_n$ are fitting coefficients. For our bi-lattice this reduces to,
\begin{equation} \label{eq:ELJ-eff3a}
E_{\text{coh}}^{(\text{E3B})}(V)=\frac{\lambda_3}{2V} \sum_{n=1}^{n_\text{max}} d_n L(t_n;B,\vec{v}_2) \quad .
\end{equation}
This approach retains the computational efficiency of the Epstein zeta function framework, allowing the three-body interactions to be treated efficiently while enabling the Burgers-Bain transition path to be mapped out. The parameters $(d_n,t_n,\lambda_3)$ were determined by a relative least-squares fit to the cohesive-energy contributions of the fcc phase, using the ATM potential as the reference.
\begin{equation} \label{eq:LeastSquare2}
f_\textrm{OBJ-E3B} =\sum_{k=1}^{k_\text{max}} \left[ \frac{E_{\text{coh}}^{(3)}(V_k) - E_{\text{coh}}^{(\text{E3B})}(V_k)}{E_{\text{coh}}^{(3)}(V_k)}\right]^2.
\end{equation}
An equidistant grid was employed over the volume interval $V\in [10,30]$ [cm$^3$/mol]. The optimized parameters (in atomic units) are  $d_1=5.938052144151344\times 10^{1}$, $d_2=9.701946933967957\times 10^{1}$, $t_1=6.03500224723525$, $t_2=5.98078135766366$, and $\lambda_3=6.228137836647347$. In principle, the present scheme can be extended to more accurate three-body potentials that correctly describe the short-range region relevant at high pressures. However, the ATM potential provides a suitable reference system for testing our approach. Moreover, with such an effective potential we can efficiently evaluate thermal contributions within the Einstein approximation using exact lattice summations.
\begin{center}
\begin{table}[hbtp!] 
\scriptsize{}
\setlength{\tabcolsep}{3pt}
\begin{tabular}{|l |l||c|c||c|c|c|c|c| } 
 \hline
 Potential  & Ref. &$r_e$ & $D_e$ &  $r_\text{NN}$ & $E_\text{coh}$& $\Delta E_\text{E}^\text{HZPV}$ & $\Delta E_\text{E}^\text{AHZPV}$ & $B$ \\ 
 \hline 
\textit{two-body} &&&&&&&&\\
(12,6)-LJ                                                                                  &TW&(3.7618)&(99.476)& 3.6536 & 856.486 & 79.342 & 1.580 & 39.468\\
(9.4294,8.014)-LJ    &[\onlinecite{Walker2022}]                                             &(3.7618)&(99.476)& 3.6798 & 759.434 & 75.683 & 1.441 & 35.950\\
ELJ/CT              &[\onlinecite{Schwerdtfeger-2006}],[\onlinecite{CybulskiToczylowski1999}]& 3.7782 & 96.975 & 3.7004 & 761.558 & 76.103 & 1.216 & 36.348\\
ELJ-CCSD(T)/CBS      &[\onlinecite{Assadollahzadeh2010}]                                    & 3.7721 & 98.698 & 3.6957 & 764.573 & 76.421 & 1.236 & 36.698\\
ELJ                                                                                    &TW& 3.7618 & 99.476 & 3.6848 & 768.089 & 76.691 & 1.256 & 37.035\\
Aziz                 &[\onlinecite{Aziz-1993}]                                              & 3.7570 & 99.555 & 3.6808 & 768.767 & 76.587 & 1.170 & 36.998\\
PS/SR                &[\onlinecite{PatkowskiSzalewicz2010}                                  & 3.7624 & 99.351 & 3.6865 & 768.042 & 76.805 & 1.214 & 37.175\\
PS/SR+SO             &[\onlinecite{PatkowskiSzalewicz2010}],[\onlinecite{Schwerdtfeger-2016}] & 3.7621 & 99.377 & 3.6862 & 768.244 & 76.819 & 1.214 & 37.192\\
JHBV/SR              &[\onlinecite{Jager_MolPhys107}]                                       & 3.7618 & 99.476 & 3.6860 & 769.064 & 76.899 & 1.215 & 37.275\\
JHBV/SR+SO           &[\onlinecite{Jager_MolPhys107}],[\onlinecite{Schwerdtfeger-2016}]       & 3.7618 & 99.483 & 3.6860 & 769.117 & 76.902 & 1.215 & 37.278\\
LPL/NR               &[\onlinecite{Lang2024}]                                               & 3.7671 & 98.708 & 3.6913 & 762.419 & 76.358 & 1.207 & 36.695\\
LPL/SR               &[\onlinecite{Lang2024}]                                               & 3.7639 & 99.224 & 3.6879 & 766.319 & 76.548 & 1.205 & 36.903\\
LPL/SR+QED           &[\onlinecite{Lang2024}]                                               & 3.7640 & 99.189 & 3.6880 & 766.030 & 76.535 & 1.205 & 36.890\\
LPL/SR+QED+CP        &[\onlinecite{Lang2024}]                                               & 3.7640 & 99.189 & 3.6880 & 766.030 & 76.535 & 1.205 & 36.890\\
LPL/SR+QED+CP+SO     &[\onlinecite{Lang2024}],[\onlinecite{Schwerdtfeger-2016}]               & 3.7640 & 99.195 & 3.6880 & 766.083 & 76.538 & 1.205 & 36.893\\
\textit{DFT} &&&&&&&&\\
PBE                                                                                        &TW& 3.9983 & 49.281 & 4.1004 & 150.962 &-&-&10.539\\
PBE-D2                                                                                     &TW& 3.7456 & 135.824 & 3.7942 & 708.370 &-&-&40.379\\
PBE-D3                                                                                     &TW& 3.8331 & 127.416 & 3.8637 & 698.036 &-&-&29.942\\
PBE-D4                                                                                     &TW& 3.8313 & 130.033 & 3.8672 & 700.308 &-&-&30.360\\
\textit{many-body} &&&&&&&&\\
ELJ+ATM                                                                                    &TW& 3.7618 & 99.476  & 3.7164 & 713.508 &-&-& 33.747\\
ELJ+ATM+ZPV(2B,AH)                                                                        &TW& 3.7618 & 99.476  & 3.7603 & 709.940 & 64.230 & 1.386 & 28.342 \\
ELJ+E3B                                                                                    &TW& 3.7618 & 99.476  & 3.7604 &713.520&72.330&1.287&28.335\\
\textit{other work} &&&&&&&&\\
MB(2+3)-JHBV+QD&[\onlinecite{Jager_MolPhys107}],[\onlinecite{Schwerdtfeger-2016}]          & - & - &3.7161& 713.747&-&-&-\\
MB(2+3)+SO+QD+ZPV(2+3B,PD+AH)         &[\onlinecite{Schwerdtfeger-2016}]                       &    -   &   -    & 3.7581 & 645.130 & 66.494 &1.331& 28.361\\
exp.                 &[\onlinecite{Herman1988}],[\onlinecite{Barrett-1964},],[\onlinecite{Schwalbe-1977}],[\onlinecite{Batchelder-1979}] & 3.761(3) & 99.2(10)& 3.7560 & 645.5(9)& -      &-& 26.4\\
\hline
\end{tabular}
\caption{\footnotesize{Calculated equilibrium distances $r_e$ (in Å) and dissociation energies $D_e$ (in cm$^{-1}$) for Ar$_2$, together with nearest-neighbor distances $r_{\mathrm{NN}}$ (in Å), cohesive energies $E_{\mathrm{coh}}$ (in cm$^{-1}$), and bulk moduli (in kbar) for fcc argon. The argon mass was taken as $M=39.9480$ amu. The LJ results for the fcc phase were obtained using exact lattice summations.\cite{burrows-2020} For the two LJ potentials and the ELJ potential employed in this work (TW), the values of $r_e$ and $D_e$ were fixed to those reported by Jaeger \textit{et al.}\cite{Jager_MolPhys107} and are shown in parentheses. Solid-state results obtained from VASP (version 6.4.0 version\cite{VASP1,VASP2,VASP3,VASP4}) calculations using the PBE functional with three different dispersion corrections are also included. Additional many-body (MB) results incorporating two- and three-body interactions, as well as corrections arising from the four-body quadruple-dipole (QD) term and zero-point vibrational energy (ZPV), are reported. The latter include two-body phonon-dispersion (PD) effects and harmonic and anharmonic (AH) two- and three-body vibrational contributions. SR: scalar relativistic; SO: spin-orbit; QED: quantum electrodynamics; CP: Casimir-Polder term.}}
\label{tab:PropPot}
\end{table}
\end{center}

The performance of the various potentials is summarized in Table \ref{tab:PropPot} for both the argon dimer and the fcc solid. For the two-body potentials, all predictions are close to the experimentally measured bond length and dissociation energy of the argon dimer.\cite{Herman1988} For the solid phase, larger deviations from experiments are observed for the ELJ+ATM and E3B models, primarily due to the neglect of zero-point vibrational effects and higher-order many-body contributions. However, upon including harmonic and anharmonic two-body zero-point vibrational corrections within the Einstein approximation (ELJ+ATM+ZPV), excellent agreement with experiment is already obtained. This is consistent with our previous work, in which J\,mol$^{-1}$ accuracy was achieved for the cohesive energy of solid argon.\cite{Schwerdtfeger-2016}

\subsection{Density Functional Calculations}
\label{sec:DFT}

Density functional theory (DFT) calculations were included for solid argon to provide a comparison with the many-body expansion results. The calculations employed the Perdew-Burke-Ernzerhof (PBE) exchange-correlation functional \cite{GGA.PBE1996} together with the scalar-relativistic projector augmented-wave (PAW) method as implemented in VASP version 6.4.0.\cite{Blochl1994PAW,VASP1,VASP2,VASP3,VASP4,VASP5} Dispersion interactions were accounted for using the Grimme corrections,\cite{Grimme-2011,Grimme-D3-2010,Goerigk2011,Caldeweyher2017,Caldeweyher2019} namely PBE-D$n$ ($n=2,3,4$) with Becke-Johnson damping,\cite{Johnson2006} for the cohesive-energy calculations. We also employed the local density approximation\cite{Kohn1965} (LDA) for comparison.

The higher-level PBE-D4 results are very close to the corresponding PBE-D3 values as shown in Table \ref{tab:PropPot}. For mapping the Burgers-Bain transition path, we therefore focused on the newly developed PBE-D4 method. The variable-cell nudged elastic band (VC-NEB) method \cite{HANNES1998,Henkelman2000,Henkelman2000a,QIAN2013} was initially employed at constant pressure to determine the transition pathway with a total of 25 images between the hcp and fcc structures. However, this method encountered difficulties in accurately describing the region around the bifurcation point because the lattice parameters exhibit apparent kinks as functions of the distortion parameter $\alpha$. These non-smooth features hinder proper convergence of the band to the minimum-energy path.\cite{Henkelman2011,Zhang2016} Consequently, we used instead the lattice parameters obtained from the (12,6)-LJ potential as initial guesses for subsequent enthalpy optimizations at fixed values of $\alpha$. This procedure provided a more accurate representation of the minimum-energy path and the location of the bifurcation point. Further details of the optimization procedure are given in Appendix \ref{NewtonRaphson}.

The electronic energy convergence criterion was set to $10^{-6}$ eV/cell, and an energy cutoff of 750 eV was employed. A $k$-point grid convergence test was performed, which shows that argon converges smoothly even at relatively low density of $k$-points. In the optimization process, a $27\times27\times15$ grid was used for all bulk DFT calculations. Errors intrinsic to the density-functional approximation \cite{Teale2022} are expected to largely cancel when considering energy differences along the Burgers-Bain path. However, we should mention that despite applying tight convergence criteria, the activation energy for the hcp$\rightarrow$fcc phase transition for argon is less than 1 kJ/mol at low pressures causing numerical problems in converging out the total energy and corresponding lattice parameters in the geometry optimizations. This made the search for the bifurcation point at the Burgers-Bain path transition a difficult task in contrast to our calculations using model potentials together with exact lattice summations.

\subsection{Vibrational and Thermal Contributions}

For argon, the electronic band gap is predicted to close only at extremely high pressures, with density functional theory calculations placing the metallization pressure at approximately 510 GPa.\cite{Kwon1995} Electronic contributions to the free energy can therefore be neglected, and only vibrational contributions need to be considered in the pressure range used in our work. To investigate vibrational and thermal effects along the Burgers-Bain path, we adopt the harmonic Einstein model for the ELJ+ATM potential and neglect phonon dispersion.\cite{Smits2021} Despite its simplicity, this approach allows a qualitative assessment of vibrational and thermal effects along the Burgers-Bain phase transition at constant pressure $P$ while retaining the advantage of exact lattice summations. A more quantitative description would require accurate many-body potentials together with phonon-dispersion effects treated using direct summation techniques, which is computationally demanding. Such an approach will be explored in greater detail in a forthcoming study.

The Gibbs free energy of the system at constant pressure $P$ and temperature $T$ for the ELJ+ATM potential for solid argon is given by,
\cite{Schwerdtfeger-2006,Smits2021}
\begin{equation} \label{eq:GELJ}
G(B,\vec{v};P,T)=U(B,\vec{v};T) + PV(B) -TS(B,\vec{v};T),
\end{equation}
with the internal energy
\begin{equation} \label{eq:GELJ1}
U(B,\vec{v};P,T)=E_\text{coh}(B,\vec{v})+E_\text{ZPV}(B,\vec{v})+ E_\text{th}(B,\vec{v};T).
\end{equation}
For the zero-point vibrational energy contribution to the cohesive energy within the isotropic Einstein approximation, we take both the harmonic and anharmonic corrections into account,
\begin{equation} \label{eq:ZPE}
E_\text{ZPV}(B,\vec{v}) = E_\text{HZPV}(B,\vec{v}) + E_\text{AHZPV}(B,\vec{v})
\end{equation}
with the dominant harmonic contribution 
\begin{equation} \label{eq:HZPE}
E_\text{HZPV}(B,\vec{v}) = \frac{3}{2} \hbar \omega_E= \frac{\hbar}{2r_e}\sqrt{\frac{3}{M}}\left[ \sum_{n} s_n\left( s_n-1\right)C_nr_e^{s_n+2}   L\left(s_n+2;B,\vec v_2\right) \right]^{\frac{1}{2}}
\end{equation}
where $M$ is the atomic mass taken for 40-Ar (39.9623831238 amu). We notice here that the Epstein zeta function carries the units in power of the lattice constant $a$ because of the definition of our $B$-matrix in Eq.\eqref{eq:Bmatrix2}, whilst the coefficients $C_n$ carry the units of energy. The anharmonic zero-point vibrational (AZPV) energy contribution was included only within first-order perturbation theory,\cite{Henkel1955,Dobbs-1957,Smits2021}
\begin{equation} \label{eq:AHZPE}
E_\text{AHZPV}(B,\vec{v}) = \frac{3\hbar^2}{32Mr_e^2} \frac{\sum_{n} \left( s_n+2 \right) \left( s_n+1 \right) s_n \left( s_n -1 \right) C_nr_e^{s_n+4} L\left(s_n+4;B,\vec v_2\right) } {\sum_{n} s_n\left( s_n-1\right)C_nr_e^{s_n+2} L\left(s_n+2;B,\vec v_2\right) } 
\end{equation}
This approximation is reasonable since the cubic force constants vanish for the fcc structure by symmetry, and for hcp they are roughly an order of magnitude smaller than the quartic force constants. Although these cubic terms are expected to become somewhat larger near the transition state along the Burgers–Bain path, they are nevertheless anticipated to remain subdominant. The two Eqs.~\eqref{eq:HZPE} and \eqref{eq:AHZPE} can easily be extended to the effective three-body E3B term.
 
For the thermal contributions to the entropy and internal energy we take the Einstein approximation in the harmonic approximation,
\begin{equation} \label{eq:EinsteinEntropy}
S(B,\vec{v};T) = 3k_B\left[ { \frac{\beta \omega_E}{e^{\beta \omega_E}-1} \rm -ln}\left( 1-e^{-\beta \omega_E}\right) 
\right], \quad E_\text{th}(B,\vec{v};T) =  3k_BT\frac{\beta \omega_E}{e^{\beta \omega_E}-1}  \:.
\end{equation}
with $\beta=\hbar / k_BT$. We notice that the thermal term for the internal energy, $E_\text{th} $, is canceled out by the identical term in the entropy expression for the free energy. It is convenient to re-express the Gibbs free energy in terms of the Einstein ZPV energy term, which for the harmonic approximation reads
\begin{equation} \label{eq:Gibbsfreeharmonic}
G(B,\vec{v};P,T)=E_\text{coh}(B,\vec{v})+E_\text{HZPV}(B,\vec{v}) + PV(B) +3k_BT {\rm ln}\left( 1-e^{-\delta_H }\right)
\end{equation}
with the notation used by Dobbs,\cite{Dobbs-1957}
\begin{equation} \label{eq:deltaharmonic}
\delta_H(B,\vec{v};T)=\frac{2}{3k_BT} E_\text{HZPV}(B,\vec{v}).
\end{equation}
This formula has been extended by Henkel\cite{Henkel1955} to include anharmonicity effects within the Einstein approximation,\cite{Dobbs-1957}
\begin{align} \label{eq:Gibbsfreeharmonic2}
G(B,&\vec{v};P,T)=E_\text{coh}(B,\vec{v})+E_\text{HZPV}(B,\vec{v})+E_\text{AHZPV}(B,\vec{v}) + PV(B) \\
\nonumber
& +3k_BT {\rm ln}\left( 1-e^{-\delta }\right) + 3\Delta\frac{e^\delta + 1}{\left( e^\delta -1\right)^2} - \frac{3\Delta^2}{2k_BT}  \frac{e^{3\delta}+10e^{2\delta}+9e^\delta}{\left( e^\delta -1\right)^4} + ...
\end{align}
with the more general functions
\begin{equation} \label{eq:deltaanharmonic}
\delta(B,\vec{v};T)=\frac{2}{3k_BT} \left\{ E_\text{HZPV}(B,\vec{v})+E_\text{AHZPV}(B,\vec{v})\right\}  \quad , \quad \Delta(B,\vec{v}) = \frac{2}{3} E_\text{AHZPV}(B,\vec{v}).
\end{equation}
The terms to the order shown in Eq.~\eqref{eq:Gibbsfreeharmonic2} were used for the free energy. We expect anharmonic thermal contributions to the free energy to remain small at low temperatures\cite{Henkel1955} and high pressures, but they are nevertheless included in our calculations of the Burgers-Bain path for solid argon for a better description of the higher temperature range. 

In the calculations of the Burgers-Bain path, the pressure and temperature are fixed at prescribed values of $P$ and $T$, while the distortion parameter $\alpha\in[0,1]$ is varied to determine the minimum-energy path within the chosen unit cell. 
For each value of $\alpha$, the Gibbs free energy $G(\vec{x})$ (see Eq.~\eqref{eq:GELJ}) is minimized with respect to the set of lattice parameters ${x_i}$ defining the generator matrix $B$ and the shift vector $\vec{v}$,
\begin{equation} \label{eq:Gibbsderiv}
\frac{\partial G}{\partial x_i} =0,
\end{equation}
using a Newton-Raphson procedure as outlined in Appendix \ref{NewtonRaphson}. As our focus is on the hcp$\leftrightarrow$fcc transition, we adopt the experimentally observed low-temperature fcc phase as the reference structure,
\begin{equation} \label{eq:DeltaGibbs2}
\Delta G(\vec{x};P,T)=G(\vec{x};P,T) - G(\vec{x}_\text{fcc};P,T).
\end{equation}
Here we employ the ELJ plus ATM/E3B potentials at different pressures up to 1 GPa, where the ATM potential starts to deviate substantially from the correct three-body behavior.\cite{Jaeger2011} It should also be emphasized that our model accounts only approximately for anharmonicity effects and neglects any phonon dispersion. These contributions are expected to become increasingly important near the melting point, where thermal expansion is largest and the crystal structure is most weakly bound. In addition, such effects become important for accurately describing the difference in cohesive energy between the hcp and fcc phases.\cite{Schwerdtfeger-2016} Further extensions would require phonon calculations based on direct lattice summations within a Gibbs free-energy minimization scheme. This, in turn, necessitates the evaluation of derivatives of the phonon frequencies with respect to the lattice parameters. Such a treatment is considerably more demanding and lies beyond the scope of the present work, which focuses primarily on the analytical treatment of inverse power law potentials using Epstein lattice sums. A more comprehensive treatment of finite-pressure including phonon dispersion will be presented in future work.

\subsection{Accuracy of the Approximations used for Solid Argon}

We conclude the theory section by discussing the accuracy of the methods employed in the high-pressure regime for solid argon. Since the energetics along the Burgers-Bain path are evaluated relative to the fcc phase, which serves as the reference structure, errors in both the ELJ+ATM (or ELJ+E3B) many-body model and the adopted DFT approximation are expected to largely cancel. Nevertheless, it is instructive to assess these approximations against more accurate many-body calculations. For this purpose, we employ the Jaeger two- and three-body potential \cite{Jager_MolPhys107} together with the CCSD(T) four-body quadruple-dipole contribution,\cite{Schwerdtfeger2017} denoted MB/JHBV, as the reference model. Zero-point vibrational effects are neglected throughout this comparison.

\begin{figure}[htb!]
    \centering
    (a) \includegraphics[width=.45\columnwidth]{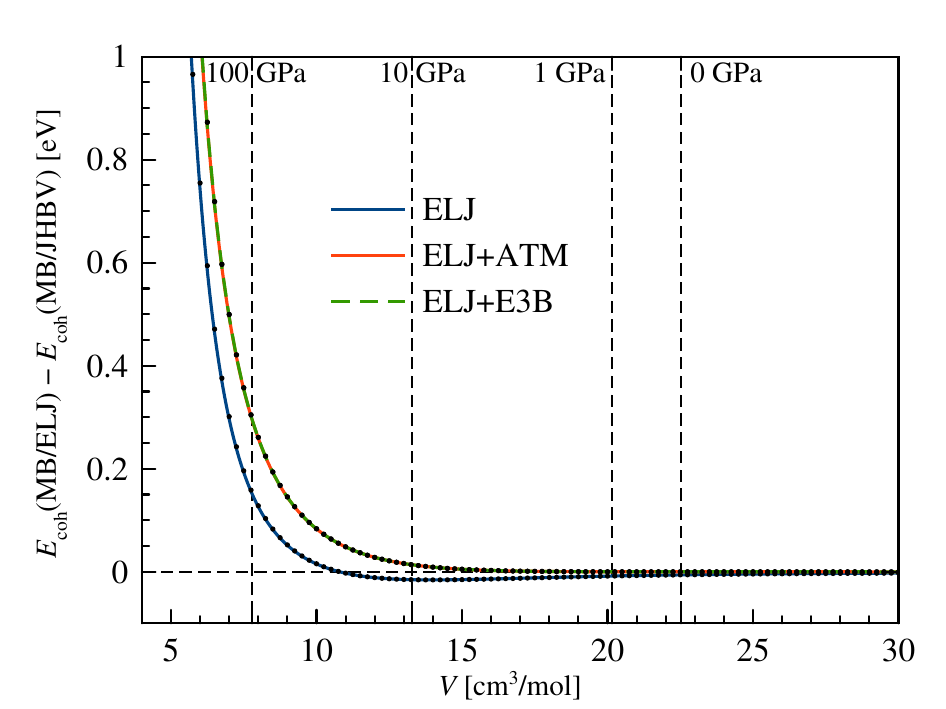}
    (b)\includegraphics[width=.45\columnwidth]{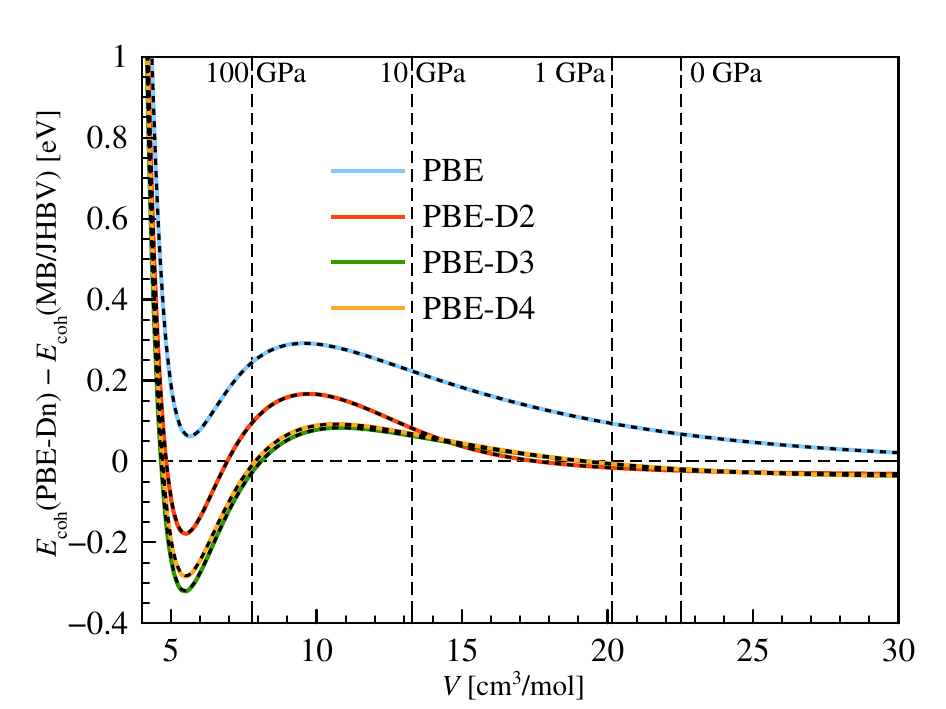}\\
    (c) \includegraphics[width=.45\columnwidth]{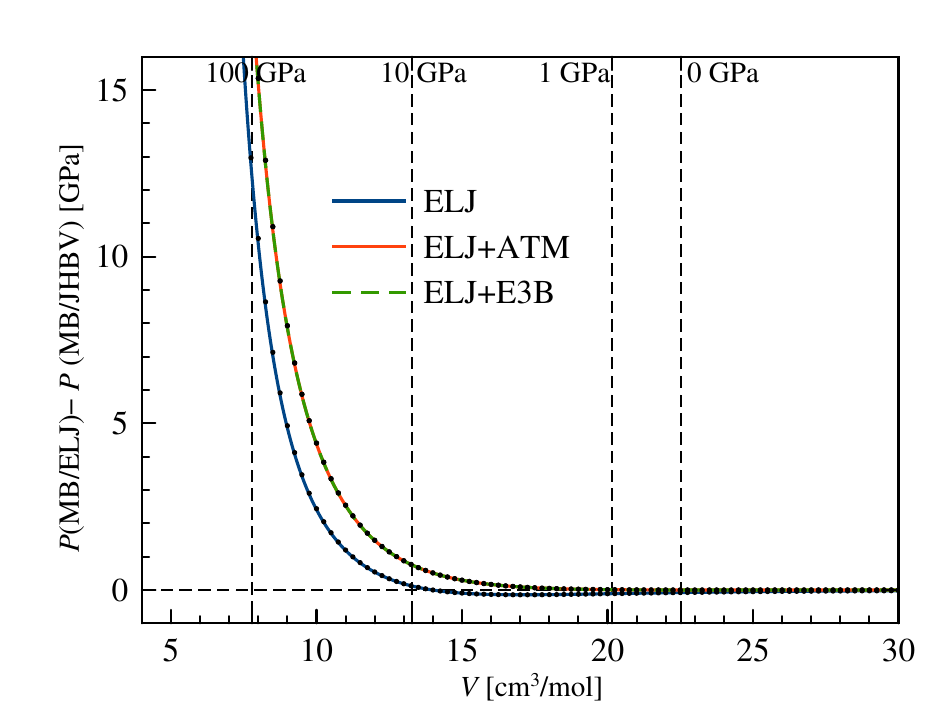}
    (d)\includegraphics[width=.45\columnwidth]{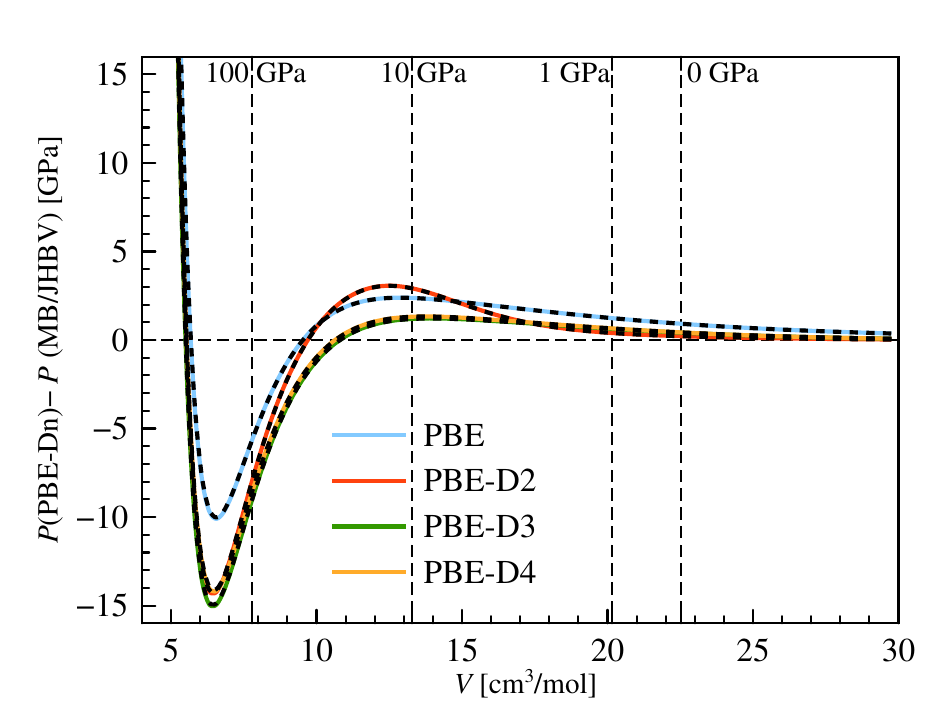}
    \caption{(a) Deviation of various ELJ+ATM/E3B and (b) PBE cohesive energies (with and without) Grimme dispersion from the many-body expansion using the Jaeger 2+3-body including the CCSD(T) 4-body potential (MB/JHBV) as a reference system for solid argon as a function of the volume $V$. (c,d) show the corresponding deviations in pressure $P$. The vertical dashed lines are pressure lines derived from a many-body expansion plus ZPV corrections from the Einstein approximation (see text). Note that the fcc and hcp curves (black dotted lines) are indistinguishable on this scale as are the two models ELJ+ATM and ELJ+E3B.}
    \label{fig:PBEdev1}
\end{figure}

Figure \ref{fig:PBEdev1} compares the ELJ+ATM/E3B potentials and several dispersion-corrected DFT approximations with the \textit{ab-initio} many-body (MB) expansion reported in our previous work,\cite{Schwerdtfeger2017} with thermal contributions omitted. As expected, the PBE functional without dispersion corrections performs poorly for solid argon, exhibiting large deviations from the MB reference results at large volumes where dispersion interactions dominate. Both the ELJ+ATM potential and its effective counterpart, ELJ+E3B, reproduce the MB results very well up to pressures of a few GPa. At shorter interatomic distances, corresponding to volumes below approximately $V=6$ cm$^3$/mol ($P>200$ GPa), appreciable deviations between the DFT results and the MB expansion begin to emerge. This behavior may indicate a slower convergence of the many-body expansion in this regime, implying that contributions beyond fourth order become increasingly important, although their explicit inclusion is computationally prohibitive. Alternatively, the deviations may originate from intrinsic errors in the DFT approximations themselves. We further find that PBE-D3 and PBE-D4 exhibit very similar levels of accuracy. The limitations of the simple repulsive ATM three-body interaction also become apparent at short range, where the DFT approximation provides a more accurate description.

\section{Results and Discussion}

We begin this section by assessing the effect of pressure on the phase-transition pathway, with particular emphasis on the bifurcation point where the Burgers path connects to the Bain path. To this end, we first consider the Lennard–Jones model, examining a range of general $(n,m)$-LJ potentials. Reduced units are used for all the calculations with the LJ potential. We then discuss the Burgers-Bain path for the ELJ potential (see Table \ref{tab:ELJparameters}), highlighting its improved performance at higher pressures. The role of three-body contributions (ATM and E3B), as well as temperature effects, is also examined. Finally, we compare these results with those obtained from DFT calculations. Detailed data for the Burgers-Bain pathway at different pressures are provided in the Supplementary Material.

\subsection{Burgers-Bain Path for a Lennard-Jones Potential at Finite Pressures}

\begin{figure*}[tb!]
  \centering
  \begin{subfigure}{0.49\textwidth}
    \centering
    (a)\includegraphics[width=\linewidth]{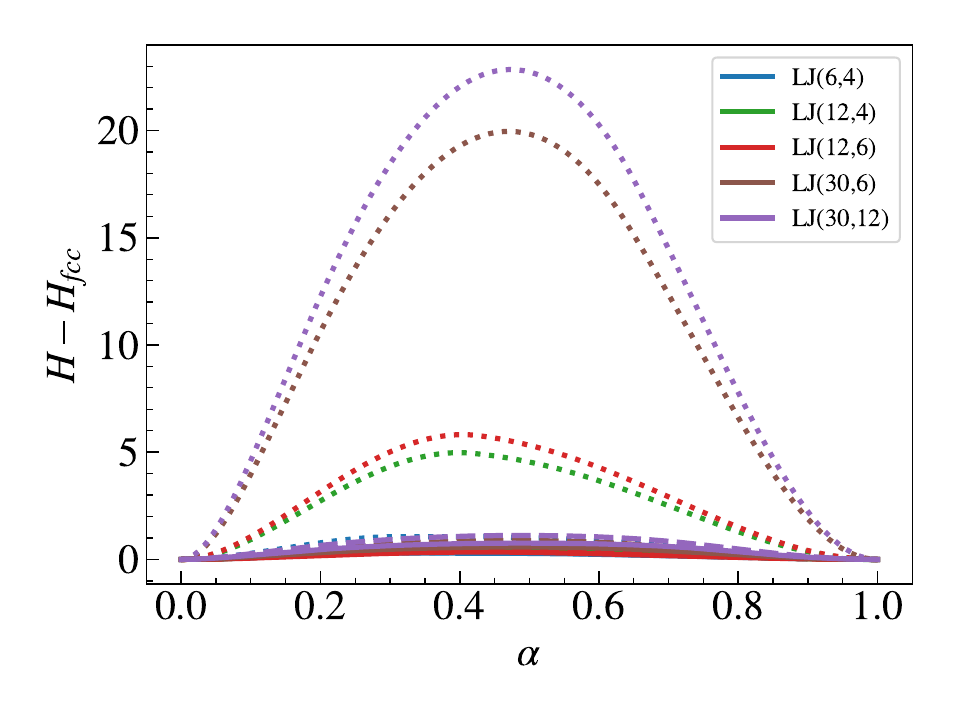}
  \end{subfigure}%
  \hfill
  \begin{subfigure}{0.49\textwidth}
    \centering
    (b)\includegraphics[width=\linewidth]{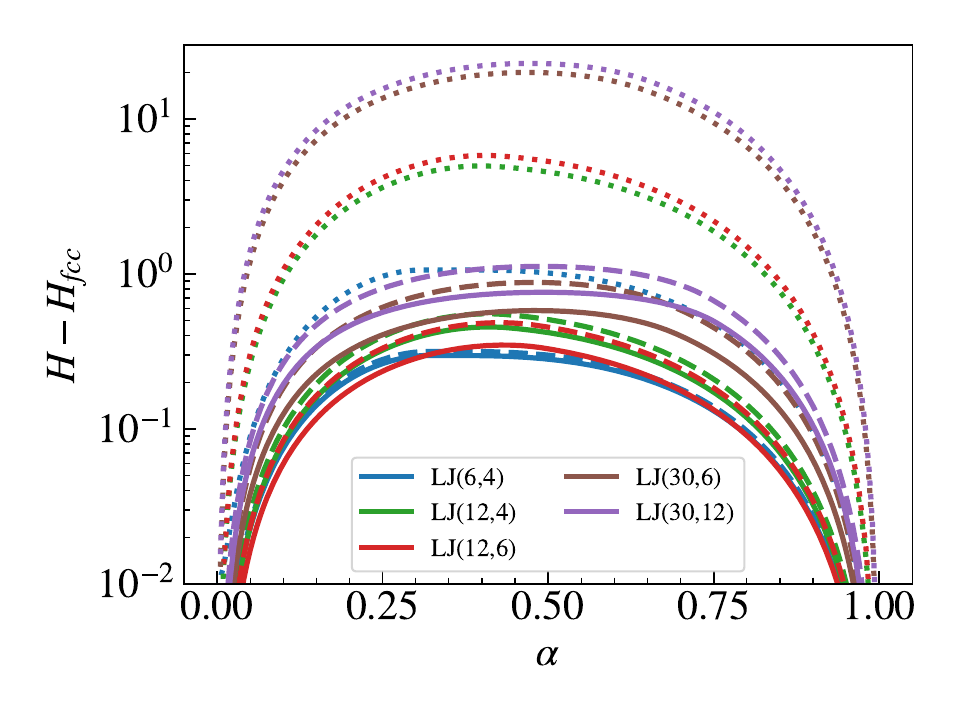}
  \end{subfigure}%
  \newline
\captionsetup{justification=raggedright,singlelinecheck=false}
  \centering
  \caption{Predicted enthalpy difference, $\Delta H = H(\alpha)-H_\mathrm{fcc},~ H=E_\textrm{coh}+PV$, along the Burgers-Bain path for $(n,m)$-LJ potentials at the isobars $P=0$, 10, and $10^3$ (reduced units are used throughout). The different $(n,m)$-LJ potentials are distinguished by color, while the three pressures are represented by different line styles. For better visibility of the curves a logarithmic scale is used in Figure (b).}\label{fig:LJnmEnthalpy}
\end{figure*}

\begin{figure*}[bt!]
  \centering
  (a)\includegraphics[width=.44\columnwidth]{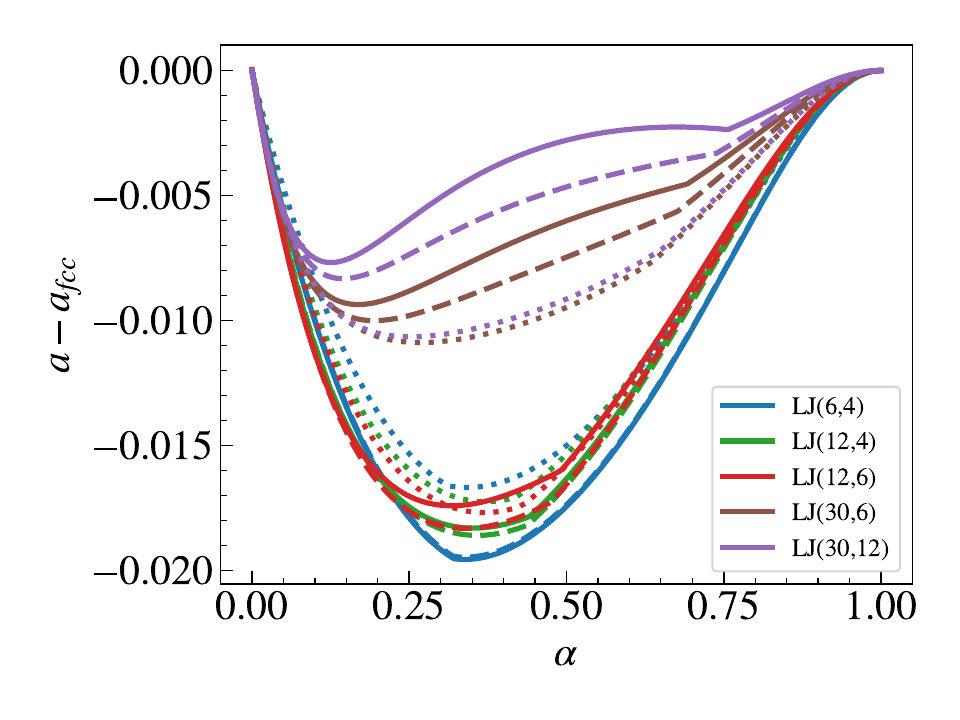}
    (b)\includegraphics[width=.44\columnwidth]{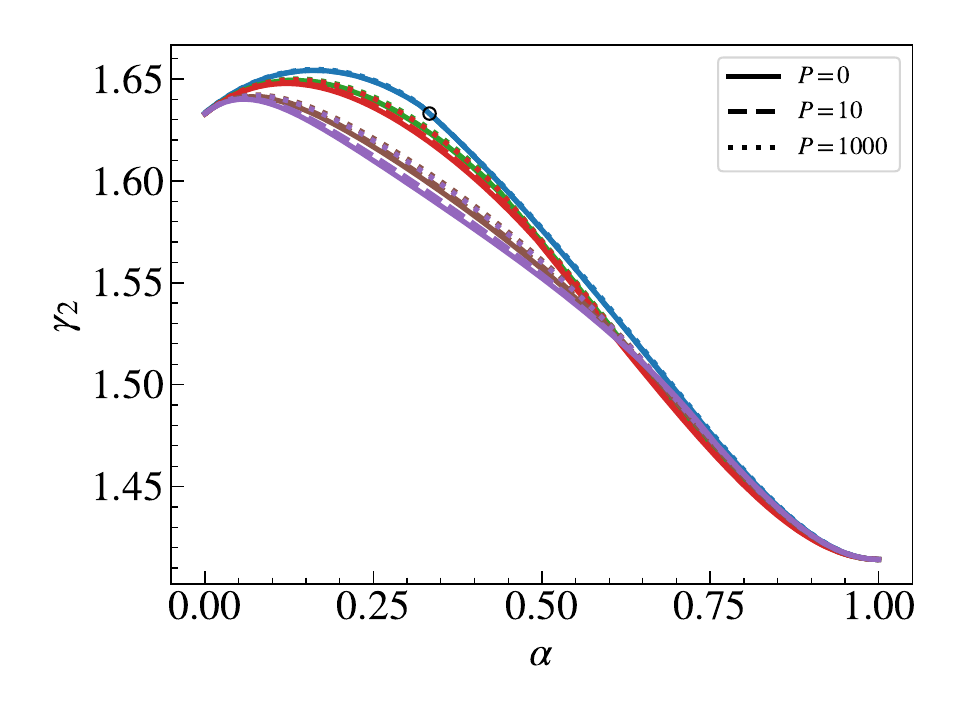}\\
    (c)\includegraphics[width=.44\columnwidth]{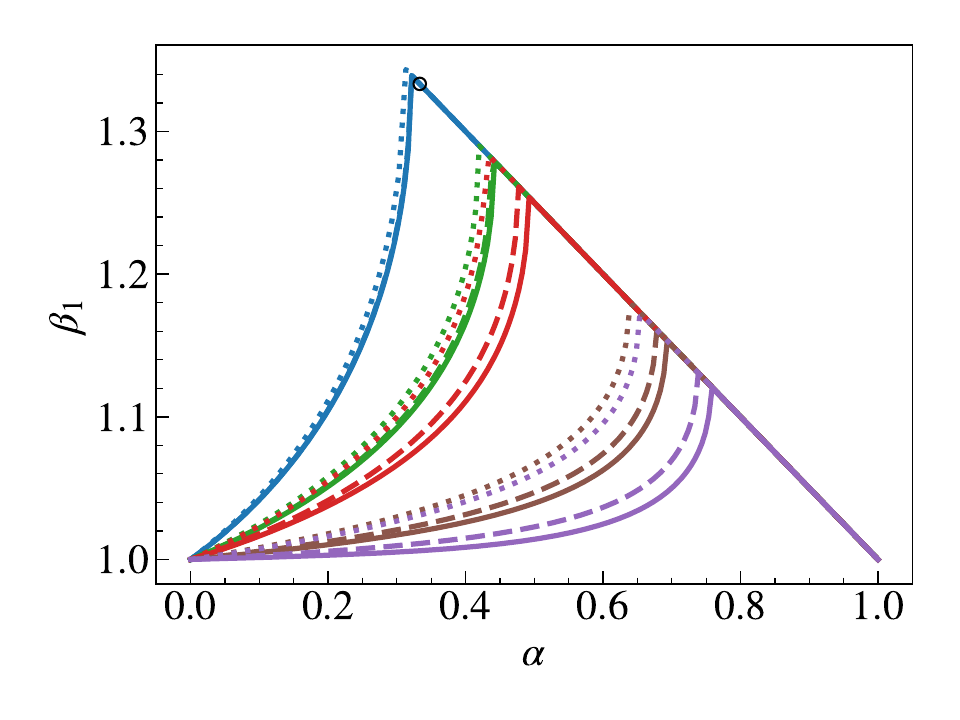}
    (d)\includegraphics[width=.44\columnwidth]{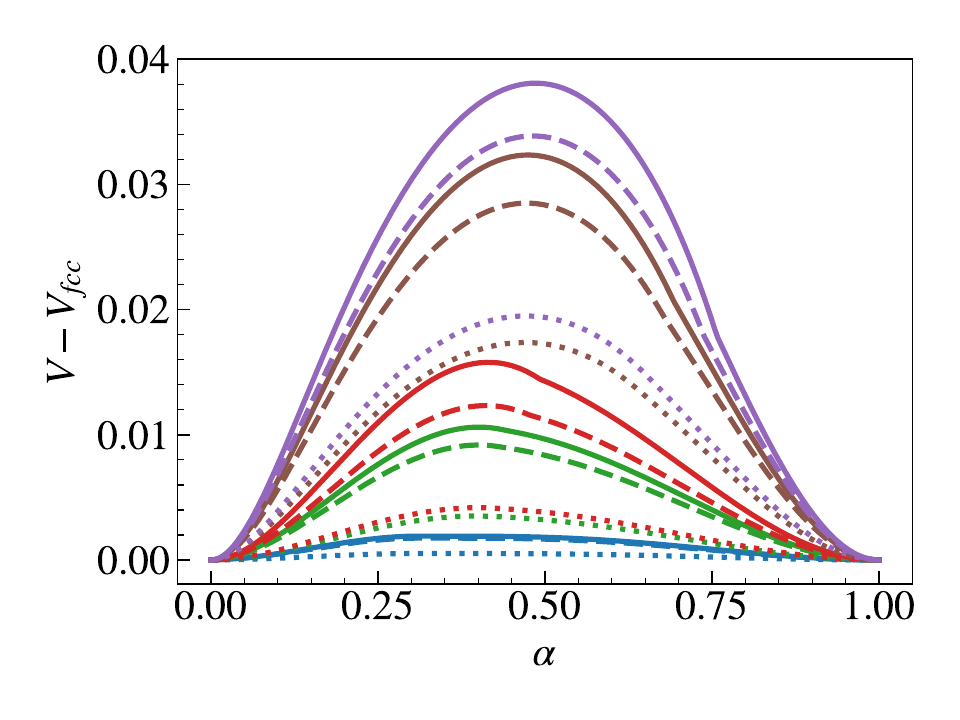}
  \caption{Predicted lattice parameters $\Delta a(\alpha) = a(\alpha)-a_\mathrm{fcc}$, $\gamma_2(\alpha)$, $\beta_1(\alpha)$, and volume difference $\Delta V(\alpha) = V(\alpha)-V_\mathrm{fcc}$ along the Burgers-Bain path for $(n,m)$-LJ potentials at the isobars $P=0$, 10, and $10^3$ (reduced units are used throughout). The different $(n,m)$-LJ potentials are distinguished by color, while the three pressures are represented by different line styles. The location of the ideal bcc structure is indicated by a round circle.}\label{fig:T=0LJnmBurgers}
\end{figure*}

The effect of pressure on the enthalpy barrier for the phase transition predicted by different $(n,m)$-LJ potentials is shown in Figures \ref{fig:LJnmEnthalpy} and \ref{fig:T=0LJnmBurgers}. As expected, increasing pressure leads to an increase in the activation enthalpy (see Fig.~\ref{fig:LJnmEnthalpy} and Table~\ref{tab:BBProperties1}). This increase is more pronounced for harder potentials, such as the $(30,12)$-LJ potential. We note, however, that in the sticky hard-sphere limit ($n,m \rightarrow \infty$, $n>m$), particle overlap is forbidden, rendering finite-pressure solid-state results physically meaningless. The cohesive-energy contribution (excluding the $PV$ term) to the activation barrier relative to the fcc (or hcp) structure remains below the sticky hard-sphere limit,\cite{Schwerdtfeger-2026} $\Delta E_\textrm{coh}^{\#}=E_\textrm{coh}(\alpha^{\#})-E_\textrm{coh}(\alpha=1)\le 1$ (in reduced units), where the symbol $\#$ denotes the transition state. At elevated pressures, the $P\Delta V^{\#}$ term clearly dominates the activation enthalpy, as evidenced by the values of $\Delta H^{\#}$ reported in Table \ref{tab:BBProperties1}.
\begin{center}
\begin{table}[t!] 
\tiny
\setlength{\tabcolsep}{3pt}
\begin{tabular}{|l|l||c|c|c|c|c|c|c|c|c|c|c|c|c|c|c|c|} 
 \hline
 Potential  & $P$ & $\alpha^{\#}$ & $\Delta H^{\#}$ & $a^{\#}$ & $\gamma_2^{\#}$& $\beta_1^{\#}$ & $\beta_2^{\#}$ &$\Delta V^{\#} (10^{-2})$&$\alpha_c$ & $\Delta H_c$ & $a_c$ & $\gamma_2^c$& $\beta_1^c$ & $\beta_2^c$ & $\Delta H^{\alpha=\tfrac{1}{3}}$& $\beta_1^{\alpha=\tfrac{1}{3}}$ & $\beta_2^{\alpha=\tfrac{1}{3}}$ \\ 
 \hline 
(6,4) & 0 & 0.3568 & 0.2981 & 0.7357 & 1.6258 & 1.3216 & 0.9469 & 0.1902 & 0.3216 & 0.2981 & 0.7357 & 1.6365 & 1.3392 & 0.9407 & 0.2981 & 1.3333 & 0.9428 \\
 & 10 & 0.3568 & 0.3163 & 0.7290 & 1.6258 & 1.3216 & 0.9469 & 0.1745 & 0.3216 & 0.3163 & 0.7290 & 1.6365 & 1.3392 & 0.9407 & 0.3163 & 1.3333 & 0.9428 \\
 & 100 & 0.3618 & 0.4406 & 0.6923 & 1.6243 & 1.3191 & 0.9477 & 0.1153 & 0.3189 & 0.4406 & 0.6923 & 1.6372 & 1.3309 & 0.9336 & 0.4405 & 1.3333 & 0.9428 \\
 & 1000 & 0.3737 & 1.0619 & 0.6027 & 1.6206 & 1.3131 & 0.9497 & 0.0511 & 0.3131 & 1.0619 & 0.6027 & 1.6388 & 1.3434 & 0.9392 & 1.0611 & 1.3333 & 0.9428 \\
 & 10000 & 0.3869 & 3.6737 & 0.4940 & 1.6165 & 1.3065 & 0.9519 & 0.0214 & 0.3081 & 3.6740 & 0.4940 & 1.6402 & 1.3356 & 0.9312 & 3.6680 & 1.3333 & 0.9428 \\
(9.4294,8.014) & 0 & 0.4322 & 0.3269 & 0.9609 & 1.5918 & 1.1371 & 0.8492 & 1.5282 & 0.4882 & 0.3140 & 0.9616 & 1.5725 & 1.2505 & 0.9627 & 0.2929 & 1.0796 & 0.7634 \\
 & 10 & 0.4221 & 0.4574 & 0.9349 & 1.5967 & 1.1524 & 0.8559 & 1.1517 & 0.4693 & 0.4447 & 0.9355 & 1.5809 & 1.2576 & 0.9584 & 0.4147 & 1.0915 & 0.7718 \\
 & 100 & 0.4050 & 1.1711 & 0.8630 & 1.6052 & 1.1861 & 0.8728 & 0.6366 & 0.4350 & 1.1568 & 0.8632 & 1.5955 & 1.2825 & 0.9593 & 1.0892 & 1.1168 & 0.7897 \\
 & 1000 & 0.3920 & 4.8768 & 0.7585 & 1.6114 & 1.2207 & 0.8918 & 0.3284 & 0.4070 & 4.8563 & 0.7586 & 1.6068 & 1.2965 & 0.9551 & 4.6365 & 1.1413 & 0.8070 \\
 & 10000 & 0.3800 & 24.4750 & 0.6507 & 1.6162 & 1.2331 & 0.8950 & 0.1763 & 0.3937 & 24.4126 & 0.6508 & 1.6120 & 1.2996 & 0.9505 & 23.5700 & 1.1581 & 0.8189 \\
(12,4) & 0 & 0.4121 & 0.4551 & 0.8940 & 1.6023 & 1.1817 & 0.8729 & 1.0596 & 0.4422 & 0.4487 & 0.8942 & 1.5926 & 1.2789 & 0.9603 & 0.4212 & 1.1107 & 0.7854 \\
 & 10 & 0.4070 & 0.5533 & 0.8796 & 1.6041 & 1.1808 & 0.8698 & 0.9179 & 0.4388 & 0.5461 & 0.8799 & 1.5940 & 1.2725 & 0.9539 & 0.5133 & 1.1135 & 0.7873 \\
 & 100 & 0.4020 & 1.1898 & 0.8258 & 1.6065 & 1.1909 & 0.8749 & 0.6000 & 0.4296 & 1.1779 & 0.8260 & 1.5979 & 1.2777 & 0.9530 & 1.1120 & 1.1214 & 0.7930 \\
 & 1000 & 0.4000 & 4.9783 & 0.7335 & 1.6079 & 1.2047 & 0.8840 & 0.3515 & 0.4200 & 4.9443 & 0.7336 & 1.6017 & 1.2899 & 0.9570 & 4.6830 & 1.1287 & 0.7981 \\
 & 10000 & 0.3950 & 27.7118 & 0.6359 & 1.6096 & 1.2026 & 0.8801 & 0.2140 & 0.4184 & 27.5148 & 0.6360 & 1.6024 & 1.2863 & 0.9536 & 26.1399 & 1.1318 & 0.8003 \\
(12,6) & 0 & 0.4322 & 0.3477 & 0.9545 & 1.5915 & 1.1333 & 0.8463 & 1.5733 & 0.4925 & 0.3329 & 0.9553 & 1.5709 & 1.2538 & 0.9673 & 0.3114 & 1.0783 & 0.7625 \\
 & 10 & 0.4271 & 0.4849 & 0.9307 & 1.5945 & 1.1472 & 0.8544 & 1.2296 & 0.4774 & 0.4691 & 0.9314 & 1.5776 & 1.2613 & 0.9653 & 0.4380 & 1.0872 & 0.7688 \\
 & 100 & 0.4121 & 1.2820 & 0.8630 & 1.6017 & 1.1698 & 0.8641 & 0.7321 & 0.4503 & 1.2587 & 0.8634 & 1.5893 & 1.2692 & 0.9574 & 1.1796 & 1.1055 & 0.7817 \\
 & 1000 & 0.4040 & 5.8176 & 0.7643 & 1.6058 & 1.1920 & 0.8766 & 0.4176 & 0.4343 & 5.7400 & 0.7646 & 1.5960 & 1.2828 & 0.9592 & 5.4309 & 1.1205 & 0.7923 \\
 & 10000 & 0.3970 & 32.5600 & 0.6637 & 1.6087 & 1.1972 & 0.8771 & 0.2505 & 0.4221 & 32.3049 & 0.6638 & 1.6009 & 1.2889 & 0.9573 & 30.6070 & 1.1279 & 0.7976 \\
(30,6) & 0 & 0.4824 & 0.5805 & 0.9861 & 1.5616 & 1.0366 & 0.7954 & 3.2343 & 0.6935 & 0.4106 & 0.9878 & 1.4941 & 1.1533 & 0.9882 & 0.5087 & 1.0198 & 0.7211 \\
 & 10 & 0.4774 & 0.8819 & 0.9729 & 1.5643 & 1.0428 & 0.7980 & 2.8509 & 0.6784 & 0.6422 & 0.9750 & 1.5001 & 1.1608 & 0.9870 & 0.7710 & 1.0240 & 0.7241 \\
 & 100 & 0.4724 & 3.0588 & 0.9343 & 1.5678 & 1.0539 & 0.8044 & 2.2158 & 0.6533 & 2.3365 & 0.9366 & 1.5103 & 1.1734 & 0.9849 & 2.6714 & 1.0310 & 0.7290 \\
 & 1000 & 0.4724 & 19.9573 & 0.8781 & 1.5688 & 1.0599 & 0.8089 & 1.7376 & 0.6432 & 15.4869 & 0.8804 & 1.5144 & 1.1784 & 0.9840 & 17.4365 & 1.0344 & 0.7314 \\
 & 10000 & 0.4673 & 155.0099 & 0.8202 & 1.5704 & 1.0601 & 0.8069 & 1.3978 & 0.6382 & 121.8208 & 0.8223 & 1.5164 & 1.1809 & 0.9835 & 135.4543 & 1.0352 & 0.7320 \\
(30,12) & 0 & 0.4925 & 0.7622 & 0.9966 & 1.5542 & 1.0136 & 0.7820 & 3.8088 & 0.7588 & 0.4715 & 0.9970 & 1.4708 & 1.1206 & 0.9927 & 0.6820 & 1.0061 & 0.7114 \\
 & 10 & 0.4874 & 1.1192 & 0.9867 & 1.5573 & 1.0216 & 0.7860 & 3.3873 & 0.7387 & 0.7163 & 0.9882 & 1.4781 & 1.1307 & 0.9914 & 0.9912 & 1.0112 & 0.7150 \\
 & 100 & 0.4774 & 3.6770 & 0.9541 & 1.5638 & 1.0399 & 0.7958 & 2.5756 & 0.6935 & 2.5655 & 0.9566 & 1.4950 & 1.1533 & 0.9882 & 3.2184 & 1.0227 & 0.7232 \\
 & 1000 & 0.4724 & 22.8434 & 0.9014 & 1.5678 & 1.0536 & 0.8042 & 1.9497 & 0.6583 & 17.1782 & 0.9039 & 1.5086 & 1.1709 & 0.9853 & 19.9543 & 1.0310 & 0.7291 \\
 & 10000 & 0.4724 & 172.4826 & 0.8439 & 1.5687 & 1.0592 & 0.8084 & 1.5417 & 0.6432 & 134.1699 & 0.8461 & 1.5144 & 1.1784 & 0.9840 & 150.6971 & 1.0341 & 0.7312 \\
ELJ & 0 & 0.4322 & 0.3459 & 0.9620 & 1.5916 & 1.1350 & 0.8476 & 1.4716 & 0.4904 & 0.3313 & 0.9627 & 1.5710 & 1.2485 & 0.9621 & 0.3091 & 1.0779 & 0.7622 \\
 & 10 & 0.4221 & 0.4683 & 0.9359 & 1.5971 & 1.1586 & 0.8605 & 1.0503 & 0.4637 & 0.4568 & 0.9364 & 1.5828 & 1.2604 & 0.9576 & 0.4252 & 1.0935 & 0.7732 \\
 & 100 & 0.3920 & 1.0358 & 0.8609 & 1.6114 & 1.2189 & 0.8905 & 0.4404 & 0.4074 & 1.0313 & 0.8609 & 1.6065 & 1.2927 & 0.9525 & 0.9828 & 1.1395 & 0.8058 \\
 & 1000 & 0.3367 & 2.5571 & 0.7436 & 1.6320 & 1.3296 & 0.9419 & 0.0795 & 0.3367 & 2.5571 & 0.7436 & 1.6320 & 1.3296 & 0.9419 & 2.5567 & 1.3018 & 0.9205 \\

\hline
\end{tabular}
\scriptsize{}
\caption{Parameters for the transition state ($\#$) and critical (bifurcation) point ($c$) of the Burgers-Bain path at different pressures (in reduced units). Values for the energy and lattice parameters at $\alpha=\tfrac{1}{3}$ are also shown for comparison with the ideal bcc structure given in Table \ref{tab:latticeparameters}. Differences are with respect to the fcc structure. For the ELJ potential, the pressure of $P=10,000$ is outside the validity range and is therefore not reported.}
\label{tab:BBProperties1}
\end{table}
\end{center}

\begin{figure}[htb!]
\centering
\includegraphics[width=.66\columnwidth]{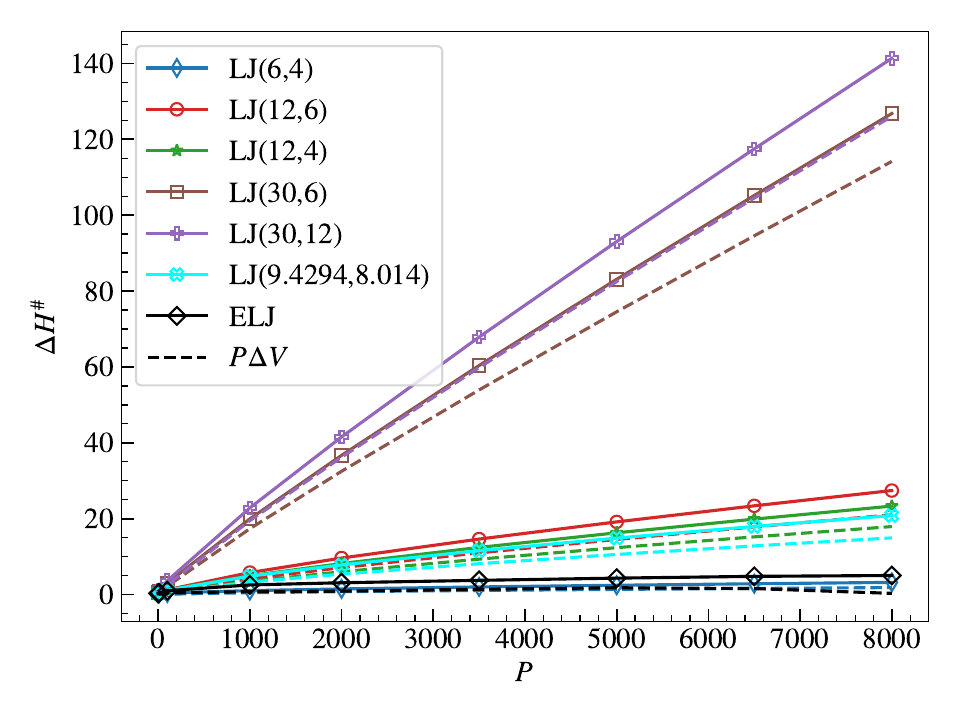}
\caption{Enthalpy difference $\Delta H^{\#}(P) = H^{\#}(P)-H_\textrm{fcc}(P)$ for the transition state along the Burgers-Bain path (solid lines) in comparison to the pressure term $P\Delta V^{\#}(P)$.}
\label{fig:HtransP}
\end{figure}

It is well known that the enthalpy change due to pressure is relatively small for a solid (or liquid) and related to the small volume decrease in $(\partial H / \partial P)_T=V-T(\partial V / \partial T)_P$ with increasing pressure. Figure \ref{fig:HtransP} shows that the enthalpy difference with respect to the fcc reference structure, $\Delta H^{\#} (P)= H^{\#}(P)-H_\mathrm{fcc}(P)$, exhibits an almost linear dependence on pressure. Hence, the change in the activation barrier $\Delta H^{\#} (P)$ remains relatively small even at extreme pressures. In the high-pressure regime, the $PV$ contribution clearly dominates over the internal (cohesive) energy, and the volume difference, $\Delta V^{\#}=V^{\#}-V_\mathrm{fcc}$, varies only weakly with pressure for a given $(n,m)$-LJ potential (see Table \ref{tab:BBProperties1}). Consequently, the increase in the term $P\Delta V^{\#}$ arises primarily from the increase in pressure itself, and we obtain an almost linear behavior of $\Delta H^{\#}(P)$ at high pressures. This rationalizes why even at high pressures solid-state phase transitions can have relatively low energy barriers as the volume change is small. This is also the case for more realistic potentials such as the ELJ potential or for DFT as discussed in the next sections.

The volume change along the Burgers-Bain path is dominated by the evolution of the lattice parameters $a$, $\gamma_2$ and $\alpha$. In the kissing hard-sphere limit, the change in volume relative to the fcc (or hcp) structure is given by a simple expression derived from basic geometry\cite{Schwerdtfeger-2026}
\begin{equation}
\Delta V(\alpha)=\frac{1}{2}\left(\sqrt{2+\alpha(1-\alpha)}-\sqrt{2}\right),
\end{equation}
per atom in the unit cell. This function reaches its maximum at exactly $\alpha=\tfrac{1}{2}$, corresponding to a change in volume of $\Delta V^{\#}=\tfrac{3}{4}-\tfrac{1}{2}\sqrt{2}=0.042893$ at the transition state. In this limit the volume change is described entirely by the change in angle between the lattice vectors $\vec{b}_1$ and $\vec{b}_2$. We also notice that the volume for the bcc structure relative to fcc with $\Delta V=\tfrac{1}{2}\left( \tfrac{8}{3\sqrt{3}}-\sqrt{2}\right)=0.062694$ lies above $\Delta V(\alpha)$ in the hard-sphere limit. As an increase in volume costs energy, the bcc structure is bypassed for the Burgers-Bain path in the hard-sphere limit. As pointed out previously, this hard-sphere result represents an upper bound for the $(n,m)$-LJ potentials, where a finite overlap of the spheres ($a<1$) is permitted,\cite{Schwerdtfeger-2026} and also for all pressures, since the volume change decreases with increasing pressure (see Figure~\ref{fig:T=0LJnmBurgers}(d)). For the most strongly repulsive potential considered, the (30,12)-LJ potential, we obtain $\Delta V^{\#}=0.038088$, which lies just below the hard-sphere limit. 

\begin{figure}[htb!]
\centering
\includegraphics[width=.66\columnwidth]{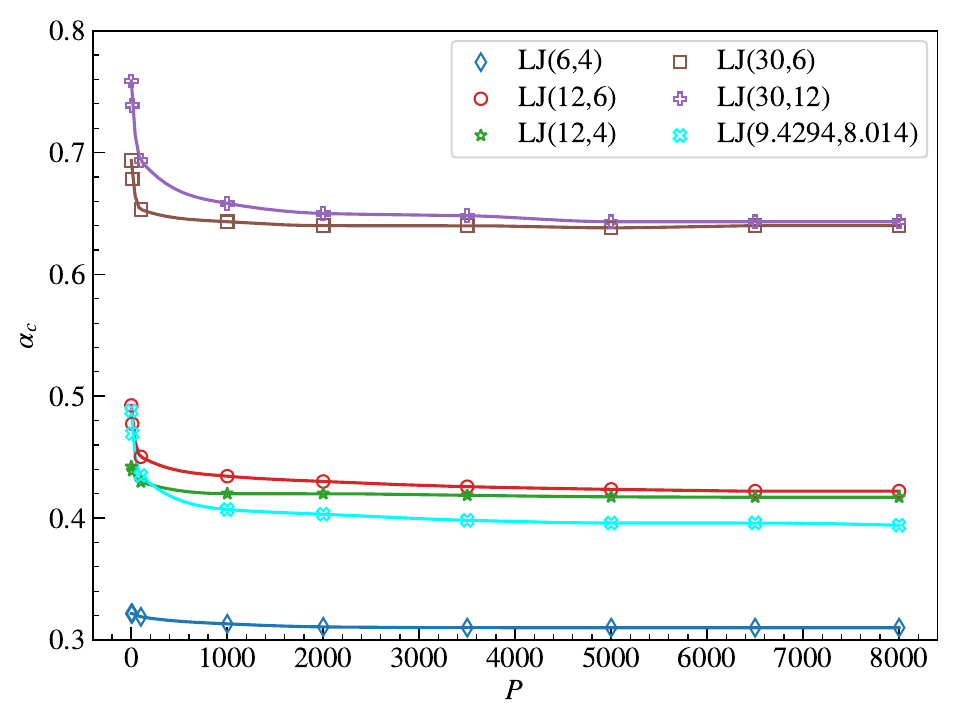}
\caption{Bifurcation point $\alpha_c$ as a function of pressure for various $(n,m)$-LJ potentials.}
\label{fig:bifurcationP}
\end{figure}

We previously reported zero-pressure results for the hcp-to-fcc/bcc Burgers minimum-energy path.\cite{Schwerdtfeger-2026} There, we found that the Burgers path merges with the Bain cuboidal-distortion path through a symmetry breaking process involving a pitchfork bifurcation point at $\alpha_c$. At this critical point, the distorted hcp-type bilattice transforms into a simple cuboidal Bravais lattice located on a Bain path connecting both the bcc and fcc structures. Unlike the original Burgers path, the Bain path exhibits a nice linear relationship for $\beta_1(\alpha)$ as pointed out in our previous work,\cite{Schwerdtfeger-2026}
\begin{equation}\label{eq:beta1linear}
\beta_1(\alpha)=\frac{1}{2}\left(3-\alpha\right)
\end{equation}
with the bifurcation point $\alpha_c$ located exactly on that line. This basic mechanism is obviously preserved at all pressures, as shown in Figure~\ref{fig:T=0LJnmBurgers}. However, increasing pressure shifts the bifurcation point $\alpha_c$ to lower values of $\alpha$, i.e. towards the bcc structure along the Bain path. The magnitude of this shift depends on the interaction potential, with the more repulsive (30,12)-LJ and (30,6)-LJ potentials exhibiting a stronger pressure dependence than the more attractive (6,4)-LJ potential, as illustrated in Figure~\ref{fig:bifurcationP}. This trend cannot persist in the sticky hard-sphere limit, where $\alpha_c$ is expected to converge to the limiting value of 1. Furthermore, for all LJ potentials considered here, the pressure-induced shift in $\alpha_c$ becomes progressively smaller at high pressures. As in the zero-pressure case, the lattice parameters as a function of $\alpha$ exhibit a kink (discontinuity in the first derivative) at the critical point $\alpha_c$, reflecting the underlying symmetry-breaking transition. We should mention that the small differences in the lattice parameters reported in Table~\ref{tab:BBProperties1}, compared with our previous zero-pressure results,\cite{Schwerdtfeger-2026} arise from minor differences in the numerical treatment.

Similar to the bifurcation point, the location of the transition state at $\alpha^{\#}$ generally decreases with increasing pressure moving more towards the bcc structure, except for the (6,4)-LJ potential where the bcc phase becomes a shallow minimum along the Burgers-Bain path with the corresponding lattice parameters listed in Table~\ref{tab:latticeparameters}.  For this potential, the bifurcation point satisfies $\alpha_c<\alpha^{\#}$, in contrast to the behavior observed for the other LJ potentials studied here. Otherwise, at $\alpha=\tfrac{1}{3}$ the Burgers-Bain minimum energy path bypasses the ideal bcc structure, as for example the deviation from the ideal bcc lattice parameter $\beta_2=2\sqrt{2}/3=0.942809$ at $\alpha=\tfrac{1}{3}$ shows. In this case, the bcc structure becomes a second-order saddle point with distortions in the lattice parameters $\alpha$, $\beta_1$, $\beta_2$ and $\gamma_2$. Consistent with this observation, the derivative $\partial H / \partial \alpha$ evaluated at $\alpha=\tfrac{1}{3}$ increases with increasing pressure. Only for the (6,4)-LJ potential does the bcc structure become a very shallow local minimum along the Burgers-Bain path, which is barely discernible in the enthalpy profiles shown in Figure~\ref{fig:LJnmEnthalpy}. Hence, by looking at the energy only one can easily miss the symmetry breaking mechanism, especially when using DFT as discussed further below. This shows the advantage of exact lattice summations for the Epstein zeta function.

Figure \ref{fig:T=0LJnmBurgers}(b) suggests that, at the two endpoints of the Burgers-Bain path corresponding to the fcc and hcp structures, the lattice parameter $\gamma_2$ approaches its ideal value, as listed in Table \ref{tab:latticeparameters}. While this is indeed the case for the fcc structure, the hcp structure exhibits small deviations from the ideal close-packed (hard-sphere) geometry. Such deviations at zero pressure are well documented (see, for example, Ref.~\onlinecite{Cooper2023}).

\begin{figure*}[bt!]
  \centering
    \centering
    (a)\includegraphics[width=.45\columnwidth]{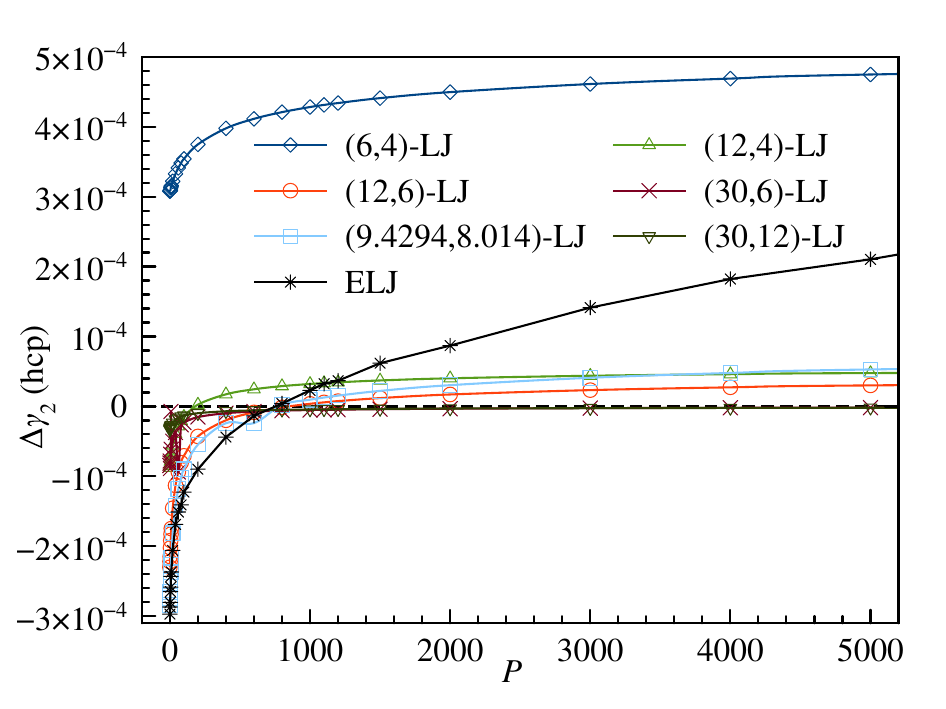}
    (b)\includegraphics[width=.45\columnwidth]{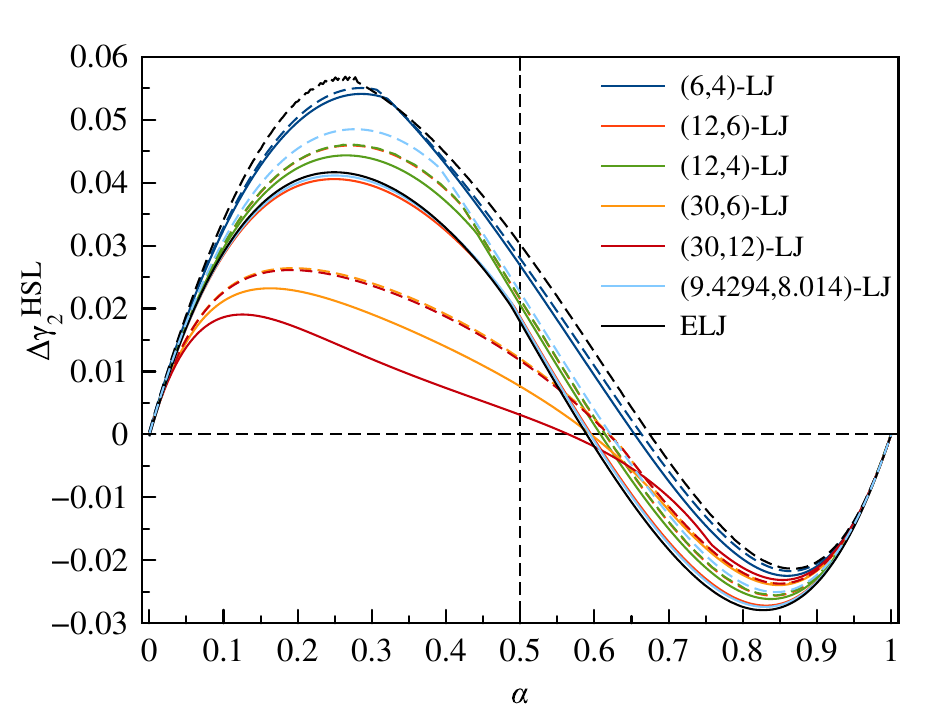}
  \caption{(a) Deviation from the ideal $\gamma_2$ value, $\Delta\gamma_2(P)=c(P)/a(P)-\sqrt{8/3}$, as a function of pressure for the hcp structure using various $(n,m)$-LJ as well as the ELJ potential. (b) Deviation from the hard-sphere limit (HSL) result, $\Delta\gamma_2^\textrm{HSL}(\alpha,P)=\gamma_2(\alpha,P)-2\sqrt{(2-\alpha)/(3-\alpha)}$, for the two pressures (in reduced units) $P=0$ (solid line) and $P=10000$ (dashed line) for various $(n,m)$-LJ potentials.}\label{fig:hcpgamma2}
\end{figure*}

Figure~\ref{fig:hcpgamma2} shows the pressure dependence of the deviation from the ideal hcp axial ratio,
\begin{equation}
    \Delta\gamma_2(P)=\frac{c(P)}{a(P)}-\sqrt{\frac{8}{3}},
\end{equation}
for a range of $(n,m)$-LJ potentials. The largest variation occurs at low pressures, after which $\Delta\gamma_2(P)$ gradually levels off with increasing pressure. With the exception of the soft $(6,4)$-LJ and the ELJ potentials, $\Delta\gamma_2(P)$ appears to approach zero asymptotically. This behavior reflects the increasing dominance of the short-range repulsive interactions under compression, which drive the crystal towards the ideal close-packed geometry. For the more realistic ELJ potential developed for solid argon, $\Delta\gamma_2(P)$ changes sign at approximately $P=$750.9 in reduced units ($P=27.9$ GPa), although its magnitude remains exceedingly small over the entire pressure range considered. Consequently, the hcp structure remains extremely close to the ideal close-packed geometry at all pressures, and deviations from the ideal axial ratio can be regarded as negligible for practical purposes.

Figure \ref{fig:T=0LJnmBurgers}(b) shows that $\gamma_2(\alpha)$ varies rather smoothly with $\alpha$, making the symmetry-breaking point barely discernible and easily mistaken for numerical noise. To highlight this feature more clearly, Figure \ref{fig:hcpgamma2}(b) presents the deviation of $\gamma_2(\alpha)$ from the hard-sphere limit (HSL) model, for which an analytical expression is available,
\begin{equation}
\gamma_2^{\mathrm{HSL}}(\alpha,P)=2\sqrt{\frac{2-\alpha}{3-\alpha}}.
\end{equation}
The resulting deviations,
$\Delta\gamma_2^{\mathrm{HSL}}(\alpha,P)$, are small, demonstrating that the HSL relation is followed to an excellent approximation throughout the transformation path. The point of symmetry breaking is more clearly seen in this figure. Furthermore, pressure-induced changes in $\Delta\gamma_2^{\mathrm{HSL}}(\alpha,P)$ become more pronounced for potentials with steeper short-range repulsion. In this respect, the ELJ potential behaves more similarly to the softer $(6,4)$-LJ potential than to the other LJ potentials considered.

While the LJ model provides valuable insight into the mechanism of symmetry breaking along the Burgers-Bain path and serves as a useful starting point for understanding the underlying physics, more accurate many-body potentials are required for real systems such as solid argon in the high-pressure regime. In this context, the standard (12,6)-LJ potential is known to become inadequate, motivating the use of more realistic interaction models such as the ELJ potential including the 3-body ATM or E3B term as discussed in the next section.

\subsection{Burgers-Bain Path for a ELJ+ATM/E3B Potential}

\begin{figure*}[bt!]
  \centering
  \begin{subfigure}{0.49\textwidth}
    \centering
    \includegraphics[width=\linewidth]{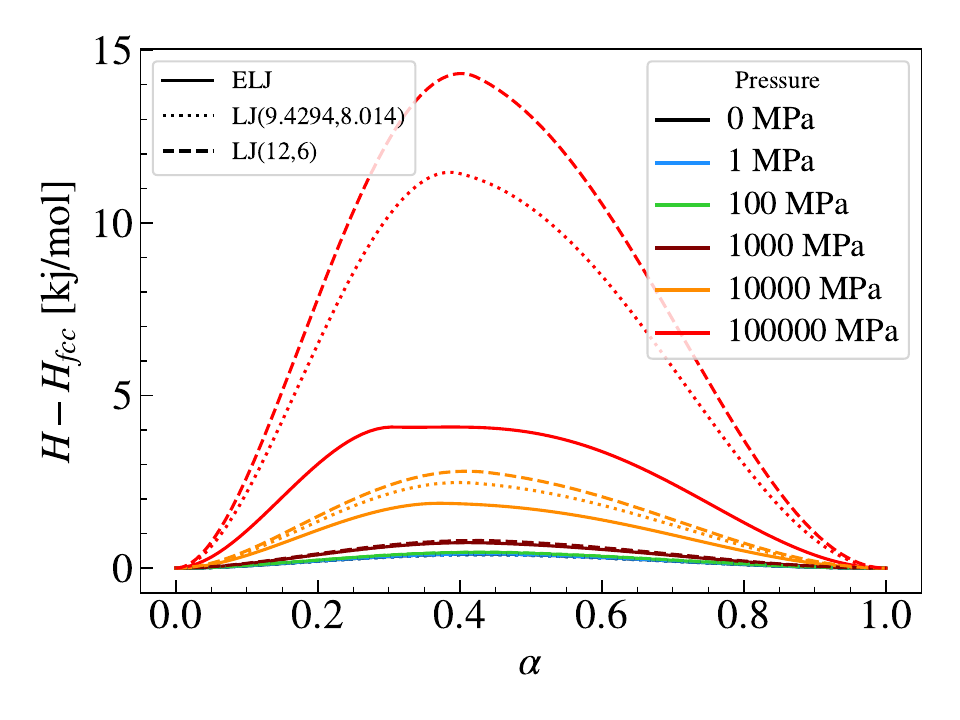}
  \end{subfigure}%
  \caption{Enthalpy differences along the Burgers-Bain path for the (12,6)-LJ, (9.4294,8.014)-LJ and ELJ potentials at isobars of 0, 0.001, 0.1, 1, 10 and 100  GPa. Black line corresponds to 0  GPa and red line to 100 GPa.}\label{fig:T=0LJELJBurgers}
\end{figure*}

\begin{figure*}[bt!]
  \centering
  (a) \includegraphics[width=.45\columnwidth]{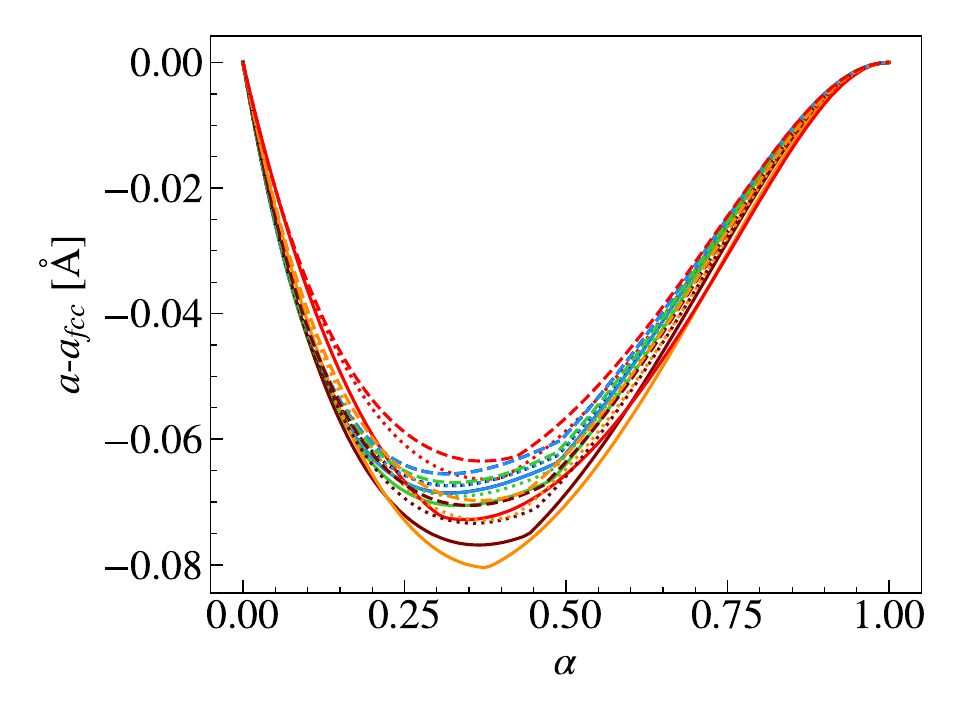}
    (b) \includegraphics[width=.45\columnwidth]{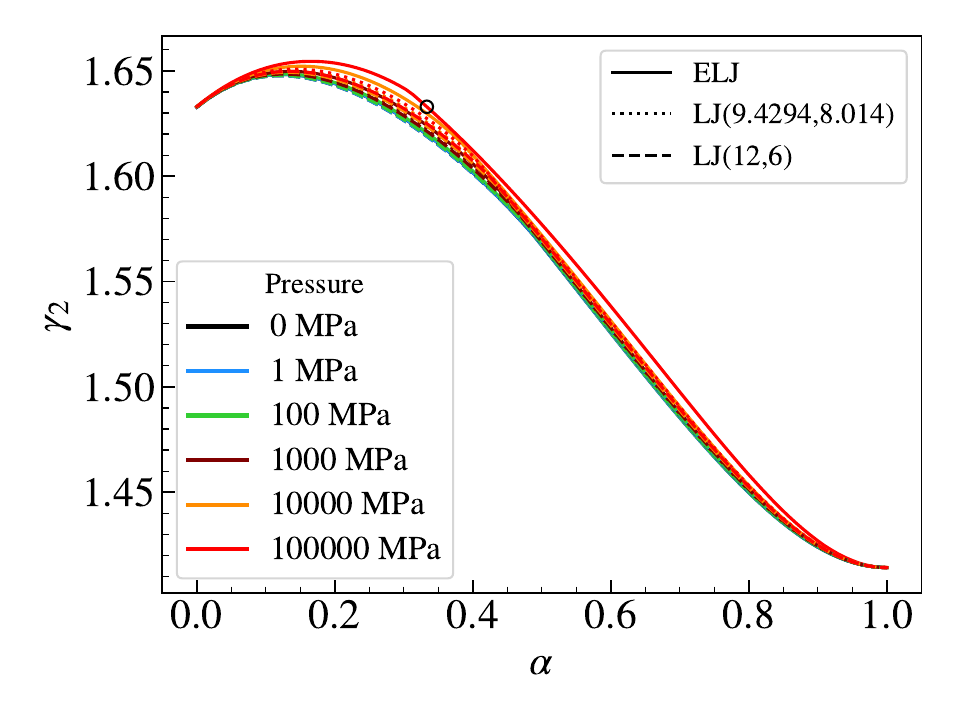}\\
    (c)\includegraphics[width=.45\columnwidth]{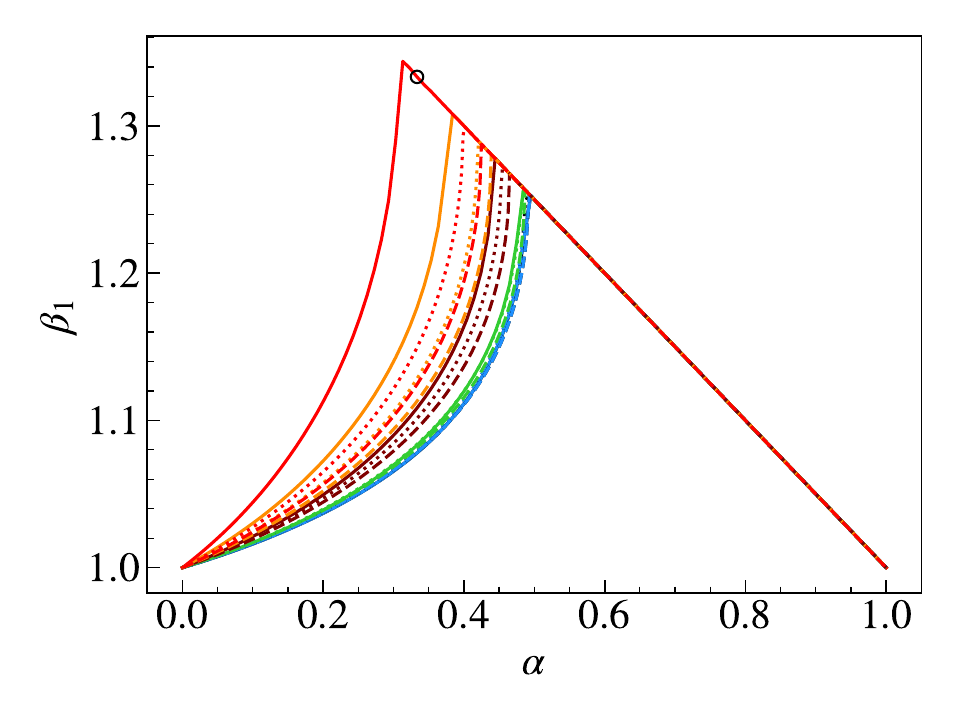}
    (d)\includegraphics[width=.45\columnwidth]{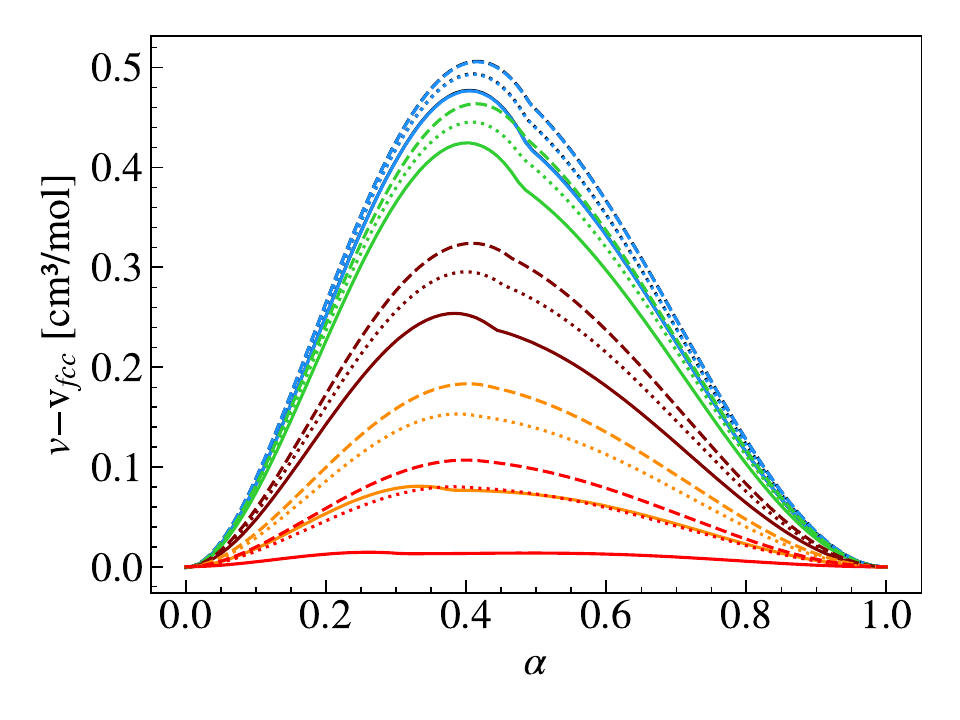}
  \caption{Calculated lattice parameters along the Burgers-Bain path for the (12,6)-LJ, (9.4294,8.014)-LJ and ELJ potentials at isobars of 0, 0.001, 0.1, 1, 10 and 100  GPa, black corresponds to 0 GPa and red to 100 GPa. The open circle marker represents the location of the bcc phase.}\label{fig:T=0LJELJBurgersc}
\end{figure*}

We first discuss the two-body ELJ potential before considering the effect of three-body corrections for solid argon. The enthalpy difference, lattice parameters $\gamma_2$, $\beta_1$, and $\beta_2$, together with the volume difference $\Delta V$ relative to the fcc structure, are shown  in Figures~\ref{fig:T=0LJELJBurgers} and \ref{fig:T=0LJELJBurgersc} for the isobars from 0 to 100 GPa as functions of the distortion parameter $\alpha$. The ELJ results are compared with those obtained using the conventional (12,6)-LJ potential and the optimized (9.4294,8.014)-LJ potential of Walker \textit{et al.}\cite{Walker2022} Overall, the ELJ potential exhibits the same qualitative behavior as the LJ potentials, i.e. compare Figures~\ref{fig:T=0LJnmBurgers} and \ref{fig:T=0LJELJBurgersc}. Quantitative differences, however, become increasingly pronounced at high pressures. This is evident from the activation enthalpies along the Burgers-Bain path (Table~\ref{tab:BBProperties2}), where the widely used (12,6)-LJ potential can deviate substantially from the ELJ results, becoming excessively repulsive above approximately 10 GPa. The resulting discrepancy is reflected in the enthalpy profiles shown in Figure~\ref{fig:T=0LJELJBurgers}. This clearly demonstrates that a more elaborate expansion in terms of inverse power law potentials is required to describe both the short- and long-range regions of the interaction potential.

\begin{figure}[htb!]
\centering
\includegraphics[width=.66\columnwidth]{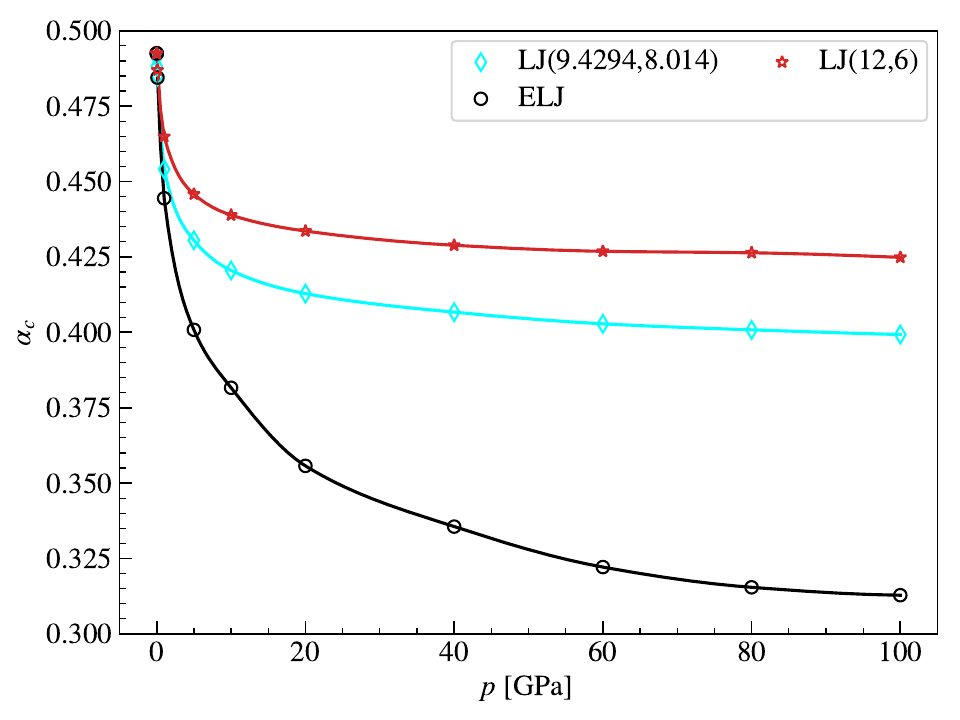}
\caption{Bifurcation point as a function of pressure for the tuned ELJ potential in addition to Lennard Jones potentials (12,6)-LJ and (9.4294,8.014)-LJ. Lines are interpolations added for visibility.}
\label{fig:bifurcationPELJ}
\end{figure}

Concerning the location of the bifurcation point (cf. Figure~\ref{fig:bifurcationPELJ}), we find that $\alpha_c$ shifts towards lower values with increasing pressure, as also observed for the LJ potentials. However, the ELJ potential exhibits a substantially stronger pressure dependence even compared to the softer LJ potential by Walker \textit{et al.}\cite{Walker2022} Furthermore, at high pressures, the location of the transition state becomes identical to the bifurcation point for the ELJ potential (see Table~\ref{tab:BBProperties2}). At approximately $P = 39.4$ GPa, the bifurcation point coincides with the unstable bcc phase at $\alpha_c = \alpha_{\mathrm{bcc}} = \tfrac{1}{3}$ with the $\beta$- and $\gamma$-parameters given in Table \ref{tab:latticeparameters}. This again highlights the importance of an accurate description of the repulsive part of the interaction potential at elevated pressures. $\alpha_c(P)$ decreases approximately exponentially with increasing pressure, consistent with the behavior found for the various LJ potentials. 

An important consequence is that the metastable bcc structure is bypassed along the Burgers-Bain minimum-energy path for pressures below $44.5$ GPa, whereas at $P \approx 44.5$ GPa the Burgers-Bain transition passes directly through the bcc phase. As noted previously, the bcc phase can become thermodynamically stable at sufficiently high temperatures. For example, for the $(12,6)$-LJ potential, the bcc phase appears immediately prior to melting.\cite{Travesset-2014} Similarly, DFT calculations by McMahan predicted a pressure-induced fcc$\rightarrow$hcp$\rightarrow$bcc phase sequence in argon which needs to be further investigated.\cite{McMahan-1986}
 
The volume change associated with the phase transition is small throughout the whole pressure range investigated in accordance with the LJ results. At the transition state, the relative volume change amounts to only 2.17\% at zero pressure and decreases to less than 0.2\% at the highest pressure considered here. We notice the kink in $\Delta V(\alpha)$ due to the symmetry breaking effect at the bifurcation point, which is more difficult to spot for the enthalpy change $\Delta H(\alpha)$. The volume difference between fcc and hcp remains very small for all pressures considered, for example we observe a change of only $2\times 10^{-3} \%$ compared to the total volume at 100 GPa. We note that $V_\textrm{hcp}<V_\textrm{fcc}$ below about 8.05 GPa and $V_\textrm{hcp}>V_\textrm{fcc}$ above that pressure for the ELJ potential. Interestingly, the condition that $V_\textrm{hcp}=V_\textrm{fcc}$ is in the pressure range of 2-13 GPa where the hcp-to-fcc phase transition takes place.\cite{dewaele2021stability}

\begin{figure*}[htb!]
  \centering
  \begin{subfigure}{0.49\textwidth}
    \centering
    \includegraphics[width=\linewidth]{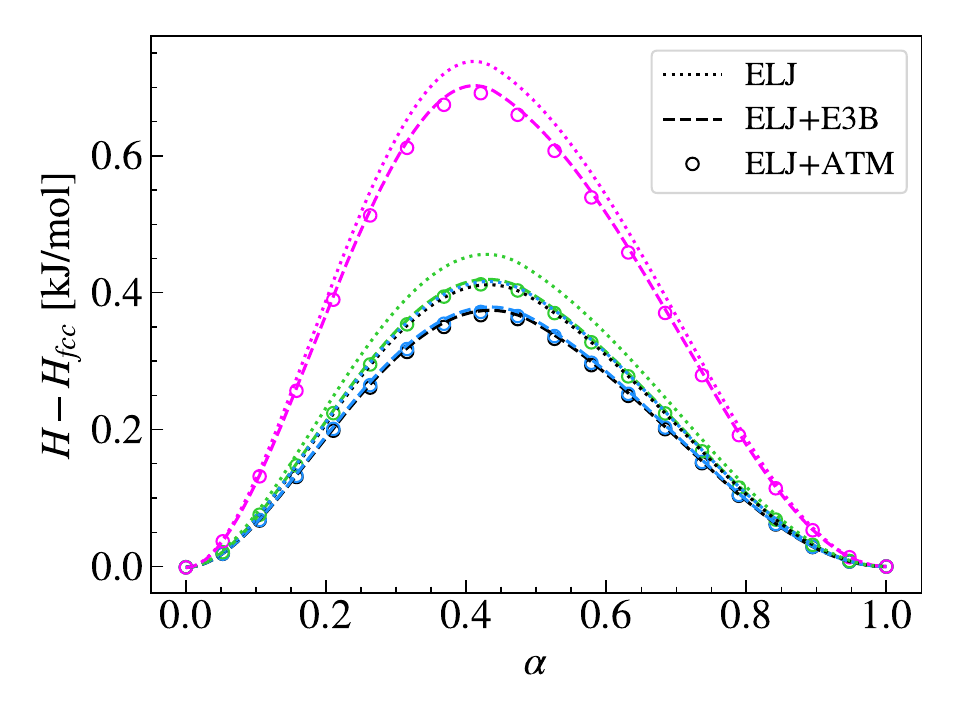}
  \end{subfigure}%
  \hfill
  \begin{subfigure}{0.49\textwidth}
    \centering
    \includegraphics[width=\linewidth]{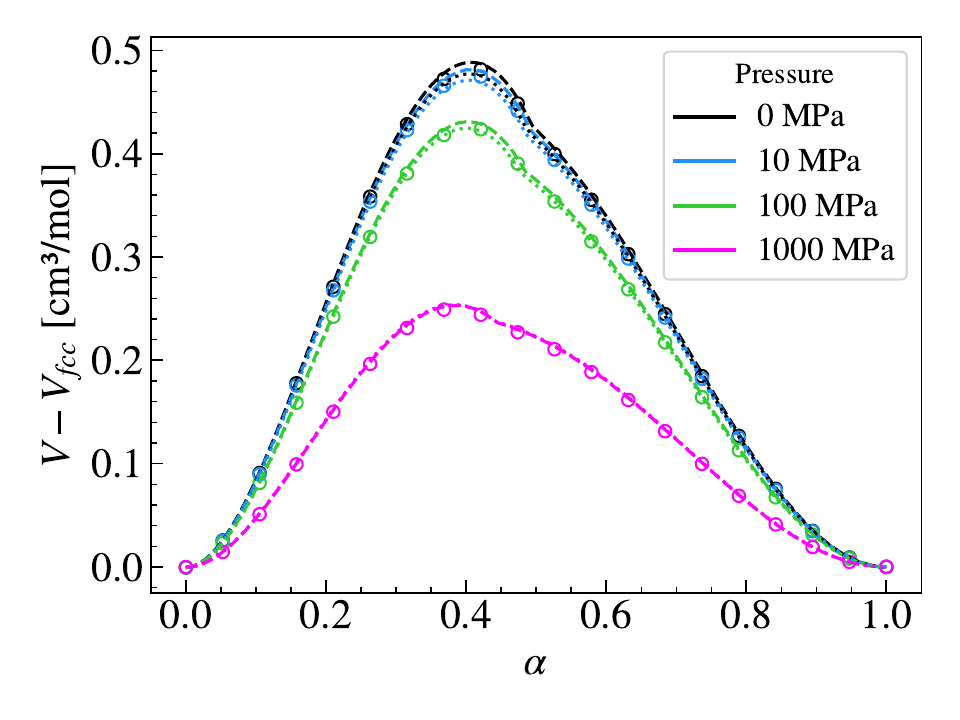}
  \end{subfigure}%
  \newline
  \captionsetup{justification=raggedright,singlelinecheck=false}
  \centering
  \caption{Enthalpy and volume difference along the Burgers-Bain path for the ab-initio tuned ELJ potential with the ATM or E3B 3-body potential contribution at different isobars.}\label{fig:ATMisobars}
\end{figure*}

Concerning the influence of three-body interactions on the transition path, Figure~\ref{fig:ATMisobars} compares results obtained using the ELJ potential alone with those including the ATM and the E3B three-body interaction respectively. Here we restrict the pressure range to $P \leq 1$ GPa because the ATM potential, being an asymptotic long-range approximation, is not expected to remain quantitatively correct at higher densities.\cite{Hermann-2009} Nevertheless, it can be effectively treated through Epstein zeta functions (see appendix \ref{sec:EpsteinATM}) and provides a useful model for assessing the qualitative impact of three-body interactions on the Burgers-Bain minimum-energy path.

\begin{figure*}[htb!]
  \centering
  \begin{subfigure}{0.49\textwidth}
    \centering
    \includegraphics[width=\linewidth]{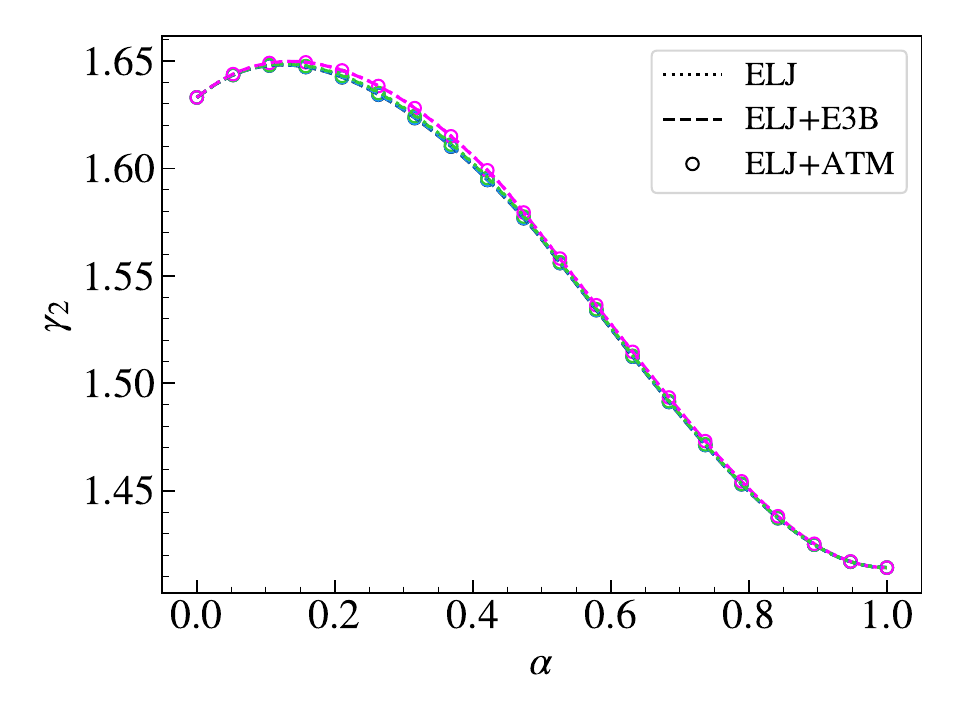}
  \end{subfigure}%
  \hfill
  \begin{subfigure}{0.49\textwidth}
    \centering
    \includegraphics[width=\linewidth]{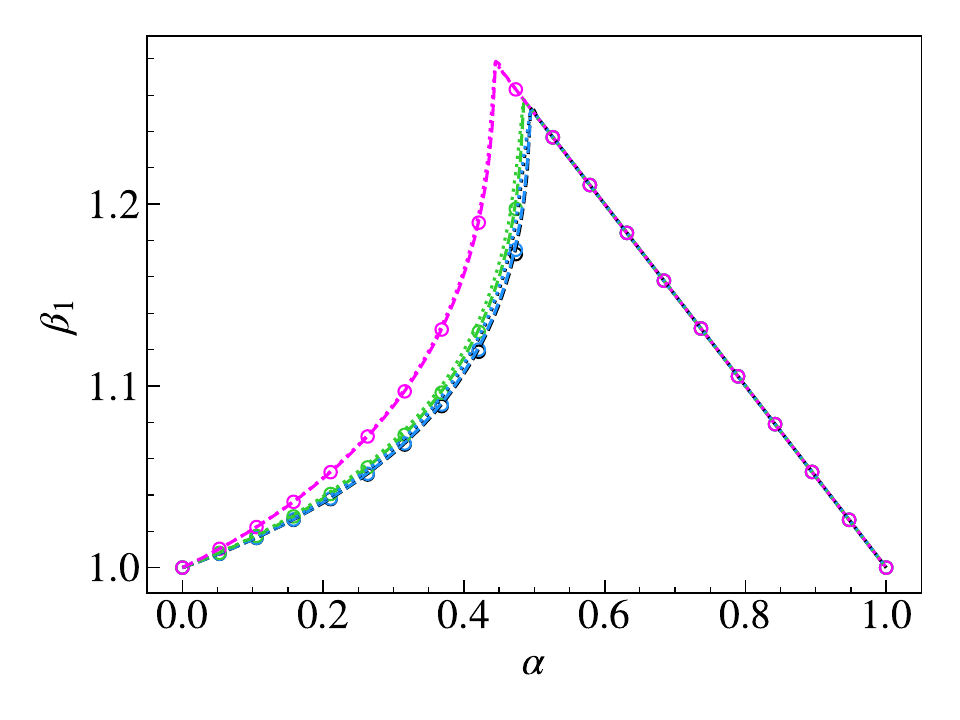}
  \end{subfigure}%
  \newline
  \captionsetup{justification=raggedright,singlelinecheck=false}
  \centering
  \begin{subfigure}{0.49\textwidth}
    \centering
    \includegraphics[width=\linewidth]{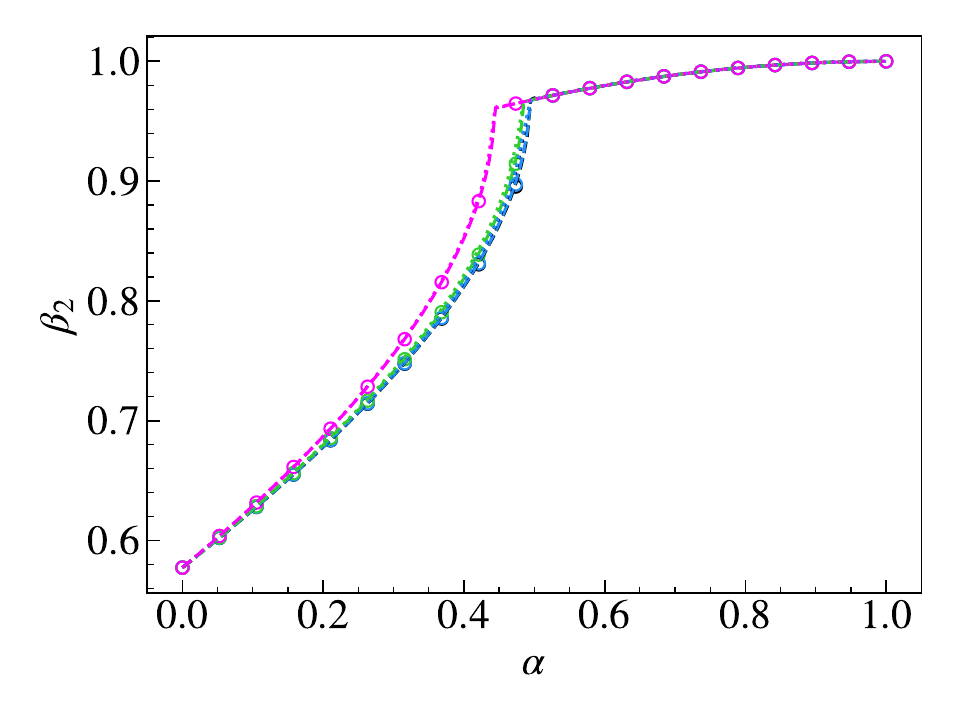}
  \end{subfigure}%
  \centering
  \begin{subfigure}{0.49\textwidth}
    \centering
    \includegraphics[width=\linewidth]{T=0_ATM_alpha_v.pdf}
  \end{subfigure}%
  \caption{Predicted lattice parameters along the Burgers-Bain path for the ELJ potential with and without an effective 3-body potential (E3B) at different isobars.}\label{fig:eff3isobars}
\end{figure*}

The enthalpy profiles in Figure~\ref{fig:ATMisobars} show that the inclusion of the three-body term lowers the activation barrier along the Burgers-Bain path at all pressures considered, including in the low-pressure regime. The effect on the volume change is more subtle and exhibits a noticeable pressure dependence. At low pressures, the three-body contribution increases the volume difference $\Delta V^{\#}$, whereas at $P=1$ GPa it leads to a slightly smaller value of $\Delta V^{\#}$. The remaining lattice parameters are affected only weakly, although the bifurcation point shifts to larger values of the distortion parameter $\alpha$ when the ATM or E3B three-body interaction is included. Overall, the effective three-body (E3B) potential reproduces the ATM results with high accuracy, indicating that it captures the essential effects of three-body interactions while avoiding the additional computational cost associated with explicit three-body calculations.

In our previous work,\cite{Andres2025a} we demonstrated that sufficiently strong ATM interactions can destabilize the fcc structure and favor the formation of bcc-like phases. For solid argon, however, the ATM interaction is too weak to induce such behavior, and neither a stable nor a metastable bcc phase is found within the pressure range investigated here.

\subsection{Density Functional Calculations}

\begin{figure}[ht!]
    \centering
    a)\includegraphics[]{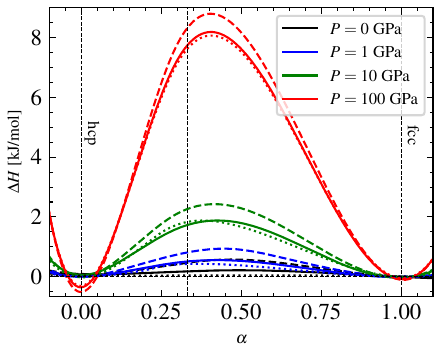}
    b)\includegraphics[]{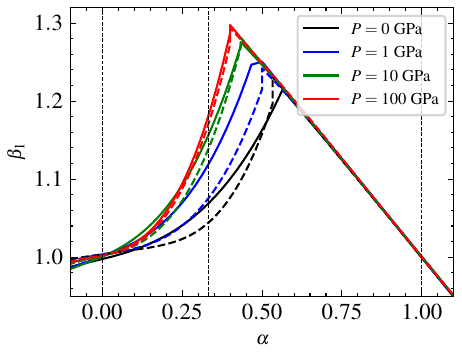}
    \caption{DFT transition paths for argon at various pressures. a) Difference in enthalpy with respect to fcc in kJ/mol. b) $\beta_{1}$ parameter showing the symmetry breaking joining the cuboidal path. The solid line corresponds to the PBE-D4 results, dashed line to PBE and the dotted line to LDA. PBE-D3 yield similar results to PBE-D4 and thus omitted from the plot. The vertical dashed line denotes the position of hcp, fcc and $\alpha=1/3$.}
    \label{fig:DFTtrans}
\end{figure}

It is useful to compare our results obtained from exact lattice summations using inverse power–law potentials with DFT. The DFT approach, which includes many-body effects to a certain level of accuracy, is also among the most widely used methods in solid-state physics. In agreement with the $(n,m)$-LJ and ELJ results, the DFT-calculated enthalpy barrier associated with the phase transition increases with pressure (see Figure \ref{fig:DFTtrans}). However, while the activation barriers obtained from DFT agree well with those from ELJ - for example, at 10 GPa we find 1.938 kJ/mol from PBE-D4 compared to 1.879 and 1.812 kJ/mol from ELJ and ELJ+E3B, respectively - the agreement deteriorates at higher pressures. At 100 GPa, DFT yields 8.184 kJ/mol (PBE-D4) whereas ELJ gives 4.088 kJ/mol. Since 100 GPa lies already outside the validity range of the ELJ and E3B potentials and the results reported above this pressure have to be taken with care, DFT is better suited for this extreme-pressure regime. The barrier decreases substantially upon lowering the pressure to the range where the hcp-to-fcc phase transition occurs. This reduction makes it difficult to solidify the hcp phase at zero pressure.

At zero pressure, the fcc and hcp phases are nearly degenerate and the stability of the fcc over the hcp phase is known to be governed primarily by phonon dispersion effects, with many-body interactions providing a secondary contribution.\cite{Schwerdtfeger-2016} The PBE, PBE-D3, and PBE-D4 calculations all reproduce the near-degeneracy of the fcc and hcp phases at low pressure. At 100 GPa, however, the hcp phase becomes significantly more stable than the fcc phase in accordance with the experimentally observed phase transition from fcc to hcp at pressures around 10 GPa.\cite{dewaele2021stability} Here the $PV$ term dominates all other contributions even at pressures as low as 1 GPa. For example, for hcp at 100 GPa, the electronic and $PV$ contributions are (per atom) $E_{\text{coh}}=2.4845$ eV, $PV=7.8081$ eV, respectively, whereas for fcc they are $E_{\text{coh}}=2.4730$ eV, $PV=7.8237$ eV. Although the fcc phase has a slightly lower electronic energy, this advantage is more than compensated by its larger $PV$ contribution, resulting in the greater thermodynamic stability of the hcp phase under compression. 

Regarding Grimme's dispersion interactions,\cite{Grimme-2011} the energy profile along the transformation path differs significantly between PBE and PBE-D3/D4 at low pressures. This is to be expected as the bonding between the two argon atoms comes mainly from dispersion and PBE severely underestimated the cohesive energy (while LDA overestimates it). At high pressure, however, the enthalpies obtained with the pure PBE functional approach those calculated with PBE-D3 and PBE-D4 as one expects. This behavior reflects the increasing dominance of short-range repulsive electronic interactions and the $PV$ contribution under compression, which is even well described at the LDA level of theory. These effects largely determine the lattice geometry, equation of state, and elastic properties of the crystal,\cite{Iitaka2001} thereby reducing the relative importance of dispersion interactions compared with their role at ambient or low pressures. At high pressures we see that all density functionals give similar profiles for the phase transition enthalpy, with LDA overestimating the energy barrier with respect to PBE-D3/D4.

The cell parameters $\gamma_{2}(\alpha)$, $\beta_{1}(\alpha)$, $\beta_{2}(\alpha)$, and $\beta_{3}(\alpha)=\gamma_2(\alpha)$ exhibit the same behavior as observed for the LJ and ELJ calculations. However, the optimization of cell parameters $a$ and $\gamma_{2}(\alpha)$ is numerically noisier compared to $\beta_{1}$ and $\beta_{2}$. In particular, the ideal relations between the $\beta_{i}$ parameters and $\alpha$ are satisfied almost exactly (within numerical accuracy) for the Bain path, as expected for the cuboidal transformation, i.e. Eq.\eqref{eq:beta1linear} and
\begin{equation}
    \beta_{2}(\alpha)=\frac{1}{2}\sqrt{4-(1-\alpha)^2} \quad \textrm{and}\quad \beta_{3}(\alpha)=\gamma_{2}(\alpha).
\end{equation}
This behavior observed on the high-$\alpha$ side of the bifurcation point persists for all pressures considered up to 100 GPa. Regarding $\beta_{3}$, since it is related to $\gamma_{2}$, it is hard to fit to a well-defined curve. The main source of the numerical noise in the cell parameters is related to the very shallow energy minima in the hypersurfaces obtained from the PBE functional, even when dispersion is included. The cell parameters are better behaved with the LDA functional, however, the region near the bifurcation point is still problematic. The above relations, that follow from the centering conditions on the central atom, are only applicable to the cuboidal transformation region, hence the low-$\alpha$ region is more flexible during the optimization and leads to a higher numerical noise. Nonetheless, the relation\cite{Schwerdtfeger-2026}
\begin{equation}
    \frac{\beta_{1}}{\beta_{2}} = \sqrt{\frac{3-\alpha}{1+\alpha}}
\end{equation}
holds approximately (within numerical errors) for the whole Burgers-Bain transformation path $\alpha\in [0,1]$ at any pressure up to $100$ GPa regardless of the levels of theory included here. 

The position of the bifurcation point, $\alpha_{c}$, shifts toward the hcp region with increasing pressure, in agreement to the LJ and ELJ results. However, the displacement of $\alpha_{c}$ is not as large at $100$ GPa compared to the ELJ as shown in Figure \ref{fig:bifurcationPELJ}. The LJ potential highly overestimates the energy barrier as can be seen by comparing Figures \ref{fig:LJnmEnthalpy}, \ref{fig:T=0LJELJBurgers} and Figure \ref{fig:DFTtrans}. Although the two-body potentials agree on some qualitative aspects with the DFT calculations, this also shows the contrast between a more accurate description of the repulsive part of the interaction potential. 

The quantitative difference at high pressure suggests that many-body interactions play an important role in the location of $\alpha_{c}$, as well as the height of the energy barrier. Such behavior is consistent with the observed deviations from the Cauchy relation, $C_{12}=C_{44}+2P_{c}$ (with $P_{c}$ being the Cauchy pressure), in compressed solid argon, which indicate that non-central many-body forces become increasingly important at high pressures.\cite{Shimizu-2001,Troitskaya2016,Iitaka2001}

It should also be noted that the choice of density functional approximation may influence the predicted location of the bifurcation point at lower pressures, while at high pressure the results are very close regardless of whether an LDA or GGA functional is used. The energy differences between the optimized structures are comparable to the intrinsic accuracy of current density functional approximations, while the description of the short-range repulsive region becomes increasingly important under compression. At 100 GPa, the bifurcation point lies close to the bcc structure, and the hcp-to-fcc transformation bypasses bcc by only about 0.2 kJ/mol with PBE-D4.

\subsection{Burgers-Bain Path at Finite Temperatures}

Although the Einstein approximation is not sufficiently accurate to resolve the small free-energy difference between the two nearly degenerate close-packed structures, fcc and hcp,\cite{Schwerdtfeger-2016} it nevertheless permits the vibrational free-energy contributions to be evaluated through exact lattice summations using Epstein zeta functions, which is particularly advantageous when very small energy differences are involved. Within this framework, we investigate the influence of temperature on both the transition state and the bifurcation point along the Burgers–Bain path using the ELJ potential by including three-body interactions for solid argon. To assess the reliability of our treatment of thermal effects at finite pressures, we first examine the temperature dependence of the volume, $V(T)$ at ambient pressure, and compare the results with available experimental data.

\begin{figure}[htb!]
    \centering
     \includegraphics[width=.75\columnwidth]{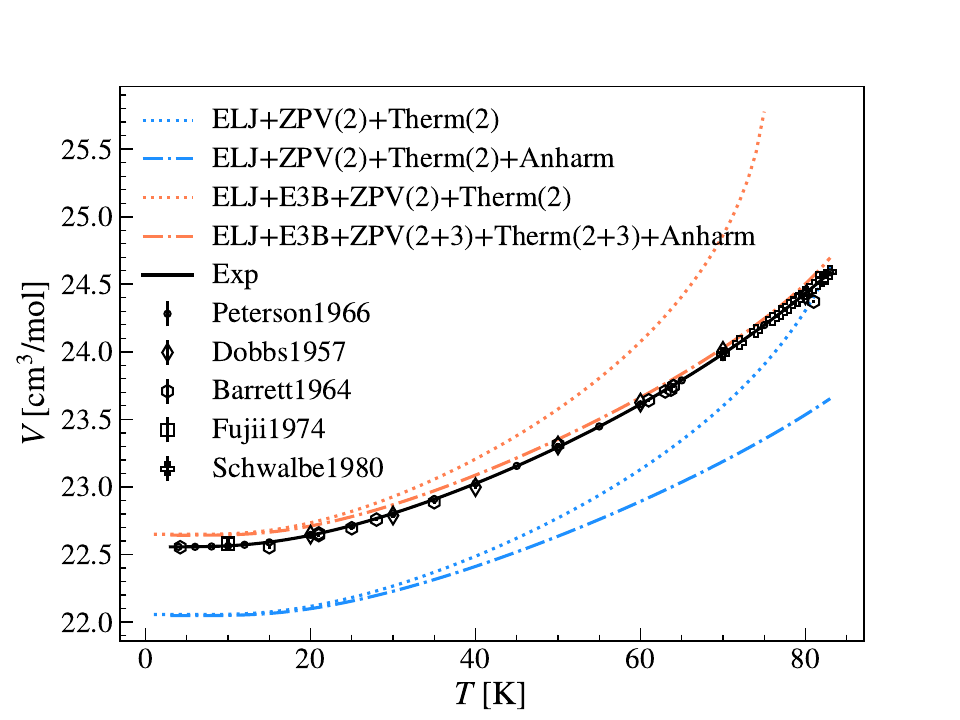}
    \caption{Comparison for the equation of state $V(T)$ between our ELJ, ELJ+E3B results and including ZPV and thermal contributions with various experimental data from x-ray and neutron diffraction measurements from Refs.\onlinecite{Dobbs-1957,Barrett-1964,PetersonBatchelderSimmons1966,Fujii1974,Schwalbe-1977}. The equation of state derived from experimental data is taken from Ref.\onlinecite{Maltby2024}. Inclusion of anharmonicity effects are for both the ZPV and thermal corrections including three-body effects through the E3B potential. Numbers set in parentheses indicate if two-body and three-body terms are included in the ZPV or thermal contributions.}
    \label{fig:V(T)}
\end{figure}

Figure~\ref{fig:V(T)} shows the equation of state for the molar volume at standard pressure (101.325 kPa). It is well known that zero-point vibrational (ZPV) and thermal vibrational effects increase the nearest-neighbor distances in solid argon with increasing temperature as one expects.\cite{Dobbs-1957} The importance of three-body corrections, even at low pressures, is evident, as previously noted,\cite{Schwerdtfeger-2016} leading to very good agreement with experiment at 0 K. Further, the harmonic Einstein approximation reproduces the experimental equation of state\cite{Maltby2024} remarkably well at low temperatures. However, above approximately 20 K, significant deviations from experiment start to appear necessitating the inclusion of anharmonicity effects. Indeed, if we consider anharmonic contributions to both the ZPV energy and the thermal free energy we substantially improve our results, yielding very good agreement with the experimental data up to close to the melting point. This is surprising as temperatures approaching and even exceeding the melting point, the static model constrained to the fcc lattice symmetry becomes less accurate, and a more realistic description requires classical molecular dynamics simulations\cite{Li-2017} employing larger supercells and adding phonon dispersion to ZPVE and thermal contributions.  Furthermore, close to melting the bcc phase becomes thermodynamically favored in argon, in agreement with the predictions from Landau theory\cite{Alexander1978} and molecular dynamics simulations.\cite{Travesset-2014}

\begin{figure*}[ht]
  \centering
(a)    \includegraphics[width=.46\linewidth]{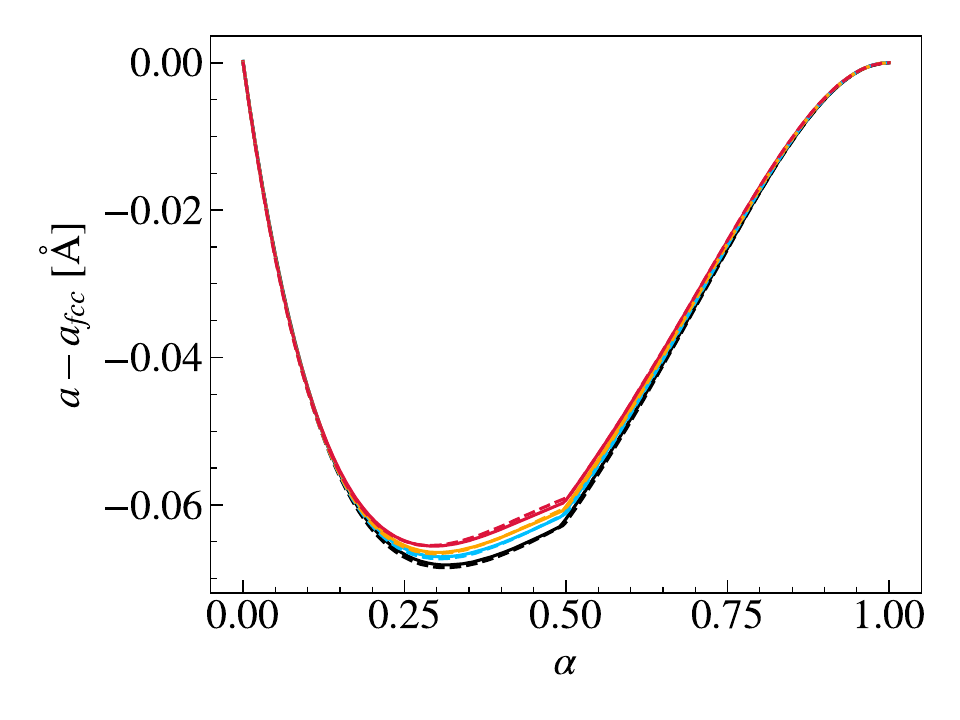}
(b)    \includegraphics[width=.46\linewidth]{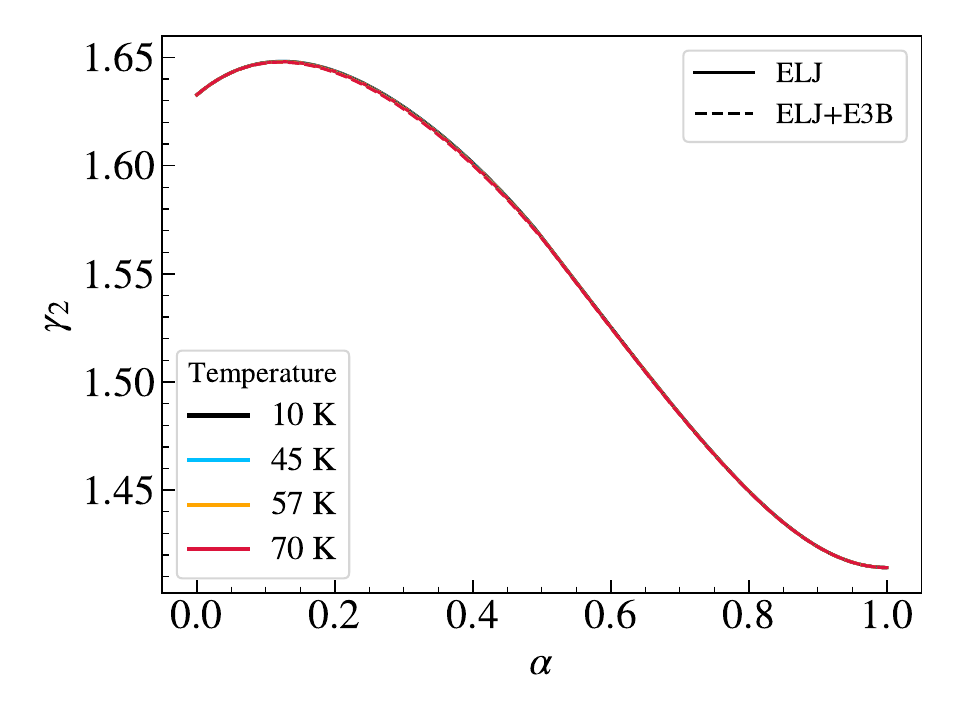}\\
(c)    \includegraphics[width=.46\linewidth]{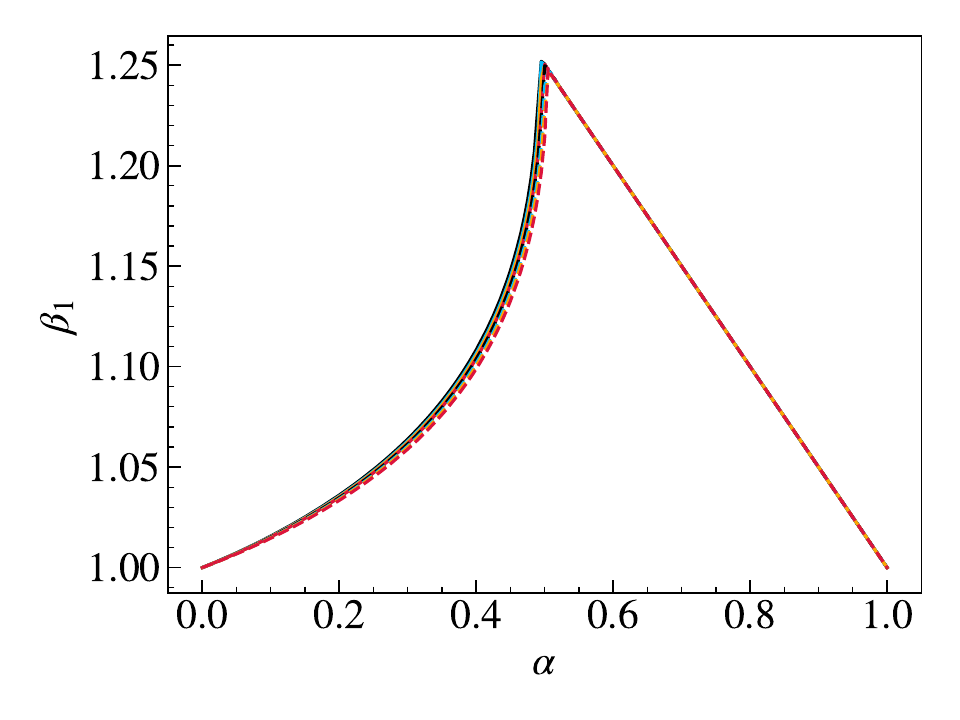}
(d)    \includegraphics[width=.46\linewidth]{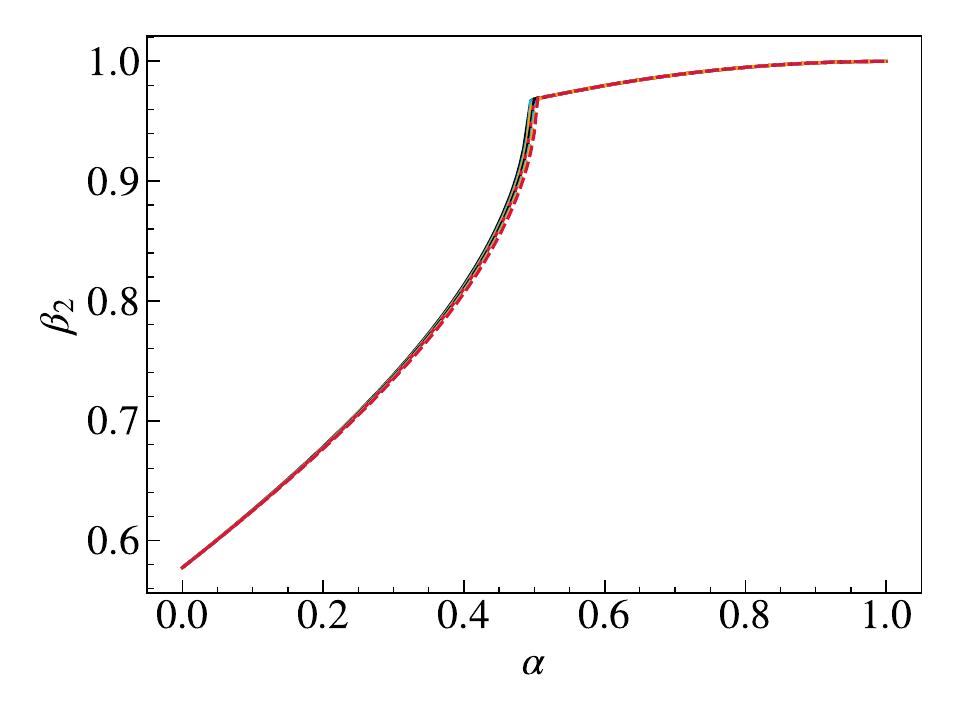}
  \caption{Lattice parameters along the Burgers-Bain path for the ELJ potential with (dotted lines) and without (solid lines) the E3B potential. Results are at zero pressure and isotherms of 10, 45, 57, and 70 K.}\label{fig:T=varELJP=0LC}
\end{figure*}

For the Burgers–Bain path, we focus on calculations employing the effective three-body (E3B) potential. Test calculations at finite temperatures showed only minor differences compared with results obtained using the full ATM potential, while the latter is computationally considerably more demanding for finite temperature optimizations of the lattice parameters despite the efficient treatment through Epstein zeta functions (see Appendix \ref{sec:EpsteinATM}). Figures~\ref{fig:T=varELJP=0} and \ref{fig:T=varELJP=0LC} compare the Burgers–Bain paths for the ELJ potential with and without the E3B correction, including anharmonic contributions, for several selected temperatures at zero pressure. The corresponding data for the critical points are collected in Table \ref{tab:BBProperties2}. Although the triple-point temperature of argon is $T_{\mathrm{tp}}=83.8$ K at $P_{\mathrm{tp}}=68.891$ kPa, the highest temperature considered at zero pressure was limited to 70 K, remaining safely below the melting point where we expect not too large distortions within the solid. As the melting temperature is approached, the crystal becomes mechanically unstable and develops spontaneous distortions accompanied by symmetry breaking, rendering a meaningful analysis of the Burgers–Bain transformation path impossible without inclusion of dynamic effects. 
\begin{figure*}[ht]
  \centering
(a)    \includegraphics[width=.46\linewidth]{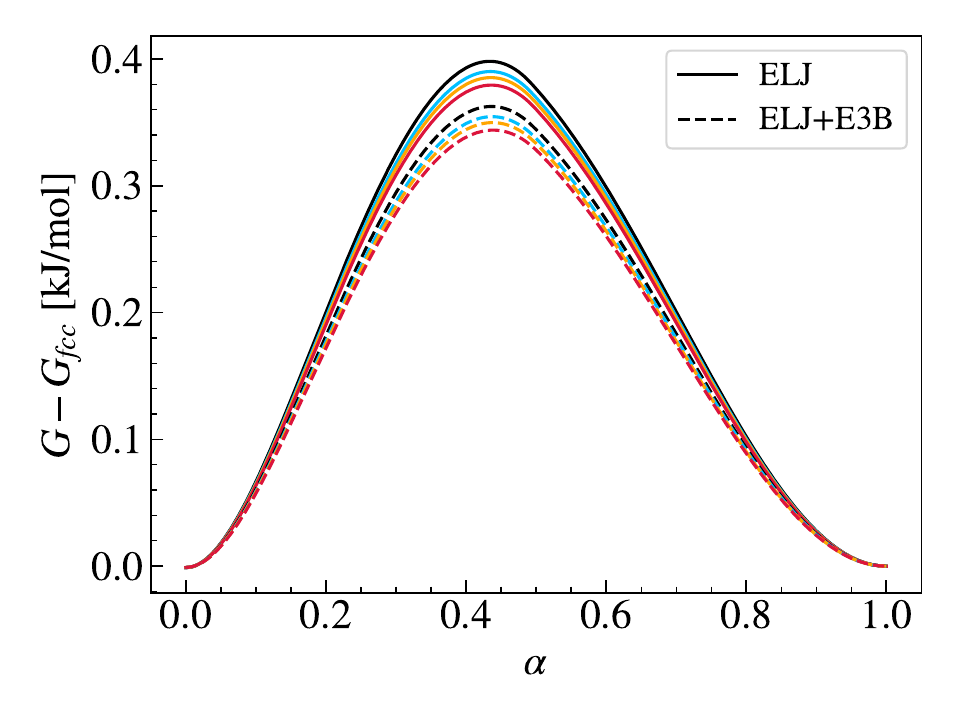}
(b)    \includegraphics[width=.46\linewidth]{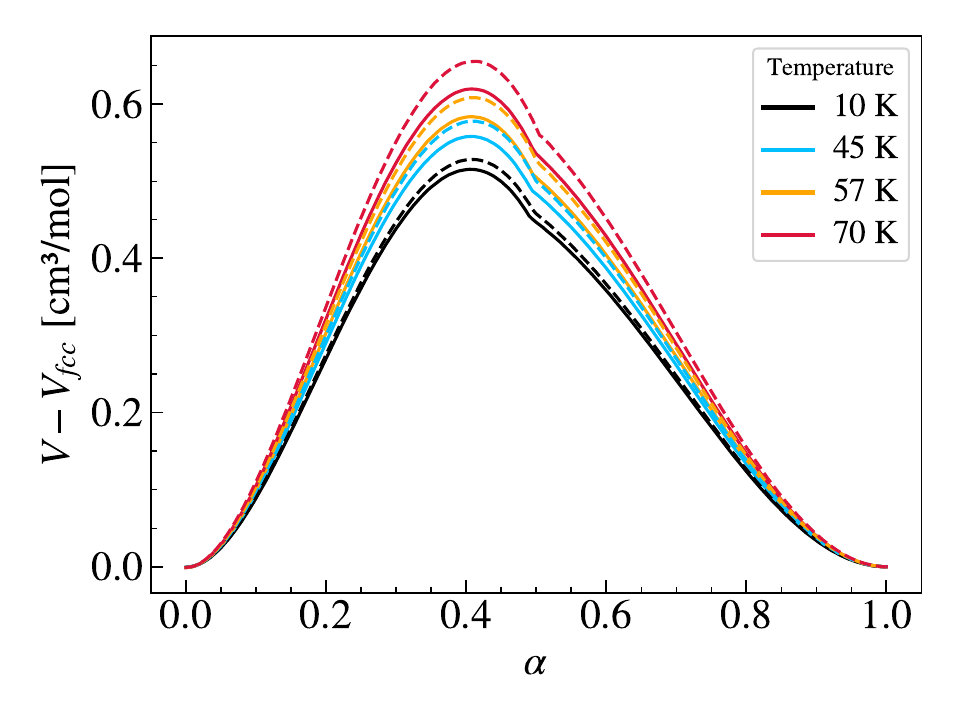}
  \caption{Gibbs free energy (a) and volume change (b) along the Burgers-Bain path for the ELJ potential with (dotted lines) and without (solid lines) the E3B potential. Results are at zero pressure choosing isotherms of 10, 45, 57, and 70 K.}\label{fig:T=varELJP=0}
\end{figure*}

\begin{table}[t!]
\tiny
\setlength{\tabcolsep}{3pt}
\begin{tabular}{|l|c|c||c|c|c|c|c|c|c|c|c|c|c|c|c|c|c|c|}
\hline
Potential &$P$ &$T$ &$a^{\#}$ &$\Delta G^{\#}$ &$\Delta E_{\mathrm{ZPV}}^{\#}$ &$\gamma_2^{\#}$ &$\beta_1^{\#}$ &$\beta_2^{\#}$ &$\Delta V^{\#}$ &$\alpha^{c}$ &$\Delta G^{c}$ &$a^{c}$ &$\gamma_2^{c}$ &$\beta_1^{c}$ &$\beta_2^{c}$ &$\Delta G^{\alpha=\tfrac{1}{3}}$ &$\beta_1^{\alpha=\tfrac{1}{3}}$ &$\beta_2^{\alpha=\tfrac{1}{3}}$ \\
\hline
ELJ & 0 & 0 & 3.6619 & 0.3982 & -0.0130 & 1.5904 & 1.1335 & 0.8479 & 0.5085 & 0.4939 & 0.3803 & 3.6646 & 1.5694 & 1.2502 & 0.9653 & 0.3552 & 1.0761 & 0.7609 \\
ELJ & 0 & 10 & 3.6620 & 0.3981 & -0.0131 & 1.5904 & 1.1335 & 0.8478 & 0.5088 & 0.4938 & 0.3803 & 3.6647 & 1.5695 & 1.2497 & 0.9648 & 0.3552 & 1.0761 & 0.7609 \\
ELJ & 0 & 45 & 3.6896 & 0.3901 & -0.0146 & 1.5902 & 1.1311 & 0.8460 & 0.5513 & 0.4950 & 0.3725 & 3.6922 & 1.5689 & 1.2516 & 0.9669 & 0.3476 & 1.0749 & 0.7601 \\
ELJ & 0 & 57 & 3.7067 & 0.3853 & -0.0154 & 1.5901 & 1.1295 & 0.8449 & 0.5770 & 0.4961 & 0.3677 & 3.7091 & 1.5684 & 1.2454 & 0.9628 & 0.3431 & 1.0742 & 0.7596 \\
ELJ & 0 & 70 & 3.7284 & 0.3794 & -0.0165 & 1.5899 & 1.1276 & 0.8434 & 0.6128 & 0.4992 & 0.3607 & 3.7310 & 1.5671 & 1.2494 & 0.9676 & 0.3375 & 1.0733 & 0.7589 \\
ELJ+E3B & 0 & 0 & 3.6949 & 0.3626 & -0.0114 & 1.5899 & 1.1273 & 0.8432 & 0.5222 & 0.4998 & 0.3445 & 3.6980 & 1.5668 & 1.2499 & 0.9681 & 0.3224 & 1.0731 & 0.7588 \\
ELJ+E3B & 0 & 10 & 3.6951 & 0.3625 & -0.0114 & 1.5899 & 1.1272 & 0.8431 & 0.5225 & 0.4997 & 0.3445 & 3.6982 & 1.5669 & 1.2496 & 0.9678 & 0.3224 & 1.0731 & 0.7588 \\
ELJ+E3B & 0 & 45 & 3.7279 & 0.3545 & -0.0130 & 1.5897 & 1.1245 & 0.8411 & 0.5722 & 0.5013 & 0.3367 & 3.7308 & 1.5661 & 1.2426 & 0.9633 & 0.3148 & 1.0718 & 0.7579 \\
ELJ+E3B & 0 & 57 & 3.7479 & 0.3498 & -0.0138 & 1.5879 & 1.1268 & 0.8451 & 0.5999 & 0.5034 & 0.3315 & 3.7507 & 1.5652 & 1.2430 & 0.9645 & 0.3105 & 1.0710 & 0.7573 \\
ELJ+E3B & 0 & 70 & 3.7743 & 0.3439 & -0.0153 & 1.5877 & 1.1245 & 0.8434 & 0.6469 & 0.5050 & 0.3256 & 3.7769 & 1.5645 & 1.2484 & 0.9696 & 0.3049 & 1.0700 & 0.7566 \\
ELJ & 0.101325 & 0 & 3.6618 & 0.3982 & -0.0130 & 1.5904 & 1.1336 & 0.8479 & 0.5084 & 0.4938 & 0.3804 & 3.6645 & 1.5695 & 1.2502 & 0.9652 & 0.3553 & 1.0761 & 0.7609 \\
ELJ & 0.101325 & 10 & 3.6619 & 0.3982 & -0.0131 & 1.5904 & 1.1336 & 0.8479 & 0.5087 & 0.4938 & 0.3804 & 3.6646 & 1.5695 & 1.2498 & 0.9649 & 0.3552 & 1.0761 & 0.7609 \\
ELJ & 0.101325 & 45 & 3.6896 & 0.3901 & -0.0146 & 1.5902 & 1.1311 & 0.8460 & 0.5511 & 0.4950 & 0.3726 & 3.6921 & 1.5689 & 1.2527 & 0.9678 & 0.3477 & 1.0749 & 0.7601 \\
ELJ & 0.101325 & 57 & 3.7066 & 0.3853 & -0.0154 & 1.5901 & 1.1296 & 0.8449 & 0.5768 & 0.4959 & 0.3679 & 3.7090 & 1.5685 & 1.2459 & 0.9631 & 0.3432 & 1.0742 & 0.7596 \\
ELJ & 0.101325 & 70 & 3.7284 & 0.3794 & -0.0165 & 1.5899 & 1.1276 & 0.8434 & 0.6129 & 0.4992 & 0.3608 & 3.7309 & 1.5671 & 1.2494 & 0.9674 & 0.3376 & 1.0733 & 0.7589 \\
ELJ+E3B & 0.101325 & 0 & 3.6949 & 0.3626 & -0.0114 & 1.5899 & 1.1273 & 0.8432 & 0.5218 & 0.4998 & 0.3445 & 3.6980 & 1.5668 & 1.2491 & 0.9675 & 0.3225 & 1.0731 & 0.7588 \\
ELJ+E3B & 0.101325 & 10 & 3.6950 & 0.3626 & -0.0114 & 1.5899 & 1.1273 & 0.8432 & 0.5227 & 0.4999 & 0.3444 & 3.6981 & 1.5668 & 1.2494 & 0.9677 & 0.3224 & 1.0731 & 0.7588 \\
ELJ+E3B & 0.101325 & 45 & 3.7278 & 0.3545 & -0.0130 & 1.5897 & 1.1246 & 0.8411 & 0.5719 & 0.5004 & 0.3372 & 3.7306 & 1.5665 & 1.2463 & 0.9657 & 0.3149 & 1.0718 & 0.7579 \\
ELJ+E3B & 0.101325 & 57 & 3.7479 & 0.3498 & -0.0138 & 1.5879 & 1.1268 & 0.8451 & 0.5998 & 0.5033 & 0.3316 & 3.7506 & 1.5652 & 1.2436 & 0.9650 & 0.3105 & 1.0710 & 0.7573 \\
ELJ+E3B & 0.101325 & 70 & 3.7742 & 0.3439 & -0.0152 & 1.5877 & 1.1245 & 0.8434 & 0.6465 & 0.5050 & 0.3256 & 3.7768 & 1.5645 & 1.2452 & 0.9671 & 0.3050 & 1.0699 & 0.7566 \\
ELJ & 1000 & 0 & 3.4466 & 0.7366 & -0.0004 & 1.6028 & 1.1758 & 0.8675 & 0.2551 & 0.4446 & 0.7247 & 3.4480 & 1.5911 & 1.2779 & 0.9610 & 0.6788 & 1.1081 & 0.7835 \\
ELJ & 1000 & 100 & 3.4693 & 0.7359 & 0.0008 & 1.6027 & 1.1749 & 0.8669 & 0.2601 & 0.4450 & 0.7240 & 3.4706 & 1.5909 & 1.2778 & 0.9609 & 0.6779 & 1.1077 & 0.7832 \\
ELJ & 1000 & 150 & 3.4896 & 0.7359 & 0.0020 & 1.6027 & 1.1740 & 0.8663 & 0.2638 & 0.4450 & 0.7241 & 3.4909 & 1.5909 & 1.2774 & 0.9607 & 0.6776 & 1.1073 & 0.7830 \\
ELJ & 1000 & 200 & 3.5120 & 0.7366 & 0.0034 & 1.6026 & 1.1730 & 0.8655 & 0.2671 & 0.4450 & 0.7251 & 3.5133 & 1.5909 & 1.2770 & 0.9603 & 0.6779 & 1.1069 & 0.7827 \\
ELJ & 1000 & 250 & 3.5370 & 0.7385 & 0.0053 & 1.6010 & 1.1787 & 0.8720 & 0.2666 & 0.4458 & 0.7268 & 3.5382 & 1.5906 & 1.2714 & 0.9567 & 0.6793 & 1.1063 & 0.7822 \\
ELJ+E3B & 1000 & 0 & 3.4657 & 0.7025 & 0.0011 & 1.6026 & 1.1732 & 0.8656 & 0.2546 & 0.4450 & 0.6915 & 3.4670 & 1.5909 & 1.2749 & 0.9588 & 0.6466 & 1.1069 & 0.7827 \\
ELJ+E3B & 1000 & 100 & 3.4894 & 0.7034 & 0.0024 & 1.6026 & 1.1723 & 0.8650 & 0.2589 & 0.4450 & 0.6926 & 3.4907 & 1.5909 & 1.2758 & 0.9594 & 0.6473 & 1.1066 & 0.7825 \\
ELJ+E3B & 1000 & 150 & 3.5106 & 0.7048 & 0.0037 & 1.6010 & 1.1785 & 0.8719 & 0.2606 & 0.4456 & 0.6938 & 3.5118 & 1.5906 & 1.2722 & 0.9572 & 0.6483 & 1.1062 & 0.7822 \\
ELJ+E3B & 1000 & 200 & 3.5337 & 0.7071 & 0.0052 & 1.6009 & 1.1775 & 0.8711 & 0.2627 & 0.4465 & 0.6957 & 3.5349 & 1.5903 & 1.2689 & 0.9552 & 0.6502 & 1.1058 & 0.7819 \\
ELJ+E3B & 1000 & 250 & 3.5596 & 0.7110 & 0.0072 & 1.6009 & 1.1760 & 0.8701 & 0.2611 & 0.4480 & 0.6988 & 3.5610 & 1.5896 & 1.2705 & 0.9573 & 0.6533 & 1.1052 & 0.7815 \\
\hline
\end{tabular}
\caption{Critical parameters for the transition state ($\#$) and bifurcation point ($c$) of the Burgers-Bain path at various pressures and temperatures for the ELJ potential with and without effective three-body corrections (E3B). Pressures are in MPa, temperatures in K, lattice constants $a$ in \AA, energies in kJ/mol, and volumes in cm$^3$/mol. All energy and volume differences are relative to FCC at the same pressure and temperature. The parameters at $\alpha=\tfrac{1}{3}$ describe the relaxed path, which need not coincide with ideal bcc. The free energies include Einstein vibrations and anharmonic corrections. The listed $\Delta E_{\mathrm{ZPV}}$ refers to the harmonic energy only and is recomputed using the original script's Einstein approximation.}
\label{tab:BBProperties2}
\end{table}

Increasing the temperature lowers the activation enthalpy and leads to an overall expansion in volume along the transition path due to an increase in the lattice parameter $a$ (and $c=a\gamma_2$) as demonstrated in Figure \ref{fig:T=varELJP=0}. Additional temperature-dependent effects are observed in all other lattice parameters, but they are relatively small for the parameters $\beta_1$, $\beta_2$ and $\beta_3=\gamma_2$ of the shift vector. The minimum observed for $(a-a_{\mathrm{fcc}})$ shifts to smaller values of $\alpha$ as the temperature increases. We also notice that the kink in $\Delta V(\alpha)$ is clearly visible at all temperatures studied here.

\begin{figure*}[htb!]
  \centering
(a)    \includegraphics[width=.46\linewidth]{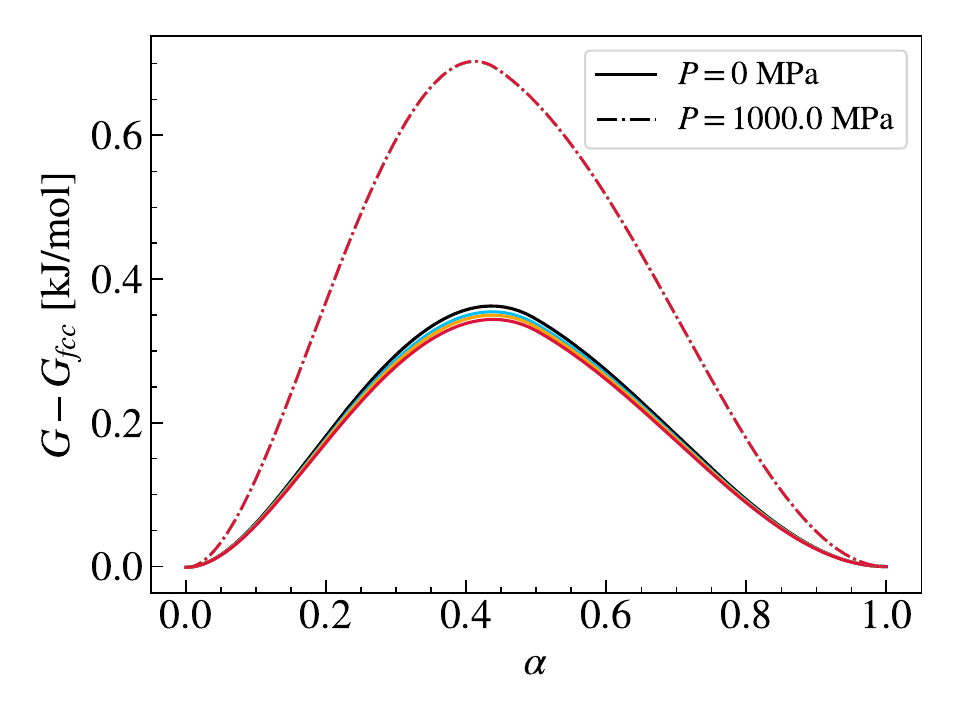}
(b)    \includegraphics[width=.46\linewidth]{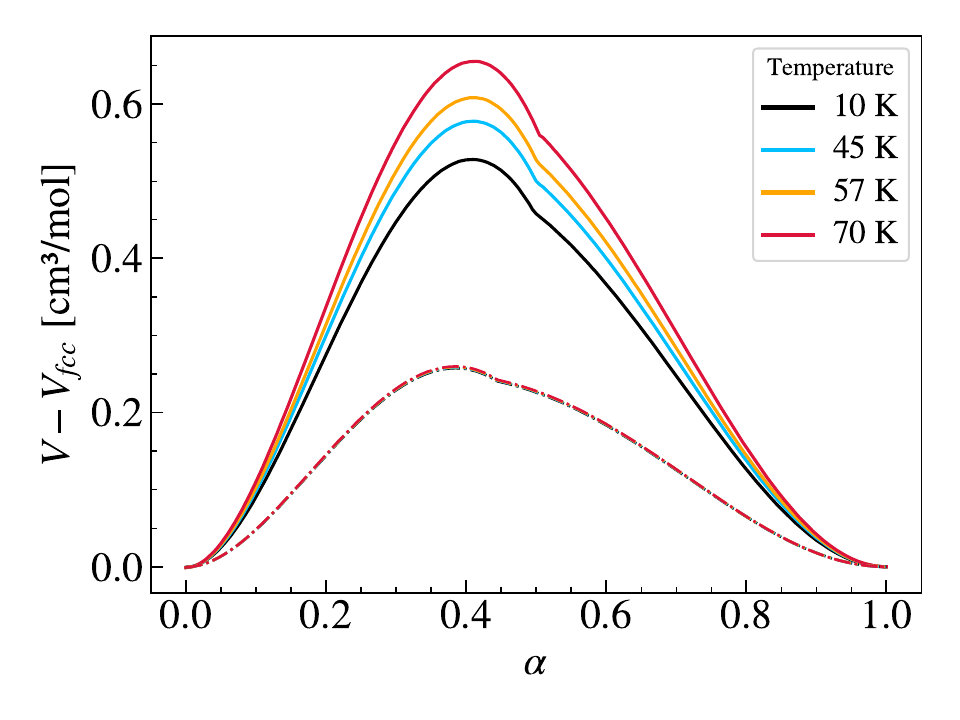}
  \caption{Comparison of the Gibbs free energy (a) and volume change (b) for the two pressures 0 and 1 GPa and isotherms 10, 45, 57, and 70 K along the Burgers-Bain path for the ELJ+E3B potential.}\label{fig:T=varELJP}
\end{figure*}

\begin{figure*}[htb!]
  \centering
(a)    \includegraphics[width=.46\linewidth]{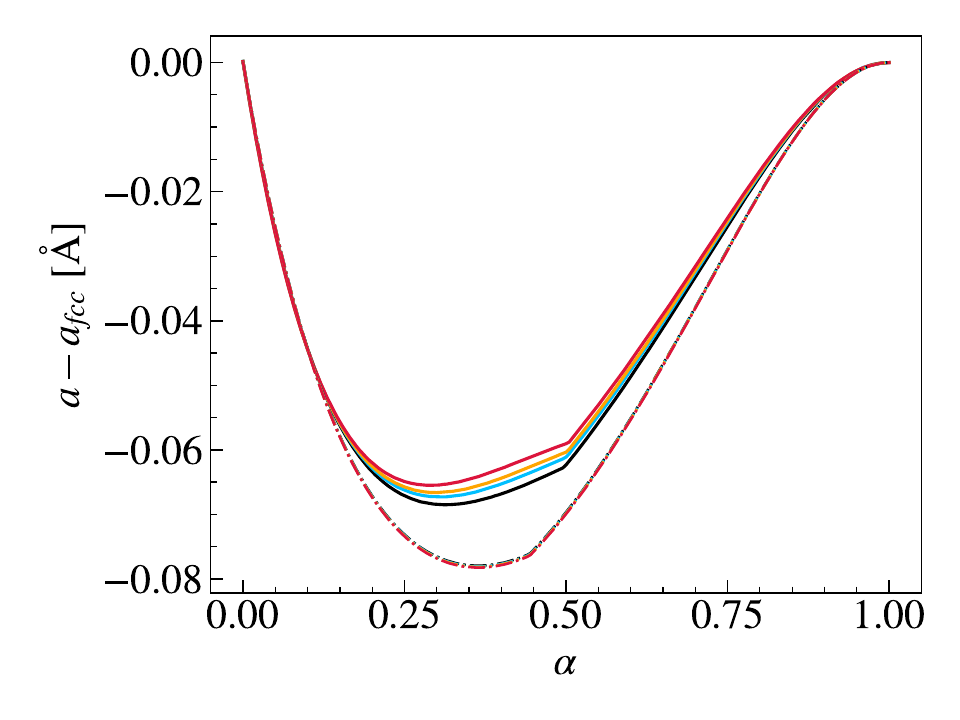}
(b)    \includegraphics[width=.46\linewidth]{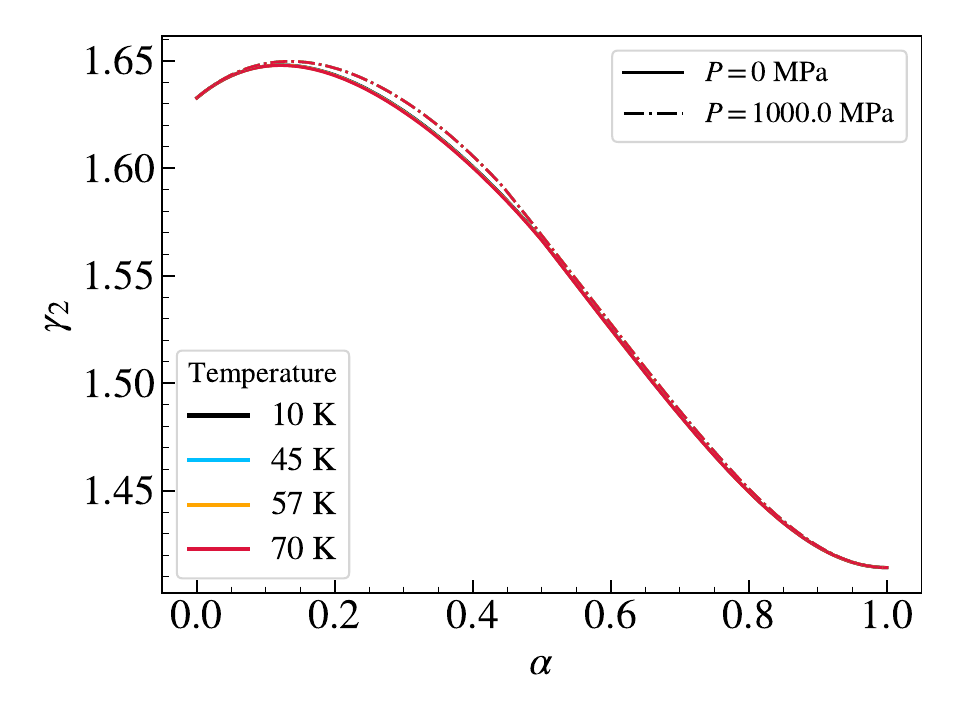}\\
(c)    \includegraphics[width=.46\linewidth]{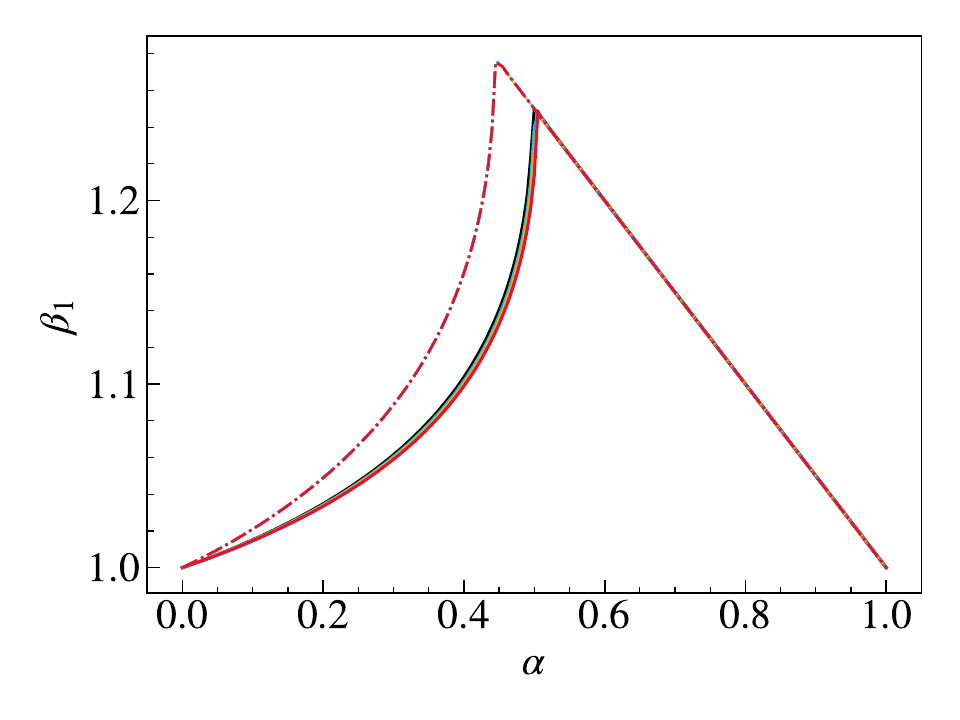}
(d)    \includegraphics[width=.46\linewidth]{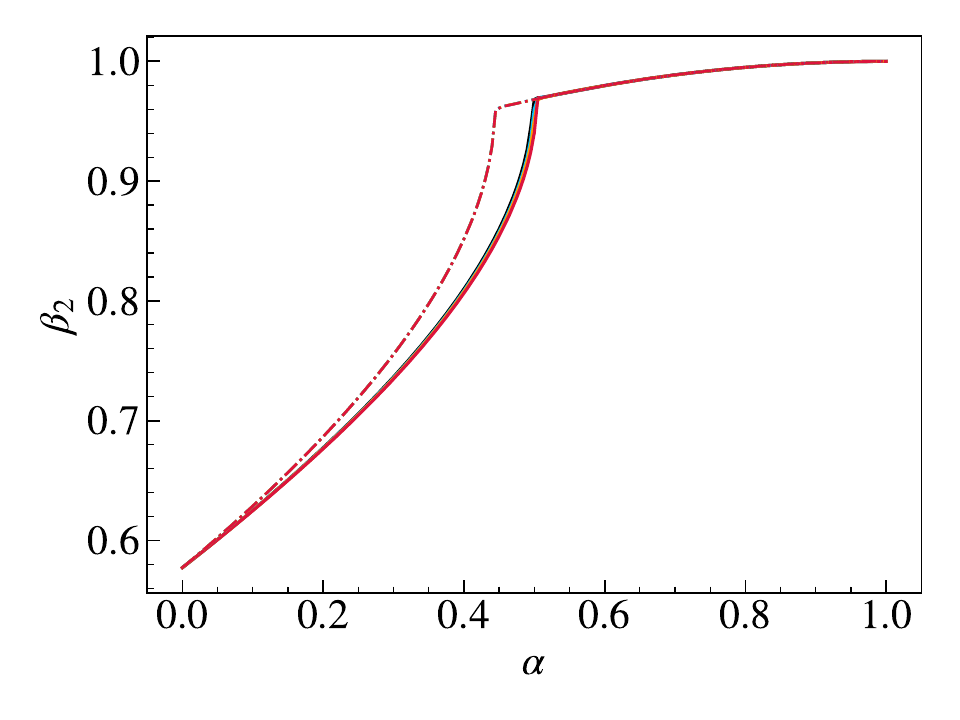}
  \caption{Comparison of lattice parameters along the Burgers-Bain path for the two pressures 0 and 1 GPa and isotherms 10, 45, 57, and 70 K along the Burgers-Bain path for the ELJ+E3B potential.}\label{fig:T=varELJP=0LC-P}
\end{figure*}

For the second pressure considered here, $P=1$ GPa, Figures \ref{fig:T=varELJP} and \ref{fig:T=varELJP=0LC-P} compare the results with those obtained at zero pressure. The influence of temperature is clearly reduced at higher pressures. This reduction due to diminishing vibrational effects compared to the dominating repulsive forces has been identified as one of the factors contributing to the phase transition from fcc to hcp at pressures above 20 GPa.\cite{Schwerdtfeger-2016,Schwerdtfeger2017}


Table \ref{tab:BBProperties2} summarizes the results obtained at $P=1$ GPa for elevated temperatures. According to the experimentally derived melting-point-pressure relation of Tegeler \textit{et al.},\cite{TegelerSpanWagner1999} the melting temperature at this pressure is approximately 254 K. At 150 K, which is well below the melting point, only small changes are observed in both the activation free energy and the activation volume associated with the barrier relative to their 0 K values. As the temperature approaches the melting point, however, significantly larger temperature effects emerge, as shown in Table \ref{tab:BBProperties2}.

\section{Conclusions}

Our analysis provided a detailed insight into the Burgers-Bain path associated with the hcp-to-fcc transition applying inverse power law model potentials, in particular the extended Lennard-Jones potential supplemented by the three-body Axilrod-Teller-Muto interaction. Despite all the short-comings of simple and most widely used $(n,m)$-LJ potential,\cite{Wales2024} the qualitative features for the Burgers-Bain phase transition remain the same compared to more elaborate methods. By treating the underlying Epstein zeta functions, exact lattice summations could be performed, enabling an accurate determination of the bifurcation point. Density functional theory (DFT) calculations yield qualitatively very similar results. However, the exact location of the bifurcation point on the minimum energy path can easily be obscured by numerical noise arising from the evaluation of the density functional and the derivatives required for the geometry optimizations. In addition, the very small enthalpy differences along the phase transition path requires very tight convergence criteria for the density functional methods. 

For pure two-body interactions, we found that increasing pressure shifts the bifurcation point towards the bcc structure at $\alpha=\tfrac{1}{3}$ along the Bain path, which is only slightly modified in the opposite direction by three-body contributions. In the case of the DFT calculations, where many-body effects are implicitly incorporated to a certain accuracy, we see that the shift of the bifurcation point due to pressure is in the same direction as the two-body potentials. However, the displacement is not as pronounced as the ELJ case at a pressure of $100$ GPa. The numerical difficulties in the optimization process may well show the limitation of current approximations in DFT for such weakly interacting systems.

For solid argon we tested our model using a more accurate extended LJ potential, adding three-body effects through the classical Axilrod-Teller-Muto term. Including temperature effects through harmonic and anharmonic vibrational terms modeled by the Einstein approximation, we obtained very good agreement for the experimentally determined thermal volume expansion up to the melting point at standard pressure.

Having established the underlying mechanism at finite pressure and temperature, future work should focus on more realistic many-body potentials and the inclusion of phonon dispersion and anharmonicity effects within large-scale molecular dynamics simulations. While Landau theory predicts the bcc phase to appear near the melting line,\cite{Alexander1978} the stability of this less densely packed phase compared to fcc or hcp at higher pressures is less than clear. For example, for iron, where many-body forces are clearly dominant, the bcc phase has recently been identified at pressures above 230 GPa.\cite{kxnf-6c5y} The use of larger supercells will also enable the investigation of possible alternative transformation pathways\cite{Li-2017} using efficient lattice summations through Epstein zeta functions including analytical derivatives of the energy with respect to the lattice parameters. Work in this direction is currently in progress.

\section*{Acknowledgment}
PS is grateful to Prof. Benoît Darquié (Université Sorbonne, Paris) for a visiting professorship at Sorbonne, to the Sir Neil Waters Fund at Massey University for financial support, and to Fabian Zehetmair (Graz) and Shaun Cooper (Auckland) for fruitful discussions. JH acknowledges support from the Quantum Fellowship Program of the German Aerospace Center (DLR), which funded their contribution to this work. TM, MH, and ØW acknowledge funding from the Research Council of Norway (RCN) through the Centre of Excellence Funding Scheme (Project No. 262644, PoreLab), as well as from the European Research Council (ERC) under the European Union’s Horizon Europe Research and Innovation Programme (Grant Agreement No. 101115669). This publication was also supported by the HYDROGENi Research Centre (\url{hydrogeni.no}), carried out under the Norwegian research program FMETEKN. TM, MH, and ØW further acknowledge the industry partners in HYDROGENi for their contributions and the Research Council of Norway for funding under Grant No. 333118. AAB acknowledges support of this work by the Klaus-Tschira Stiftung under Grant No. 00.025.2025.

\section*{Supplementary Material}

The supplementary material is available upon request and contains two excel spreadsheets called SI1.xlsx and SI2.xlsx. SI1.xlsx contains the data (lattice parameters, deviation from the hard-sphere limit, enthalpies, volumes, pressures) all in reduced units for Burgers-Bain path (at zero temperature) applying various LJ potentials including the ELJ potential. It also contains data for the critical points and the deviation of $\gamma_2=c/a$ from the ideal $\sqrt{\tfrac{8}{3}}$ value. The second file SI2.xlsx contains data for solid argon at finite pressures (in units of Pa), optimized lattice parameters at constant temperature as well as free energy data (atomic units are used, all contributions included as described in the main text) using the ELJ/ATM/E3B potentials. In addition, it contains data for the volume equation of state $V(T)$ including references to experimental data.

\section*{Appendix}
\appendix
\setcounter{table}{0}
\renewcommand{\thetable}{A\arabic{table}}
\numberwithin{equation}{section}

\section{Lattice Parameter Optimization} 
\label{NewtonRaphson}

For the inverse power law potentials, the iterative Newton–Raphson algorithm\cite{hammerlin2012numerical} was used to determine the minimum energies within the lattice parameter space $\vec{x}$ along the Burgers-Bain path to the required numerical accuracy
\begin{equation}\label{eq:mincond1}
\vec{x}_{N+1}=\vec{x}_N -  \lambda \left[ (\partial_{x_i}\partial_{x_j}) G(\vec{x}_N )\right]^{-1}\vec{\nabla} G(\vec{x}_N )   \,,
\end{equation}
where for evaluating the gradient and Hessian we used numerical methods (except trivially for the pressure term $PV(\vec{x})$), i.e. the well known numerical expressions were used for the first and second derivatives,\cite{abramowitzstegun2008}
\begin{equation}
\partial_x f(x)= \left\{ f(x-2h) - 8f(x-h) + 8f(x+h) -f(x+2h) \right\}/12h
\end{equation}
\begin{equation}
\partial_x^2 f(x)= \left\{ -f(x-2h) + 16f(x-h) -30f(x) + 16f(x+h) -f(x+2h) \right\}/12h^2
\end{equation}
with the chosen step size of $h=2\times 10^{-3}$. The convergence criterion was set to $|\vec{x}_{N+1}-\vec{x}_N|<\epsilon$ with $\epsilon=10^{-15}$. For the mixed derivatives we used the following formula,\cite{abramowitzstegun2008}
\begin{equation}
\partial_x\partial_y f(x,y)= \left\{ f(x+h,y+h) - f(x+h,y-h) -f(x-h,y+h) + f(x-h,y-h)  \right\}/4h^2
\end{equation}
These mixed derivatives are required to achieve faster convergence for Eq.\eqref{eq:mincond1}, especially for the shift vector where the lattice parameters $\beta_1$ and $\beta_2$ are strongly coupled. We note that the convergence of the Newton-Raphson procedure to the correct lattice parameters along the Burgers-Bain path is quite sensitive to the initial guess $\vec{x}_0$ and parameter $\lambda$. 
For the starting value of the lattice parameter $a$, which is sensitive to different pressure values, we used the Newton method first to find the root of the function for a LJ potential including the $PV$ term,
\begin{equation}
\partial_a \left[ E_\text{coh}(a,{\vec{x}_i}{'}) + PV(a,{\vec{x}_i}{'})\right] = 0
\end{equation}
where ${\vec{x}_i}{'}$ are all the other (fixed) lattice parameters except for $a$ and fixing $b=a$ and $c=\gamma_2a$. While the lattice sums calculated through the Epstein zeta function are accurate to computer precision, the numerical noise in the data comes from the numerical treatment in the Newton-Raphson procedure.

For the DFT calculations, obtaining the Hessian was computationally too expensive. We therefore used the BFGS algorithm\cite{Broyden1970a,Broyden1970b,Fletcher1970,Goldfarb1970,Shanno1970} that iteratively constructs an approximation to the inverse Hessian matrix, $B_{k}$. The step size was determined using a backtracking line search satisfying Armijo's condition.\cite{armijo1966m} If no acceptable step length was found after a prescribed number of backtracking iterations, a minimum step size was adopted to maintain progress while limiting the influence of numerical noise in the DFT energy evaluations. In this algorithm, the search direction is\cite{Nocedal2006}
\begin{equation}
    \vec{d}_{k} = -B_{k}\nabla H(\vec{x}_{k})
\end{equation}
where $\nabla H(\vec{x}_{k})$ is the gradient of the enthalpy at step $k$, and $B_{k}$ is the inverse Hessian that is updated according to
\begin{equation}
    B_{k+1} = \left(I-\rho_{k}\vec{s}_{k}\vec{y}_{k}^{\top}\right)B_{k}\left(I-\rho_{k}\vec{y}_{k}\vec{s}_{k}^{\top}\right) + \rho_{k}\vec{s}_{k}\vec{s}_{k}^{\top}
\end{equation}
where $\vec{s}_{k}=\vec{x}_{k+1}-\vec{x}_{k}$, $\vec{y}_{k}=\nabla H(\vec{x}_{k+1})-\nabla H(\vec{x}_{k})$, and $\rho_{k}=1/(\vec{y}_{k}^{\top}\vec{s}_{k})$. The update was performed only when the curvature condition $\vec{y}_{k}^{\top}\vec{s}_{k}>0$ holds to ensure that $B_{k}$ remained positive definite, otherwise the update is skipped and the approximate inverse Hessian resets. There is far more numerical noise in our DFT calculations compared to the results from inverse power law potentials, which mainly comes from capturing small energy differences within the numerical treatment of the DFT procedure for the solid state and the accompanying geometry optimizations.

\section{Epstein integral representation of the ATM potential}
\label{sec:EpsteinATM}

The use of the Epstein zeta function for inverse power law potentials goes back to Lennard-Jones and Ingham.\cite{Jones-1925} They used Bessel function expansions leading to fast converging series, and a summary of the different techniques can be found in Ref.~\onlinecite{burrows-2020}. For two-body inverse power law potentials such as the LJ potential, the exact lattice sum treatment through Epstein zeta functions based on Crandall's original approach\cite{crandall1998fast} was described very recently.\cite{Buchheit2026,Andres2025a,Buchheit2026a,Schwerdtfeger-2026,Buchheit2026a} In addition, Buchheit and Busse introduced an effective treatment of the ATM cohesive energy\cite{AxilrodTeller1943,Muto1943} for mono-atomic cuboidal lattices using an integral representation that involves products of Epstein zeta functions.\cite{Buchheit2026a,Schwerdtfeger-2026} Here we generalize this approach for arbitrary generator matrices and shift vectors and provide a simplified approach to the ATM potential through the recent generalization of Epstein zeta functions to anisotropic interaction kernels.\cite{buchheitComputationAnisotropicSingular2026}

In the following we use a more compact notation.
Denote by  $\Lambda=A\mathbb Z^d$, $A\in\mathds R^{d\times d}$ regular a $d$-dimensional Bravais lattice, consider an exponent $\nu>d$, a shift vector $\vec v$ and a wave vector $\vec k$, so that the Epstein zeta function reads\cite{Epstein-1903} 
\begin{equation}
Z_{\Lambda,\nu}(\vec v,\vec k) = \,\sideset{}{'}\sum_{\vec z\in \Lambda}\frac{e^{-2\pi i \vec z \cdot \vec k}}{\vert \vec z-\vec v\vert^\nu},
\end{equation}
where the lattice sum can be meromorphically continued to $\nu\in \mathbb C$. For the LJ potential, the corresponding Epstein zeta function is evaluated at $\vec k=0$. However, the inclusion of the wave vector argument becomes crucial in the evaluation of three-body and general many-body lattice sums.  

The ATM potential \eqref{eq:ATM} only depends on two relative distance vectors and the coupling parameter $\lambda\in\mathbb{R}_{+}$, and in our short-hand notation we have
\begin{align}
\label{eq:MAT1}
E^{(3)}_{\rm ATM}&(\vec x,\vec y)=\lambda \;\; \bigg( \frac{1}{\vert \vec{x} \vert^3 \vert  \vec{y}\vert^3 \vert \vec{z} \vert^3}-3 \frac{(\vec{x}\cdot \vec{y})(\vec{y}\cdot \vec{z})(\vec{z}\cdot \vec{x})}{\vert \vec{x} \vert^5 \vert  \vec{y}\vert^5 \vert \vec{z} \vert^5}\bigg)\bigg \vert_{\vec{z}=\vec{y}-\vec{x}},
\end{align}
where $\vec x = \vec R_{0i}$, $\vec y = \vec R_{0j}$ and $\vec z=\vec R_{ij}=\vec y-\vec x$, and the minus sign on the right-hand side arises due to $\vec R_{ji} =-\vec R_{ij}$.
The three-body term of the cohesive energy per particle for a mono-atomic lattice $\Lambda$ is given by
\begin{equation}
    E_{\mathrm{coh}}^{(3)} = \frac{1}{6} \,\sideset{}{'} \sum_{\vec x,\vec y\in \Lambda} E^{(3)}_{\rm ATM}(\vec x,\vec y),
\end{equation}
where the prefactor $1/6$ avoids double counting and where the primed sum excludes the singularities originating from $\vec x=0$, $\vec y=0$, and $\vec x=\vec y$. 

We now extend our treatment to multi-atomic lattices $L$ with
\[
L=\bigcup_{j=1}^{N} (\Lambda+\vec w_j),
\]
with $\vec w_j\in E=A[-1/2,1/2)^d$ the (atomic) shift vectors within the unit cell, where we set $\vec w_1= 0$. First, the cohesive energy per particle admits the form
\[
E_{\mathrm{coh}}^{(3)} = \frac{1}{6N} \,\sideset{}{'}\sum_{\vec r_0\in L\cap E}\,\sideset{}{'}\sum_{\vec r_1,\vec r_2 \in L} E^{(3)}_\mathrm{ATM}(\vec r_1-\vec r_0,\vec r_2-\vec r_0).
\]
Notice that, in contrast to mono-atomic lattices, translational invariance only allows us to restrict summation of $\vec r_0$ to the elementary lattice cell, but not to eliminate the sum completely. Here, the prefactor $1/N$ ensures that the energy per particle is evaluated, not the energy per lattice cell. We then rewrite the sum as follows 
\[
E_\mathrm{coh}^{(3)}=\frac{\lambda}{6} \frac{1}{N} \,\sideset{}{'}\sum_{i,j,\ell=1}^N \sum_{\vec x,\vec y,\vec{z}\in \Lambda} \Bigg(\frac{1}{|\vec x_{ij}|^3|\vec y_{j\ell}|^3|\vec z_{\ell i}|^3}-3\frac{(\vec x_{ij}\cdot \vec y_{j\ell})(\vec y_{j\ell}\cdot \vec z_{\ell i})(\vec z_{\ell i}\cdot \vec x_{i j})}{|\vec x_{i j}|^5|\vec y_{j\ell}|^5|\vec z_{\ell i}|^5}\Bigg) \delta_{\vec x+\vec y+\vec z,0}
    \]
    with $\vec x_{ij}=\vec x-(\vec w_{i}-\vec w_j) $ and $\vec x,\vec y, \vec z\in \Lambda$ such that $\vec x+\vec y+\vec z=\vec 0$ and thus $\vec z=-(\vec x+\vec y)$.
    The equality constraint on $\vec z$ can now be encoded in integral form
    \[
    \delta_{\vec x+\vec y+\vec z, 0} = V_\Lambda \int_{E^\ast} e^{-2\pi i (\vec x+\vec y+\vec z)\cdot \vec k}\,\mathrm d\vec k= V_\Lambda \int_{E^\ast} e^{-2\pi i (\vec x_{ij}+\vec y_{j\ell}+\vec z_{\ell i})\cdot \vec k}\,\mathrm d\vec k,\quad \vec x,\vec y,\vec z \in \Lambda.
    \]
    where $E^\ast=(A^{-1})^\top[-1/2,1/2)^d$ defines the unit cell of the reciprocal lattice $\Lambda^\ast=(A^{-1})^\top\mathbb Z^d$ centered around the $\Gamma$-point. By Fubini's theorem, we can subsequently draw the sums inside the integral. The energy contribution of the isotropic part of the ATM potential can then be computed after defining the matrix of Epstein zeta functions
    \[
    (\mathcal Z_{\Lambda,\nu})_{ij}(\vec k) = e^{2\pi i \vec k\cdot(\vec w_i-\vec w_j)}  Z_{\Lambda,\nu}(\vec w_{i}-\vec w_j, 
        \vec k),\quad i,j=1,\dots,N,
    \]
    and we have
    \[
    \frac{\lambda}{6N} V_\Lambda\int_{E^\ast} \,\sideset{}{'}\sum_{i,j,\ell=1}^N \sum_{\vec x,\vec y,\vec{z}\in \Lambda} \frac{e^{-2\pi i (\vec x_{ij}+\vec y_{j\ell}+\vec z_{\ell i})\cdot \vec k}}{|\vec x_{ij}|^3|\vec y_{j\ell}|^3|\vec z_{\ell i}|^3}\,\mathrm d \vec k 
    = \frac{\lambda}{6} \frac{1}{N} V_\Lambda\int_{E^\ast}  \mathrm{Tr}(\mathcal Z_{\Lambda,3}^3(\vec k))\,\mathrm d \vec k,
    \]
    where the exponent inside the trace is to be understood as the corresponding matrix power. 

The anisotropic part of the ATM potential can subsequently be expressed in terms of anisotropic Epstein zeta functions, the recent generalization of the Epstein zeta function to anisotropic interaction kernels.\cite{buchheitComputationAnisotropicSingular2026}
Consider a $d$-dimensional multi-index $\vec \gamma =((\vec\gamma)_1,\ldots,(\vec\gamma)_d)^\top\in\mathbb N^d$, where $\mathbb N=\{0,1,\ldots\}$ are the natural numbers, an exponent $\nu>d+|\vec \gamma|$, 
where $|\vec \gamma|=(\vec\gamma)_1+\ldots+(\vec\gamma)_d$ denotes the total order of the multi-index,
a shift vector $\vec v$ and a wave vector $\vec k$.
Then the anisotropic Epstein zeta function reads\cite{buchheitComputationAnisotropicSingular2026}
\begin{equation}
Z_{\Lambda,\nu,\vec\gamma}(\vec v,\vec k) = \,\sideset{}{'}\sum_{\vec z\in \Lambda}
(\vec{z}-\vec{v})^{\vec{\gamma}}
\frac{e^{-2\pi i \vec z \cdot \vec k}}{\vert \vec z-\vec v\vert^\nu}
,
\end{equation}
where we employ the multi-index notation for the product of powers
$\vec{z}^{\vec{\gamma}}=\prod_{\alpha=1}^d(\vec z)_{\alpha}^{(\vec \gamma)_{\alpha}}$
and
where the lattice sum can be meromorphically continued to $\nu\in \mathbb C$.
We now define the tensor
    \[
(\mathcal D_{\Lambda,\nu})_{ij}^{\alpha \beta} = 
e^{2\pi i \vec k\cdot(\vec w_i-\vec w_j)}
Z_{\Lambda,\nu,\vec e_{\alpha}+\vec e_{\beta}}(\vec w_i-\vec w_j,\vec k)
,\quad  \alpha,\beta=1,\dots,d,
    \]
    so that 
    \[
    (\mathcal D_{\Lambda,\nu})_{ij}^{\alpha \beta}(\vec k) = \sideset{}{'}\sum_{\vec z\in \Lambda} \frac{(\vec z_{ij})_\alpha (\vec z_{ij})_\beta}{\vert \vec z_{ij}\vert^\nu} e^{-2\pi i \vec z_{ij}\cdot \vec k},
    \qquad \mathrm{Re}(\nu)>d+2
    \]
where
$\vec e_{\alpha}$ is the $\alpha$-th canonical unit vector with components $(\vec e_{\alpha})_{\beta}=\delta_{\alpha\beta}$, $\alpha,\beta=1,\ldots,d$.
    Now notice that the anisotropic components in the ATM sum admit the form of a trace of matrix products,
    \[
    \frac{(\vec{x}\cdot \vec{y})(\vec{y}\cdot \vec{z})(\vec{z}\cdot \vec{x})}{\vert \vec{x} \vert^5 \vert  \vec{y}\vert^5 \vert \vec{z} \vert^5}=\sum_{\alpha,\beta,\tau=1}^d\frac{(x_\alpha y_\alpha) (y_\beta z_\beta)(z_\tau x_\tau)}{\vert \vec{x} \vert^5 \vert  \vec{y}\vert^5 \vert \vec{z} \vert^5}=\sum_{\alpha,\beta,\tau=1}^d \frac{(x_\tau x_\alpha)}{\vert \vec x\vert^5} \frac{(y_\alpha y_\beta)}{\vert \vec y\vert^5}\frac{(z_\beta z_\tau)}{\vert \vec z\vert^5}.
    \]
    In total, we find that the anisotropic contribution to the ATM cohesive energy per particle admits the tensor representation,
    \[
    \frac{\lambda}{6N} V_\Lambda\int_{E^\ast} \Big(-3\, \mathrm{Tr}(\mathcal D_{\Lambda,5}^3(\vec k))\Big)\,\mathrm d \vec k,
    \]
    where we understand the exponent within the trace as the tensor contraction
    \[
    (\mathcal D_{\Lambda,\nu} \mathcal D_{\Lambda,\nu} )^{\alpha \tau}_{i\ell} = \sum_{\beta=1}^d \sum_{j=1}^N (\mathcal D_{\Lambda,\nu})^{\alpha \beta}_{i j} (\mathcal D_{\Lambda,\nu})^{\beta \tau }_{j \ell},
    \]
    and the trace is taken with respect to all tensor dimensions except $\vec k$, i.e.,
    \[
    \mathrm{Tr}(\mathcal D_{\Lambda,\nu})=\sum_{\alpha=1}^d \sum_{i=1}^N (\mathcal D_{\Lambda,\nu})^{\alpha \alpha}_{i i}.
    \]
    In total, the ATM cohesive energy for a general $N$-atomic lattice admits the efficiently computable tensor trace integral representation 
    \begin{equation}
    \label{eq:ecoh_final}
    E_\mathrm{coh}^{(3)} = \frac{\lambda}{6N} V_\Lambda\int_{E^\ast} \Big(\mathrm{Tr}(\mathcal Z_{\Lambda,3}^3(\vec k))-3\, \mathrm{Tr}(\mathcal D_{\Lambda,5}^3(\vec k))\Big)\,\mathrm d \vec k.
    \end{equation}

The singularity of the integrand at $\vec k=0$ can now be treated in analogous fashion as in our previous work.\cite{buchheitComputationAnisotropicSingular2026}
We can separate the anisotropic Epstein zeta function into an analytic function and a singularity as follows
\begin{equation}
\label{eq:epstein_decomposition}
Z_{\Lambda,\nu,\vec\gamma}(\vec v, \vec k) = e^{-2\pi i \vec v\cdot \vec k}\left(Z_{\Lambda,\nu,\vec \gamma}^{\mathrm{reg}}(\vec v, \vec k) +\frac{1}{V_\Lambda}
\frac{
\hat s^{(\vec \gamma)}_{\nu}(\vec{k})
}{
(-2\pi i)^{|\vec \gamma|}
}
\right)
,\qquad 
\vec k \neq \vec 0
\end{equation}
and continuously extend to $\vec k = \vec 0$,
where the regularized anisotropic Epstein zeta function $Z_{\Lambda,\nu,\vec \gamma}^{\mathrm{reg}}(\vec v,\vec k)$ is,
as a function of $\vec k$, analytic in the reciprocal unit cell.\cite[Th. 2.22]{buchheitComputationAnisotropicSingular2026} The function $\hat s^{(\vec{\gamma})}_\nu(\vec{k})$ can be understood as the $\vec \gamma$-derivatives of the distributional Fourier transform
$\hat s_\nu(\vec{k})$ of $\vert \vec{z} \vert^{-\nu}$.
The latter is defined as
\begin{align*}
    \hat s_\nu(\vec{k}) =  \frac{\pi^{\nu-d/2}}{\Gamma(\nu/2)}\Gamma\big((d-\nu)/2\big)  \vert \vec{k}\vert ^{\nu - d},\quad \nu \not\in (d+2\mathbb N).
\end{align*}
In case that $\nu =d+2n$, $n\in \mathbb N$, the Fourier transform is only uniquely defined up to a polynomial of degree $2n$. We adopt the choice
\[
\hat s_{d+2n}(\vec{k}) =  \frac{\pi^{n+d/2}}{\Gamma(n+d/2)}\frac{(-1)^{n+1}}{n!} ( \pi |\vec{k}|^2 )^{n} \log (\pi  |\vec{k}|^{2}).
\]
Hence, the anisotropic Epstein zeta function, that coincides with the Epstein zeta function for $\vec \gamma = \vec 0$, equals the sum of an analytic function and a power-law or logarithmic singularity. 
Therefore, the integral can be efficiently computed using either a specialized Gauss-Legendre quadrature or a Duffy transformation.\cite{duffy1982quadrature} The implementation was checked against calculations using the alternative of Bessel function expansions of the lattice sums as detailed in Ref.~\onlinecite{Schwerdtfeger-2026}. Both methods lead to identical results within numerical accuracy.

\newpage
\bibliography{references}

\end{document}